\documentclass[twocolumn,floatfix,showpacs,prd,aps,tightenlines,letterpaper,superscriptaddress]{revtex4-1}
\usepackage{amsmath}
\usepackage{graphicx}
\usepackage{psfrag}
\usepackage{verbatim} 
\usepackage[usenames]{color}

\usepackage{dcolumn}
\usepackage{bm}
\usepackage{longtable}
\usepackage[mathscr]{eucal}
\usepackage{mathrsfs}
\usepackage{xspace}
\usepackage{leftidx}
\def\msun{\rm M_{\odot}}
\def\kms{\rm km \, s^{-1}}

\def\simlt{\mathrel{\rlap{\lower 3pt\hbox{$\sim$}}\raise 2.0pt\hbox{$<$}}}
\def\simgt{\mathrel{\rlap{\lower 3pt\hbox{$\sim$}} \raise 2.0pt\hbox{$>$}}}
\def\lsim{\mathrel{\rlap{\lower 3pt\hbox{$\sim$}}\raise 2.0pt\hbox{$<$}}}
\def\gsim{\mathrel{\rlap{\lower 3pt\hbox{$\sim$}} \raise 2.0pt\hbox{$>$}}}

\def\msunpc3{\msun~{\rm {pc^{-3}}}}
\newcommand{\be}{\begin{equation}}
\newcommand{\ee}{\end{equation}}
\def\kms{{\rm\,km\,s^{-1}}}

\newcommand{\bea}{\begin{eqnarray}}
\newcommand{\eea}{\end{eqnarray}}
\newcommand{\beq}{\begin{equation}}
\newcommand{\eeq}{\end{equation}}

\newcommand{\analytic}{analytic\xspace} 

\usepackage{hyperref}
\hypersetup{
    pdftitle={},
    colorlinks=false,           
}

\begin{document}

\def\fun#1#2{\lower3.6pt\vbox{\baselineskip0pt\lineskip.9pt
  \ialign{$\mathsurround=0pt#1\hfil##\hfil$\crcr#2\crcr\sim\crcr}}}
\def\lap{\mathrel{\mathpalette\fun <}}
\def\gap{\mathrel{\mathpalette\fun >}}
\def\kms{{\rm km\ s}^{-1}}
\def\vk{V_{\rm recoil}}

\title{Comparing an analytical spacetime metric for a
merging binary to a fully nonlinear numerical evolution using
curvature scalars} 

\author{Jam~Sadiq}
\affiliation{Center for Computational Relativity and Gravitation, 
Rochester Institute of Technology, Rochester, NY 14623, USA}

\author{Yosef~Zlochower}
\affiliation{Center for Computational Relativity and Gravitation, 
Rochester Institute of Technology, Rochester, NY 14623, USA}

\author{Hiroyuki~Nakano}
\affiliation{Faculty of Law, Ryukoku University, Kyoto 612-8577, Japan}

\begin{abstract}
We introduce a new geometrically invariant prescription for comparing
two different spacetimes based on geodesic deviation. We use this
method to compare a family of recently introduced analytical spacetime
representing inspiraling black-hole binaries to fully nonlinear
numerical solutions to the Einstein equations. Our method can be used
to improve analytical spacetime models by providing a local measure of
the effects that violations of the Einstein equations will have on
timelike geodesics, and indirectly, gas dynamics. We also discuss the
advantages and limitations of this method.
\end{abstract}

 \maketitle

\section{Introduction}\label{sec:Introduction}

Ever since the breakthroughs in numerical relativity in the early
2000s~\cite{Pretorius:2005gq, Campanelli:2005dd, Baker:2005vv}, it has
been possible to simulate black-hole binaries (BHBs) for from the rapid
inspiral phase, through the plunge and merger. 
Modern numerical relativity codes are now capable of simulating
inspiraling BHBs for over 100
orbits~\cite{Szilagyi:2015rwa}. These simulations are the most
accurate known means of generating the gravitational waveform from
such mergers. However, they are also 
computationally expensive. Recently, a family of analytic metric representing
the inspiral phase of a BHB was proposed and used
extensively to study accretion physics~\cite{Noble:2012xz,
Gallouin:2012kb, Mundim:2013vca, Zilhao:2014ida, Zlochower:2015baa,
Ireland:2015cjj, Nakano:2016klh, Bowen:2016mci, Bowen:2017oot}.

In this paper, we introduce a new technique to study the accuracy of
this family BHB spacetimes by comparing them to full numerical
evolutions starting from a set of fiducial separations. Our method is
based on analyzing a set of scalars related to geodesic deviation.
This study complements previous studies in Ref.~\cite{Zilhao:2014ida},
where the hydrodynamics and magnetohydrodynamics of accreting gas were
compared between versions of the analytical spacetime at different
approximation orders.

In this paper, we express tensors in both the more conventional
coordinate basis and in orthonormal bases. Latin indices near the
beginning of the alphabet are abstract tensor
indices~\cite{Wald:1984rg}, which indicate the type of tensors
involved in a calculation, as well as contraction. Latin indices near
the end of the alphabet denote coordinate-basis components of spatial
tensors, while Greek letters denote 4-dimensional spacetime components
in the coordinate basis. Components of tensors in an orthonormal basis
(the first element of the orthonormal basis is always timelike) are
denoted by a Greek or Latin letter surrounded by square braces.
Whether associated with coordinate bases or orthonormal bases, Greek
indices range from $0$ to $3$, while Latin indices  near the end of
the alphabet range from $1$ to $3$.  In this paper, we use the
geometric unit system, where $G = c = 1$.

This paper is organized as follows. In Sec.~\ref{sec:techniques}, we
describe the analytical and numerical techniques used in this paper.
In Sec.~\ref{sec:code_verification}, we present the tests we used to
confirm the accuracy of our results. In Sec.~\ref{sec:results}, we
describe the results of our study. In Sec.~\ref{sec:discussion} we
discuss the consequences and limitations of our study.

\section{Techniques}\label{sec:techniques}

\subsection{Analytic Black-Hole Binary
Spacetime}\label{sec:analytic_metric}

In this paper, the analytic metric we consider represents a
nonspinning, equal-mass
BHB in a quasicircular inspiral. This spacetime was first
constructed in Ref.~\cite{Mundim:2013vca} based on earlier work on binary initial
data in Refs.~\cite{Yunes:2005nn, Yunes:2006iw, JohnsonMcDaniel:2009dq}.

The analytic spacetime is constructed
by asymptotically matching metrics in three different zones
characterizing three different spacetime regions of validity for
different analytic metrics: (i) a far zone (FZ) where the spacetime
can be described by a two-body perturbed flat spacetime with outgoing
gravitational radiation and where retardation effects are fully
accounted for; (ii) a near zone (NZ) which is less than one GW length
from the center of mass of the binary [but not too close to each black hole (BH)]
that is described by a post-Newtonian metric
(this includes retardation effects at
a perturbative level and binding interactions between the two BHs);
and (iii) inner zones (IZs) that are described by perturbed
Schwarzschild (or Kerr) BHs. The full spacetime is then constructed
by smoothly transitioning from zone to zone in the so-called buffer
zones (BZs).

\begin{figure}
  \includegraphics[width=\columnwidth]{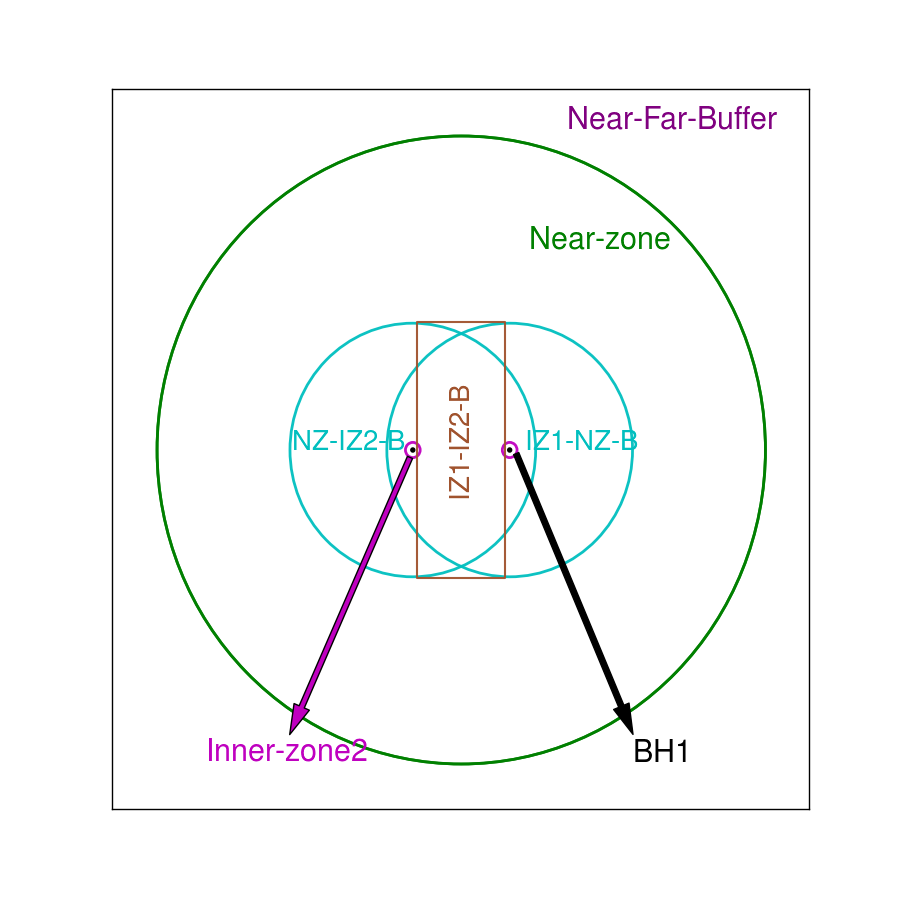}
  \caption{The zones for the analytic metric. The large (green) circle
    is the outer boundary of the near zone. Immediately inside this
    circle the metric is exclusively the post-Newtonian near-zone metric, while
    outside, it is a superposition of the near and far zone
    metrics. All points in the figure
    outside this circle are in the near-far buffer zone (the other
    boundary of this zone is not show). The smaller (cyan) circles
    denote the inner boundary of the near zone. Inside the envelope of
    these circles is the near-inner buffer zones.  The box (orange)
    denotes the region inside the near-inner buffer zone where the
    metric is a superposition of both BH1 and BH2 inner zones, as well
    as the near zone. Outside the box, the metric is a superposition
    of the near zone metric and either one of the inner zone metrics.
    Finally inside the very small (magenta) circles
    are the two inner zones, where the metric is purely
    the inner zone perturbed Schwarzschild metric.}
  \label{fig:regions}
\end{figure}

Figure~\ref{fig:regions} shows where these regions are located with
respect to the two BHs.

\subsection{Geodesic Analysis}

The primary analysis in this paper concerns how the fully nonlinear
evolution of initial data based on the \analytic metric differs from
the \analytic metric itself at some later time. In order to do this, we need
gauge invariant measurements that can elucidate to what degree two
spacetimes are locally similar.

To be precise, on some fiducial spatial slice $\Sigma_0$, which
corresponds to a surface of constant coordinate time $t=t_0$, the induced
metric and extrinsic curvatures of the \analytic metric are used as
initial data for a CCZ4 evolution. Critical to our analysis, on
$\Sigma_0$, the \analytic and numerically evolved metrics are
identical. Furthermore, if the \analytic metric solved the vacuum
Einstein equations, up to
truncation error, the numerical and \analytic metrics would only
differ by a gauge transformation at all later times (at least in the
domain of dependence on the initial numerical slice, which will be of
finite extent).

The fact that gauges are identical on $\Sigma_0$ allows us to use
geodesic dynamics to explore how the numerical and \analytic
spacetimes begin to differ with time. In particular, if we take as
initial data for a timelike geodesic some given coordinate position 
and the spatial projection of the 4-velocity, $V^a$ (from which we can
reconstruct the full 4-velocity $u^a$ at $t_0$ via $u^a = \sqrt{1 +
\gamma_{ij} V^i V^j} n^a + V^a$, where $V^0=0$ and $n^a$ is the unit
norm to $\Sigma_0$), and if the \analytic metric solved the
vacuum Einstein equations,
the resulting geodesic, as calculated on the two metrics, would be
geometrically identical. By this, we mean that the two geodesics would
only differ by a gauge transformation. The question remains though,
how do we show that geodesics in two different gauges are identical or
not if the gauge transformation is unknown?

To address this question, we consider measuring curvature scalars
along each geodesic as a function of proper time. Our construction of
these scalars is as follows.

Let $u^a(\tau)$ be the 4-velocity associated with a geodesic (and
hence unit norm). At
each point along the geodesic construct an orthonormal basis
$\{e^a_{[0]},\, e^a_{[1]},\, e^a_{[2]},\, e^a_{[3]}\}$, where
$ e^a_{[0]} = u^a(\tau)$ and
$e^a_{[\mu]} e^b_{[\nu]} g_{ab} = \eta_{[\mu][\nu]}$. The choice of components
1, 2, and 3 of this basis is arbitrary. Given any such basis, we can
define a $3\times3$ symmetric matrix of scalars ${\bf M}$, where
\begin{eqnarray}
{\bf M} = \left(\begin{array}{ccc}
M_{[1][1]} & M_{[1][2]}  & M_{[1][3]} \\
M_{[2][1]} & M_{[2][2]}  & M_{[2][3]} \\
M_{[3][1]} & M_{[3][2]}  & M_{[3][3]} \\
\end{array}\right) \,,
\label{eq:scalar_matrix}
\end{eqnarray}
and
\begin{equation}
  M_{[i][j]} = R_{a b c d} u^a e^b_{[i]} u^c e^d_{[j]}, \quad i,j = 1,2,3 \,.
\end{equation}
Importantly, the eigenvalues of this matrix are independent of how
$e^a_{[1]}$, $e^a_{[2]}$, and $e^a_{[3]}$ are constructed. This
follows because any two choices $(e^a_{[1]},\, e^a_{[2]},\, e^a_{[3]})$
only differ by an orthogonal transformation (which preserves
eigenvalues).
Consequently, if the \analytic metric and numerically evolved metric
represented the same spacetime, the eigenvalues of ${\bf M}$
constructed this way on each metric would be
identical. We will refer to these eigenvalues as {\it curvature
eigenvalues} in the sections below.

Of course, the \analytic metric, being an approximate solution,
violates the vacuum field equations to
some degree (see Refs.~\cite{Mundim:2013vca, Zilhao:2014ida} for a detailed
analysis). Thus the \analytic metric and its numerical evolution will
differ to some level. Our goal here is to demonstrate a local measure
of how the two spacetimes actually differ. To do this, we note that
the elements of $M_{[i][j]}$ have the interpretation of being the
(negative of the) acceleration of deviation vector $e^a_{[i]}$ along the direction
$e^b_{[j]}$. We can thus interpret relative differences in the
curvature eigenvalues of ${\bf M}$ as proxies for the relative differences in
the effective potentials experienced by timelike geodesics traversing
these two spacetimes.

One important limitation of our procedure is that because slightly
different geodesics can, in principle, follow very different
trajectories on secular timescales, our analysis will need to be done
at when the geodesics are relatively close to $\Sigma_0$. Otherwise, a
small difference in the two spacetimes may incorrectly be interpreted
as a large difference. We ameliorate this problem by only choosing
geodesics that are {\it stable}. By this, we mean that the trajectories are
largely insensitive to small perturbations of the initial velocity. As
such, we do not include cases where small perturbations lead to the
geodesic orbiting a different black hole, or ones where small
perturbations lead to geodesics falling into either black hole. 

We also note that our analysis here can be extended in a
straightforward manner to include all 20 independent components of the
Riemann tensor. To do this, we would need the vectors $e^a_{[i]}$
($i=1,\,2,\,3$) to obey $u^b\nabla_b e^a_{[i]} =0$. That is, evolve the
entire basis. Under this extended construction, all components of
\begin{equation}
  R_{[\mu][\nu][\rho][\sigma]} = R_{a b c d} e^a_{[\mu]}
	e^b_{[\nu]}e^c_{[\rho]}e^d_{[\sigma]}
\end{equation}
are gauge invariant. We can thus compare each component
as constructed on the \analytic and numerical spacetimes. We leave
this analysis for a later work.

\subsection{Reconstructing the 4-dimensional Riemann tensor}

As is done by in many numerical relativity codes, our numerical
evolutions uses the standard 3+1
Arnowitt-Deser-Misner~\cite{Arnowitt:1962hi} split of the Einstein
equations. In this paper, we will need to reconstruct the full
4-dimensional Riemann tensor from the three dimensional quantities
evolved by our code. In this section, we provide the details of how
this is accomplished.
In order to avoid confusion, we will use the notation $(3)$ and $(4)$
to indicate a three or four dimensional tensors, respectively.

In the standard 3+1 split, the metric on a spatial slice (given by
$t={\rm const}$) is obtained from the full 4-dimensional metric via
\begin{equation}
  \gamma_{\mu \nu} = g_{\mu \nu} + n_{\mu} n_{\nu} \,,
\end{equation}
where $n^{\mu}$ is the unit norm to the spatial
hypersurface and the spatial components of this tensor (i.e., indices 1
through 3) form the 3-dimensional metric tensor.
Note that while $\gamma_{ij} = g_{ij}$,  $\gamma^{ij} \neq
g^{ij}$. The full 4-dimensional tensor $\gamma_{\mu \nu}$
also serves as a projection operator which
takes four-dimensional tensors to three-dimensional ones. To avoid
confusion, we will use $P_{\mu \nu} = \gamma_{\mu \nu}$ to denote the
projection tensor.

In order to reconstruct the 4-dimensional Riemann tensor,
${}^{(4)}R_{\mu\nu\epsilon\delta}$,
we follow Ref.~\cite{Baumgarte:2010ndz} and write it as
\begin{eqnarray}
  {}^{(4)}R_{\mu\nu\epsilon\delta} &=& P_\mu{}^\zeta
  P_\nu{}^\tau   P_\epsilon{}^\kappa   P_\delta{}^\sigma
  \ {}^{(4)}R_{\zeta\tau\kappa\sigma} 
\cr &&
  - 2 P_\mu{}^\zeta  P_\nu{}^\tau
  P_{[\epsilon}{}^\kappa n_{\delta]} n^\sigma \ {}^{(4)}R_{\zeta\tau\kappa\sigma} 
\cr &&
  - 2 P_\epsilon{}^\zeta  P_\delta{}^\tau
  P_{[\mu}{}^\kappa n_{\nu]} n^\sigma \ {}^{(4)}R_{\zeta\tau\kappa\sigma} 
\cr &&
  + 2 P_\mu{}^\zeta P_{[\epsilon}{}^\kappa
  n_{\delta]}n_\nu n^\tau n^\sigma \ {}^{(4)}R_{\zeta\tau\kappa\sigma} 
\cr &&
  - 2 P_\nu{}^\zeta P_{[\epsilon}{}^\kappa
  n_{\delta]}n_\mu n^\tau n^\sigma \ {}^{(4)}R_{\zeta\tau\kappa\sigma}  \,,
\end{eqnarray}
where
\begin{eqnarray}
  P^\zeta{}_\mu  P^\eta{}_\nu   P^\kappa{}_\epsilon   P^\sigma{}_\delta \
  {}^{(4)} R_{\zeta \eta\kappa\sigma}
  &= & {}^{(3)}\!R_{\mu\nu\epsilon\delta}\nonumber \\ &&
  +K_{\mu\epsilon} K_{\nu\delta} -
  K_{\mu\delta}K_{\epsilon\nu} \,, \label{eq:ppppR}\\
P^\sigma{}_\mu  P^\eta{}_\nu   P^\kappa{}_\delta  n^\zeta R_{\sigma
\eta \kappa \zeta} &=& D_\nu K_{\mu\delta}  -
  D_\mu K_{\nu\delta} \,, \label{eq:pppnR}\\
P^\zeta{}_\mu P^\kappa{}_\nu  n^\delta n^\epsilon R_{\delta\kappa
\epsilon\zeta} &=& \mathcal{L}_n K_{\mu \nu}
  +\frac{1}{\alpha} D_\mu D_\nu\alpha \nonumber \\&& +K^\epsilon{}_\nu
  K_{\mu \epsilon} \,,
\label{eq:pnpnR}
\end{eqnarray}
and $D_i$ is the covariant derivative associated with $\gamma_{ij}$,
$\alpha$ is the lapse, and $K_{ij}$ is the extrinsic curvature.

Note that the left-hand sides of Eqs.~(\ref{eq:ppppR})--(\ref{eq:pnpnR}) are all
naturally defined in terms of 3-dimensional tensors. To construct a
4-dimension tensor from a 3-dimension tensor $T^{i_1 i_2\cdots}_{j_1
j_2\cdots}$, we use the following operator,
\begin{eqnarray}
  T^{\mu_1 \mu_2\cdots}_{\nu_1 \nu_2\cdots} = {\Lambda^{\mu_1}}_{i_1}
  {\Lambda^{\mu_2}}_{i_2} \cdots {\Lambda_{\nu_1}}^{j_1}
  {\Lambda_{\nu_2}}^{j_2}\cdots T^{i_1 i_2\cdots}_{j_1 j_2 \cdots} \,,
\end{eqnarray}
where
\begin{equation} 
  {\Lambda_\mu}^i  = \left (\begin{array}{ccc }
     \beta^1 &\beta^2  & \beta^3 \\
       1     &   0     &   0     \\
       0     &   1     &   0     \\
       0     &   0     &   1     \end{array} \right) \,,
\end{equation}
where $\beta^i$ is the shift,
and
\begin{equation} 
  {\Lambda^\mu}_i  = \left (\begin{array}{ccc }
     0 & 0  & 0 \\
       1     &   0     &   0     \\
       0     &   1     &   0     \\
       0     &   0     &   1     \end{array} \right) \,.
\end{equation}

Finally, the left-hand side of Eq.~(\ref{eq:pnpnR}) is evaluated by
assuming the standard ADM vacuum evolution equations are obeyed. That is,
\begin{eqnarray}
 \mathcal{L}_n K_{\mu \nu} &=&
\frac{1}{\alpha}(\mathcal{L}_t K_{\mu \nu} - \mathcal{L}_{\beta}
K_{\mu
\nu} ) \,,
\cr
 \mathcal{L}_t K_{\mu \nu} &=&
-D_\mu D_ \nu \alpha + \alpha ({}^{(3)}\!R_{\mu \nu} -2 K^\kappa{}_ \nu
K_{\mu \kappa}+
K K_{\mu \nu}) \cr
&&
-8\pi\alpha ({S_{\mu \nu} -\frac{1}{2} \gamma_{\mu \nu} (S -\rho))}  +
\mathcal{L}_{\beta} K_{\mu \nu} \,,\label{eq:ADM_EV}
\end{eqnarray}
where $S_{\mu \nu}=\gamma_\mu{}^\kappa\gamma_ \nu{}^ \sigma T_{\kappa
\sigma}$,
$S=S^\mu{}_\mu$, and $\rho=n^\mu n^\nu T_{\mu \nu}$ are all assumed to be zero.

Since we actually evolve the metric using the CCZ4
system~\cite{Alic:2011gg}, the actual
form of the evolution equation for the extrinsic curvature is
\begin{eqnarray}
\mathcal{L}_t K_{\mu \nu} &=&
-D_\mu D_ \nu \alpha + \alpha ({}^{(3)}\!R_{\mu \nu} -2 K^\kappa{}_ \nu
K_{\mu \kappa}+
K K_{\mu \nu}) \cr
&&
+ \alpha ( D_{\mu}Z_{\nu} +D_{\nu}Z_{\mu})
\cr &&
- \left(2 \alpha K_{\mu \nu}
+ \alpha  \gamma_{\mu \nu}\frac{1}{\phi^2}\kappa_1 (1 + \kappa_2) \right)\Theta
\cr &&
+ \mathcal{L}_{\beta} K_{\mu \nu} \,, 
\end{eqnarray}
where $\Theta$ and  $Z_{i}$ denote deviations from the Einstein equations,
and constants $\kappa_i$ are free parameters.
Thus using Eq.~(\ref{eq:ADM_EV}) is equivalent to assuming $\Theta$
and $Z_i$ are zero. At $t=0$ this is the case, and both variables
remain small due to the constraint damping of the CCZ4 system. In
order to make our code more general, we assume Eq.~(\ref{eq:ADM_EV}),
which means that it can be used equally well with a BSSN, CCZ4, or
other 3+1 evolution system.

To reconstruct ${}^{(4)}R_{\mu \nu \kappa \sigma}$, we interpolate
$\gamma_{ij}$, $\partial_k \gamma_{ij}$, the 3-dimensional Ricci tensor
${}^{(3)}R_{ij}$, $K_{ij}$,
$\partial_k K_{ij}$, $\alpha$, $\beta^i$, and $\partial_j \beta^i$
along each geodesic. Note that we do not need second derivatives of
the lapse because the $D_\mu D_ \nu \alpha$ cancel out. From these
quantities, we can reconstruct all terms in
Eqs.~(\ref{eq:ppppR})--(\ref{eq:pnpnR}). Note that the 3-dimensional
Riemann tensor can be reconstructed directly from the 3-dimensional
Ricci tensor.

To compute the Riemann tensor for the \analytic spacetime, we use an
eighth-order finite differencing algorithm and directly differentiate
the 4-dimensional metric.

\subsection{Numerical Evolutions}

We first explored evolving the \analytic metric using the fully
nonlinear numerical relativity codes in Ref.~\cite{Zlochower:2015baa}. We
use an identical procedure here, which we summarize below.

We evolved the BHB initial data using the {\sc
LazEv}~\cite{Zlochower:2005bj} implementation of the moving puncture
approach~\cite{Campanelli:2005dd, Baker:2005vv} with the conformal
function $W=\sqrt{\chi}=\exp(-2\phi)$ suggested by
Ref.~\cite{Marronetti:2007wz} and the Z4~\cite{Bona:2003fj,
Bernuzzi:2009ex, Alic:2011gg} and BSSN~\cite{Nakamura87, Shibata95,
Baumgarte99} evolution systems. 
Here, we use the conformal covariant Z4 (CCZ4) implementation of
Ref.~\cite{Alic:2011gg}.
Note that the same technique has been recently applied to the evolution
of binary neutron stars~\cite{Kastaun:2013mv, Alic:2013xsa}.
For the CCZ4 system, we again
used the conformal factor $W$. We used centered eighth-order finite
differencing for all spatial derivatives, a fourth-order Runge-Kutta
time integrator, and both fifth- and seventh-order Kreiss-Oliger
dissipation~\cite{Kreiss73}.

Our code uses the {\sc EinsteinToolkit}~\cite{Loffler:2011ay,
Moesta:2013dna,
einsteintoolkit} / {\sc Cactus}~\cite{cactus_web} / {\sc
Carpet}~\cite{Schnetter-etal-03b, carpet_web}
 infrastructure.  The {\sc Carpet}
mesh refinement driver provides a ``moving boxes'' style of adaptive mesh
refinement (AMR). In this approach, refined grids of fixed size are arranged
about the coordinate centers of both holes.  The {\sc Carpet} code
then moves these fine grids about the computational domain by
following the trajectories of the two BHs.

To obtain initial data, we use eighth-order finite differencing of the
analytic global metric to obtain the 4-metric and all its first derivatives at
every point on our simulation grid. The finite differencing of the
global metric is constructed so that the truncation error is
negligible compared to the subsequent truncation errors in the full
numerical simulation (here we used finite difference step size of
$10^{-4}$, which is 90 times smaller than our smallest grid size in any
of the numerical simulations discussed below). We then reconstruct the
spatial 3-metric $\gamma_{ij}$ and extrinsic curvature $K_{ij}$ from
the global metric data. Note that with the exception of the
calculation of the extrinsic curvature, we do not use the global 
metric's lapse and shift.  In order to evolve these data, we need to
remove the singularity at the two BH centers. Unlike in the puncture
formalism~\cite{Brandt97b}, the singularities here are true curvature
singularities. We {\it stuff}~\cite{Etienne:2007hr, Brown:2007pg,
Brown:2008sb} the BH interiors in order to remove the singularity.
Our procedure is to replace the singular metric well inside the
horizons with nonsingular (but constraint violating) data through the
transformations,
\begin{eqnarray}
&& \gamma_{ij} \to f(r)\ \gamma_{ij} \,, \quad i\neq j \,,
\cr
&& \gamma_{ii} \to f(r)\ \gamma_{ii} + (1-f(r)) \Xi \,,
\cr
&& K_{ij} \to f(r)\ K_{ij} \,, 
\end{eqnarray}
where
\begin{equation} 
f(r) =
 \left\{\begin{array}{lr}
  0 \,, & r < r_{\rm min} \\
  1 \,, & r > r_{\rm max} \\
  P(r) \,, & r_{\rm min} \leq r \leq r_{\rm max}
 \end{array} \right. \,.
\end{equation}
Here, $r$ is the distance to a BH center, and $P(r)$ is
a fifth-order polynomial that obeys $P(r_{\rm min}) = P'(r_{\rm
min})=P''(r_{\rm min})=0$, $P(r_{\rm max}) = 1$, $P'(r_{\rm max})
=P''(r_{\rm max})=0$, and $\Xi$ is a large number. The resulting
data are therefore $C^2$ globally. The parameters $r_{\rm min}$,
$r_{\rm max}$, and $\Xi$ are chosen such that both transitions occur
inside the BHs and so that $W$ varies smoothly with negligible
shoulders in the transition region and is small at the centers.

The grid structure for the runs below consisted of a course grid
extending to $(x,\,y,\,z) = \pm (3200,\,3200,\,3200)M$ (we exploited both the
$z$-reflection and $\pi$-rotational symmetry of the data in order to
reduce the computational volume by a factor of 4. We used 12 levels of
mesh refinement. In the sections below, we indicate the global
resolution of each simulation by indicating the number of points on
the coarsest grid from the origin to each outer face. That is, a
resolution of $N=100$ indicated that the coarsest grid spacing is
$3200M/100 = 32M$. The resolution was always set to be the same in
each direction.

To evolve timelike geodesics in the numerical spacetime, we use the
following algorithm. The 4-velocity of each geodesic is decomposed
into a component tangent to the unit norm $n^a$ and a spatial
component $V^a$. That is,
\begin{equation}
  u^\mu = \varpi n^\mu + V^\mu \,,
\end{equation}
where $\varpi = \sqrt{1+V^i V^j \gamma_{ij}}$ and
$V^0=0$~\cite{Hughes:1994ea} (note that
$V_i = u_i$ $[i=1,2,3]$). The geodesic equation then gives
\begin{eqnarray}
  \frac{dx^i}{dt} &=& - \beta^i + \frac{\alpha}{\varpi}\, V^i \,,
  \nonumber \\
  \frac{d\tau}{dt} &=& \frac{\alpha}{\varpi} \,, \nonumber \\
  \frac{dV_i}{dt} &=& -\varpi \alpha_{,i} -V_j {\beta^j}_{,i} +
  \frac{1}{2} V^jV^k{\gamma_{jk}}_{,i} \,.\label{eq:geo}
\end{eqnarray}
This form of the geodesic equation has the advantage that explicit time
derivatives of the 4-metric are not needed for the evolution and the
integration variable is $t$, which is the time coordinate used in the
code. We evolve the geodesics using the same RK4 time integrator
used to evolve the metric itself.

Since we evolve these geodesics with an adaptive-mesh code, there are
complication associated with geodesics crossing refinement level
boundaries. Our algorithm is as follows. The AMR grid is distributed
such that on a given refinement level, a single CPU will only {\it
own} a single Cartesian box. We then search for the finest resolution
box that contains that geodesic and assign the evolution of the
geodesics (at that time step) to that processor. For our purposes, a
geodesic is only contained in a given box if all points used by the
interpolation stencil are in that box (excluding buffer zones, but
including ghost zones). If a geodesic is too close to buffer zones,
then it will be evolved using the next coarsest level.

A geodesic that crosses from a coarse refinement level to a finer one
may actually be ahead, in time, of the rest of the fields on that
refinement level. In such a case,
the evolution of the geodesic is stalled until the time associated with that
refinement level catches up. On the other hand, when a geodesic moves
from a finer level to a coarser one, it is generally behind. In that
case, we use a second-order accurate algorithm to evolve the geodesic
forward in time until it is {\it caught up} with the rest of the
fields on that refinement level.

On the other hand, for the \analytic metric, we use the more
conventional formulation of the geodesic equation,
\begin{eqnarray}
  \frac{d x^\mu}{d\tau} &=& u^\mu \,, \cr
  \frac{d u^\mu}{d\tau} &=& - {\Gamma^\mu}_{\rho \sigma} u^\rho
  u^\sigma \,,
\end{eqnarray}
where ${\Gamma^\mu}_{\rho \sigma}$ is the 4-dimensional Christoffel symbols.
Here, we evolve the geodesics using an adaptive RK45 algorithm.

\section{Code verification}\label{sec:code_verification}

Our code suite consists of three parts. A stand-alone code written in
C++ that integrates geodesics and calculates the Riemann tensor given
a function that can provide $g_{\mu \nu}$ at arbitrary coordinate
positions. A Cactus Thorn that evolves geodesics alongside the metric
within the Einstein Toolkit, as well as interpolates the metric (and
derivatives) along these geodesics. Finally, our toolkit contains a
set of Python scripts that calculates the curvature eigenvalues of
Eq.~(\ref{eq:scalar_matrix}) given the data provided by the previous
two programs.

We performed several verification tests of the this code suite,  which
we will describe here. Our first test consisted of using the
stand-alone C++ code to evolve identical geodesics on Schwarzschild
backgrounds, but in very different gauges.

To do this, we started with the standard Schwarzschild metric,
\begin{eqnarray}
  d s^2 &=& -  \left(1-\frac{2M}{R} \right) d T^2
  +  \left(1-\frac{2M}{R} \right)^{-1} d R^2
  \cr &&
  + R^2 \, d\Theta^2  +R^2 \, \sin^2\Theta \, d\Phi^2 \,,
\end{eqnarray}
and used the simple coordinate transformation,
\begin{eqnarray}
T &=& t + A \,\sin(\omega t) \,\sin(\omega t) \,\cos(r) \,,
\cr
R &=& r + A \,\sin(\omega t) \,\sin(\omega t) \,,
\cr
\Theta &=& \theta \,,
\cr
\Phi &=& \phi \,,
\end{eqnarray}
where $A$ is a constant.
\begin{figure}
  \includegraphics[width=\columnwidth]{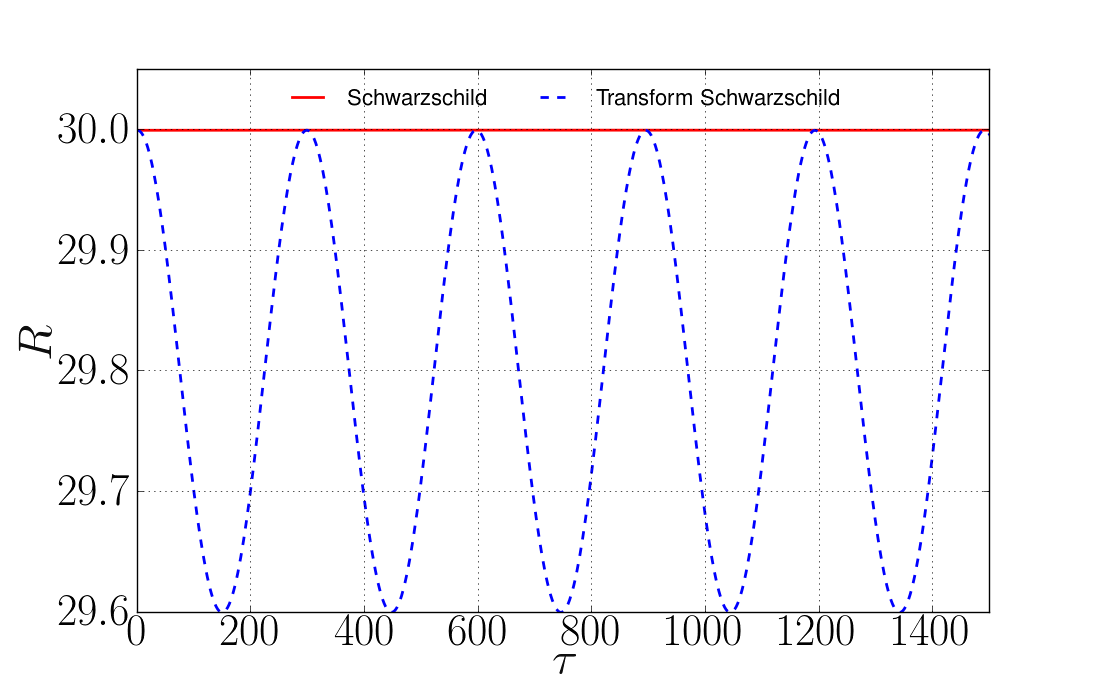}
  \includegraphics[width=\columnwidth]{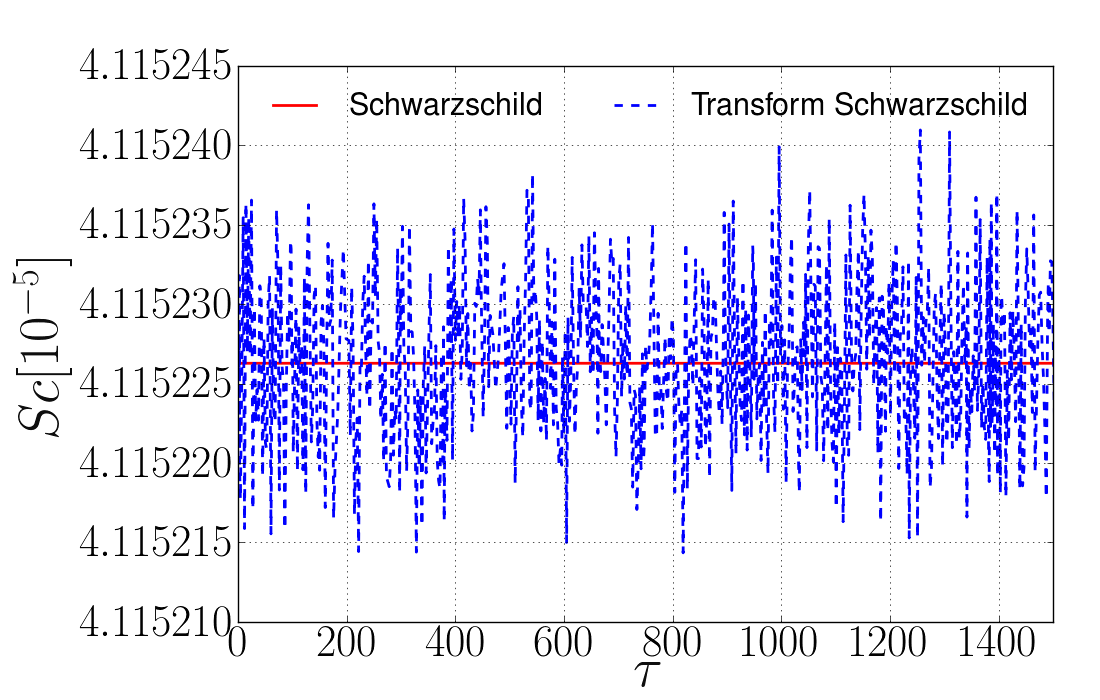}
  \caption{Circular geodesics in standard Schwarzschild and
    transformed Schwarzschild coordinates. While the trajectory is
    gauge dependent (top), the associated curvature eigenvalues (only one shown)
    are not (bottom).
    The differences between the eigenvalues (\textbf{Sc}) calculated in to the
    two gauges are consistent with roundoff errors.
}\label{fig:gauge_dep}
\end{figure} 
As is readily apparent in Fig.~\ref{fig:gauge_dep}, the coordinate
trajectory of the geodesic is quite different in the two
coordinate systems. However, the calculated curvature eigenvalues (only one
shown) are identical. There are three curvature eigenvalues, 
two are positive with very similar magnitudes and one has a negative
value, but is roughly a factor of two larger in absolution value than
the other two. When plotting the eigenvalues, we make the
fiducial choice of plotting the intermediate eigenvalue,
which we denote by \textbf{Sc} in the figures below.

\begin{figure}
  \includegraphics[width=\columnwidth]{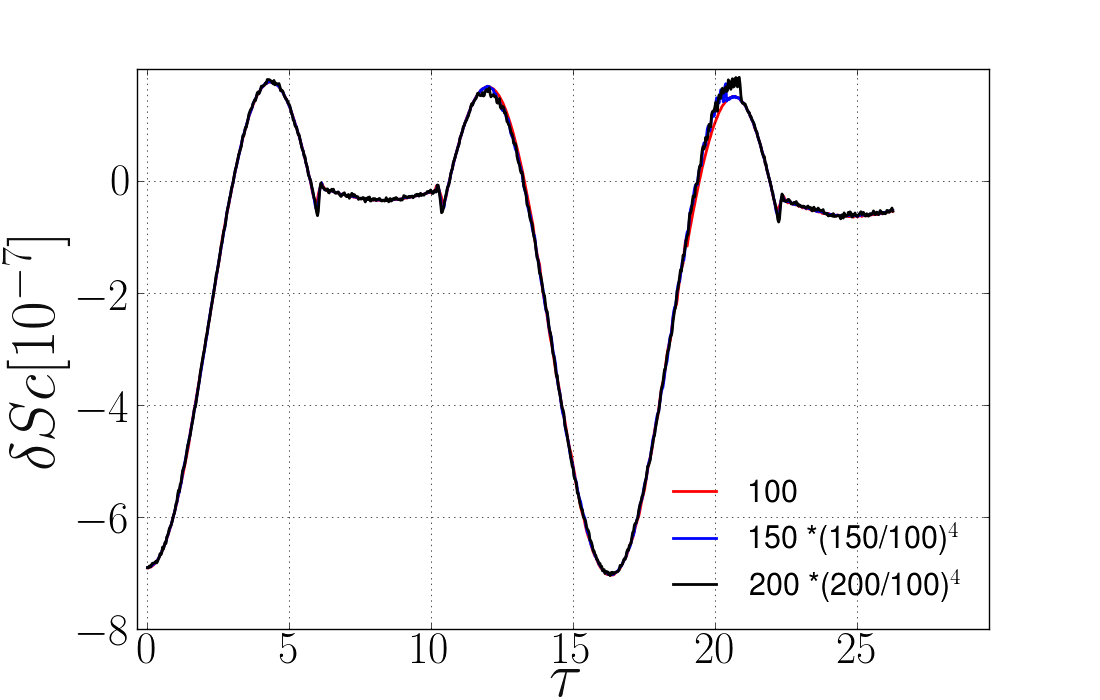}
  \caption{
    The differences between one of curvature eigenvalues(\textbf{Sc})  versus time from our new
    geodesic thorn and the exact
    values (as determined by a stand-alone code). Here, we 
    denote the resolution of a given simulation by the number of
    gridpoint, per dimension, from the origin to the outer boundary,
    and rescale the differences by the ratio of the grid resolution
    to the fourth power.
  }\label{fig:Schconv_test}
\end{figure}

Next, we repeated the same calculation using our EinsteinToolkit-based
geodesic thorn. Here we set the metric analytically, but evolved the
geodesics, and calculated the Riemann tensor (see Sec.~\ref{sec:techniques})
numerically. 
Here, three grid resolutions were used
to test the numerical convergence. As shown in  
Fig.~\ref{fig:Schconv_test},
the relative differences between the analytical and numerical evolution 
of the curvature eigenvalues shows the expected fourth-order convergence. 

To test for convergence of our geodesic thorn in the context of a
fully nonlinear numerical spacetime, we evolve the Schwarzschild
metric in trumpet
coordinates~\cite{Dennison:2014eta} (with the trumpet parameter $R_0 =
M$). For reference, the metric has the form,
\begin{eqnarray}
  && d s^2 = -  \left(\frac{R- M}{R+M} \right) d T^2 +
  \frac{2 M}{R} dT dR \cr
  && + \left(1 + \frac{M}{R}\right)^2 
   \left( d R^2 + R^2 \, d\Theta^2  + 
   R^2 \, \sin^2\Theta \, d\Phi^2 \right) \,.
\end{eqnarray}
Following Ref.~\cite{Dennison:2014sma}, we use the lapse condition
$
\partial_t \alpha = {\cal L}_\beta \alpha - \alpha  ( 1 - \alpha) K$,
for which all the metric functions are constants (up to truncation error)
as functions of time.

\begin{figure}
  \includegraphics[width=\columnwidth]{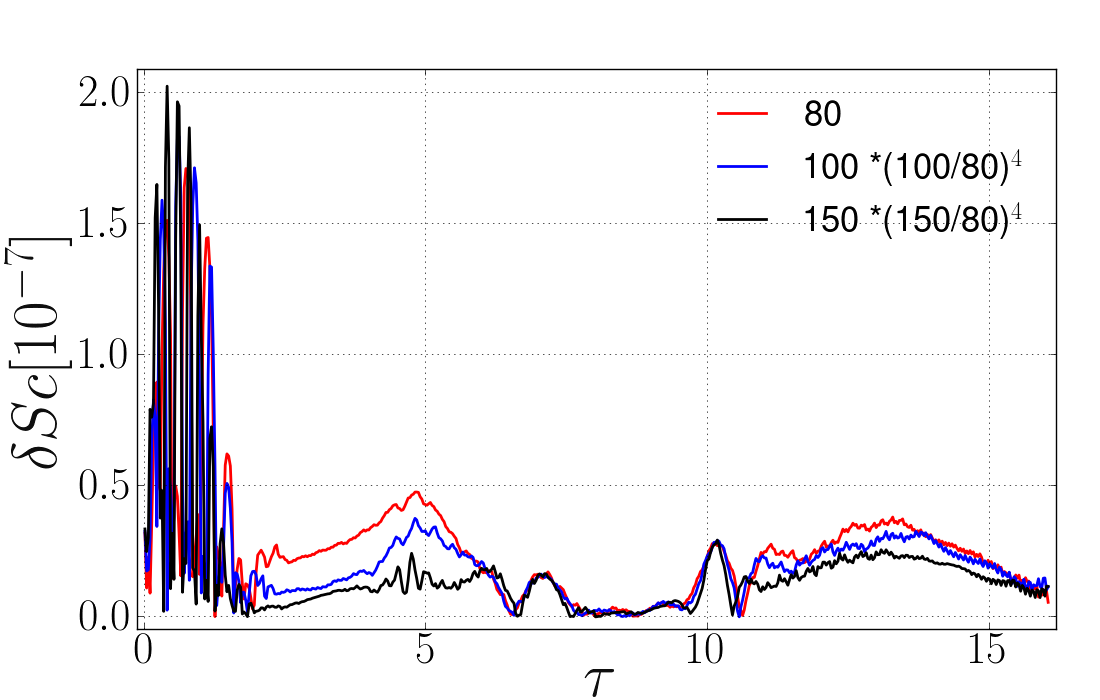}
  \caption{The difference between one of the gauge independent
    curvature eigenvalue (\textbf{Sc}) as calculated using a fully nonlinear numerical
    evolution of time independent trumpet Schwarzschild data using
    the EinsteinToolkit, and as calculated using the exact trumpet
    Schwarzschild metric with the trumpet parameter $R_0 = M$.
    Here, we rescale the
    differences by the ratio of the grid resolution to the fourth power.
}\label{fig:statictrumpet_test}
\end{figure}

A convergence plot of the curvature eigenvalues from a fiducial geodesic is
shown in Fig.~\ref{fig:statictrumpet_test}. Here, too, we find
fourth-order convergence. 

\begin{figure}
  \includegraphics[width=\columnwidth]{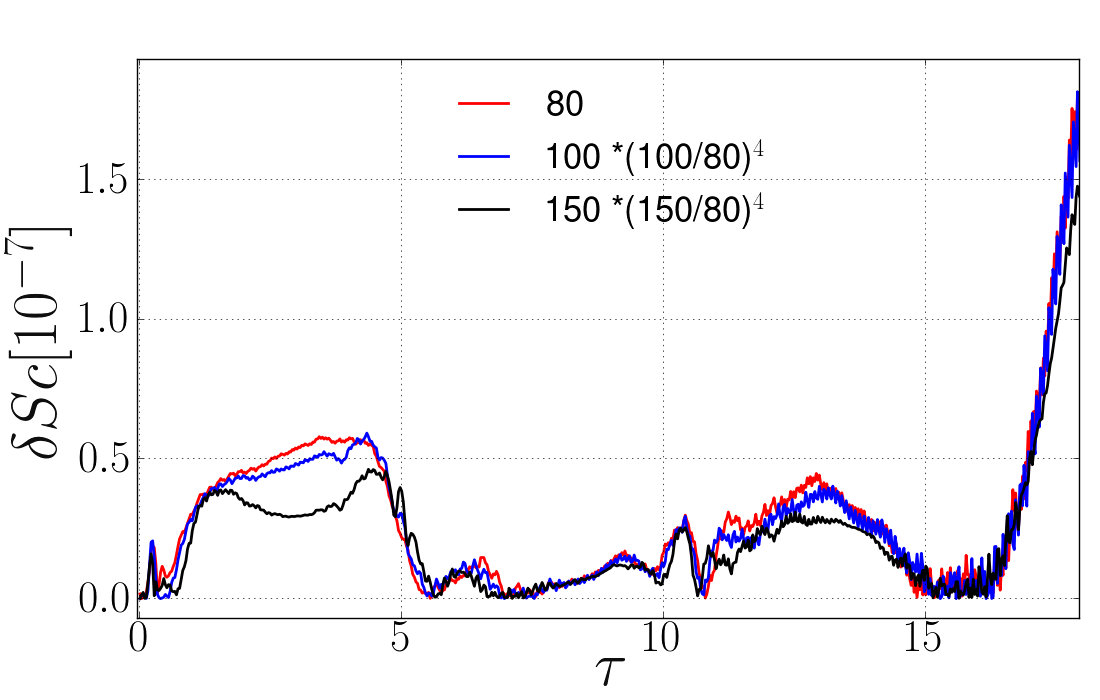}
  \caption{The convergence of the one of the gauge independent curvature
    eigenvalues (\textbf{Sc}) for a slowly time-dependent Schwarzschild trumpet.
    The convergence order is still fourth-order.}
\label{fig:dynamictrumpet_test}
\end{figure}

One aspect of numerical evolutions of a binary spacetime on AMR grids
that we will encounter is stochastic noise in the
curvature~\cite{Zlochower:2012fk}
due to unresolved gauge waves~\cite{Etienne:2014tia}. In order to test
our code with a time dependent metric, we evolved the same trumpet
data, but with a modified lapse condition $\partial_t \alpha = {\cal
L}_\beta \alpha - 1.001 \alpha  ( 1 - \alpha) K $.
This introduces a small time dependence to the metric without simultaneously
introducing an unresolved gauge wave. As seen in
Fig.~\ref{fig:dynamictrumpet_test}, the convergence is still
fourth-order. However, when using a more standard puncture-based
initial data and $1+\log$ lapse, the convergence order reduced to
second-order, which is consistent with the second-order time
prolongation we use. The reason for this drop in convergence rate is
likely the very rapid evolution of the gauge during the first few $M$
of evolution. These rapid changes can lead to the second order (in
time) prolongation error dominating the error budget.

\begin{figure}
  \includegraphics[width=\columnwidth]{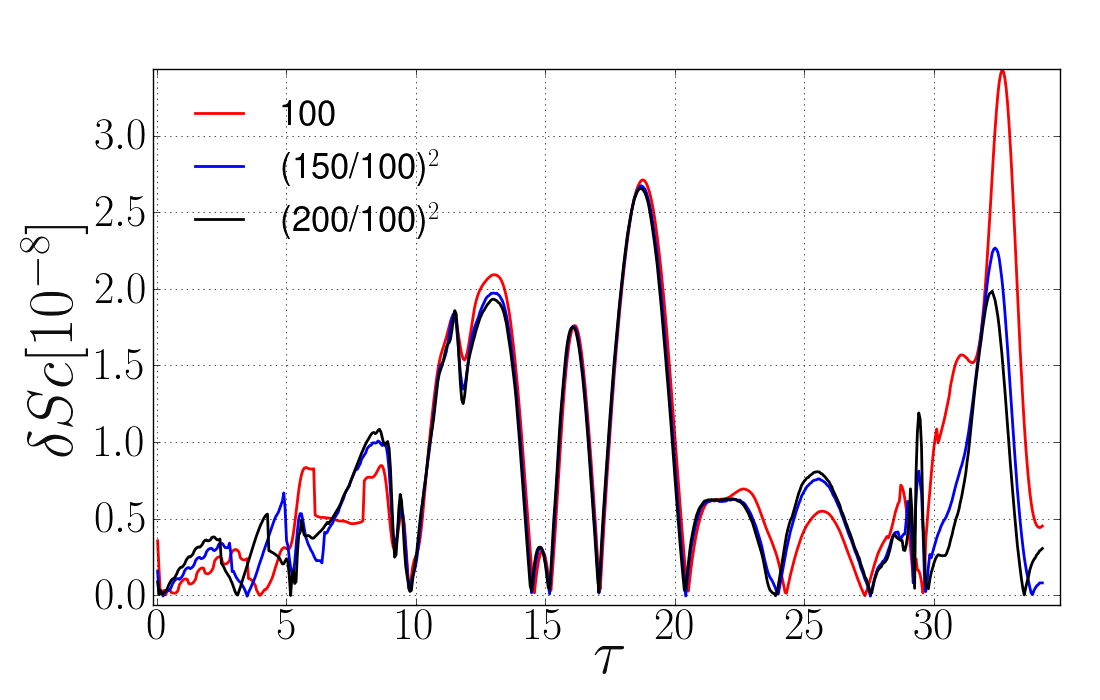}
  \caption{Second-order convergence of the gauge independent
    curvature eigenvalues (\textbf{Sc}) as calculated
    using a fully nonlinear numerical evolution
    of Kerr data using the EinsteinToolkit, and as calculated
    using the exact Kerr metric in quasi-isotropic coordinates. 
}\label{fig:kerr_test}
\end{figure}

Finally, we evolved a set of geodesics in Kerr spacetime in quasi-isotropic
coordinates~\cite{Brandt:1996si} and fully nonlinear numerical
evolutions of a Kerr BH starting with quasi-isotropic initial data. 
Here the two codes evolve the geodesics in gauges that rapidly
deviate from each other. The effects of the unresolved gauge wave are
apparent in the noise and lower-order convergence seen in
Fig.~\ref{fig:kerr_test}. We see a similar lower order convergence
when using Schwarzschild isotropic data.

\section{Results}\label{sec:results}

The main analysis of this paper concerns the dynamics of geodesics on
spacetimes obtained by numerically evolving (using CCZ4) initial data obtained from
the \analytic metric at various starting separations. In particular, we
compare those geodesics with the ones obtained by solving the geodesic
equation on the \analytic spacetime. The differences between the
numerically evolved metric and the \analytic one arise from the
differences in the Ricci tensor of the two. The CCZ4 algorithm drives
the constraint violation toward small values, at which point the
evolved metric is consistent with $T_{\mu \nu}=0$. The \analytic
metric, on the other hand, has $T_{\mu \nu}\neq0$. Differences in
$T_{\mu \nu}$ exist even at $t=0$, which means that the Riemann tensor
on the initial slice is not the same between the numerical and
\analytic metrics.

We use the EinsteinToolkit to evolve geodesics on spacetimes obtained
by using the \analytic metric, with $m_1 = m_2 = M/2$,
as initial data with separations of 
$D=50M$, $25M$, $20M$, $15M$, and $10M$. We simultaneously evolve these
geodesics using our stand-alone C++ code with the purely \analytic metric.

\begin{figure}
  \includegraphics[width=.9\columnwidth]{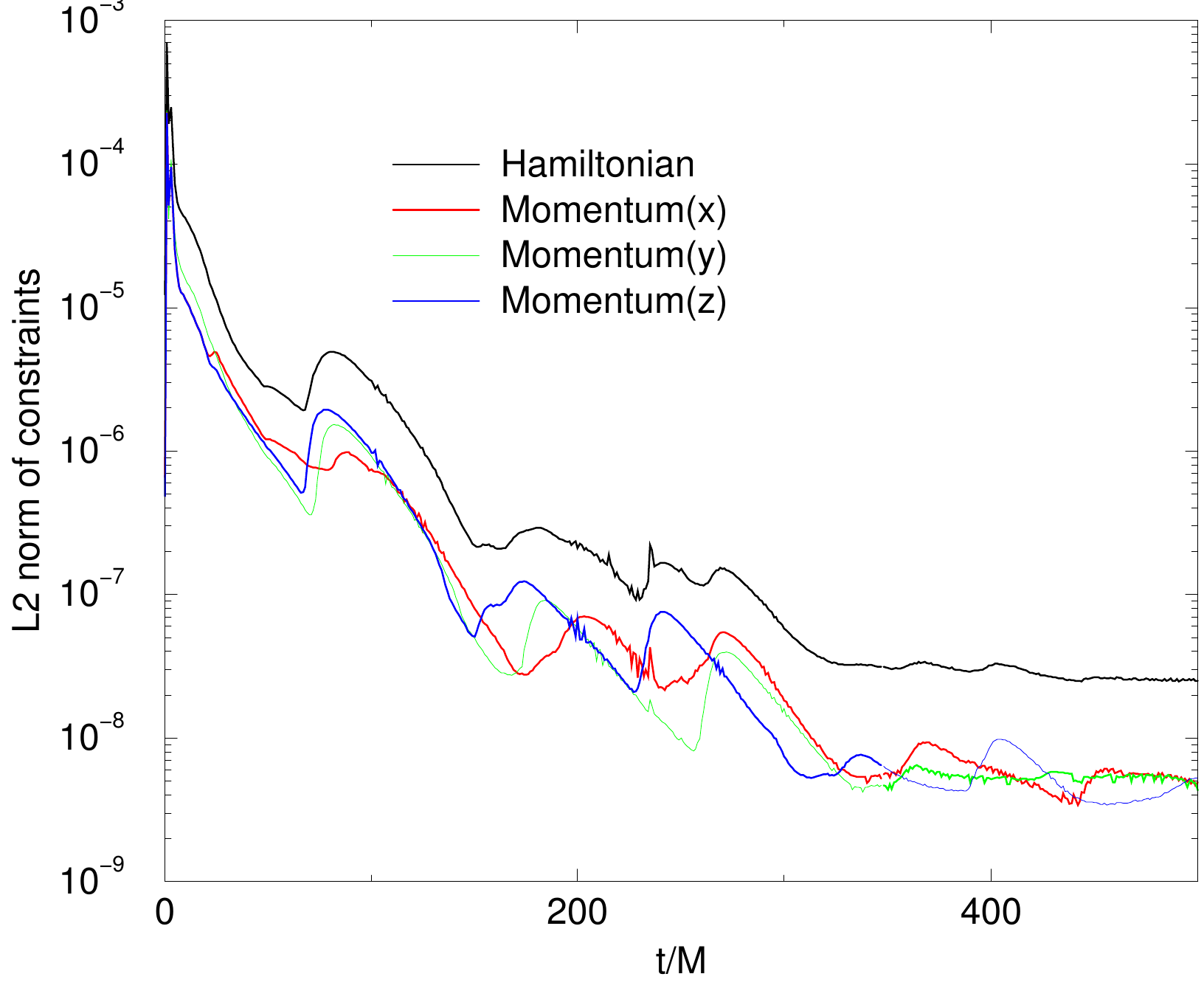}
  \caption{The L2 norm of the constraints for the $D=25M$
    configuration. Here the constraints are calculated within the
    volume outside the two horizons and inside the coordinate sphere
  $r=30M$.}\label{fig:constraints}
\end{figure}

In Fig.~\ref{fig:constraints}, we show how the constraint violations
decay with time using the CCZ4 evolution code (we see a decrease of
over three orders of magnitude).

\begin{figure*}
\includegraphics[width=.30\textwidth]{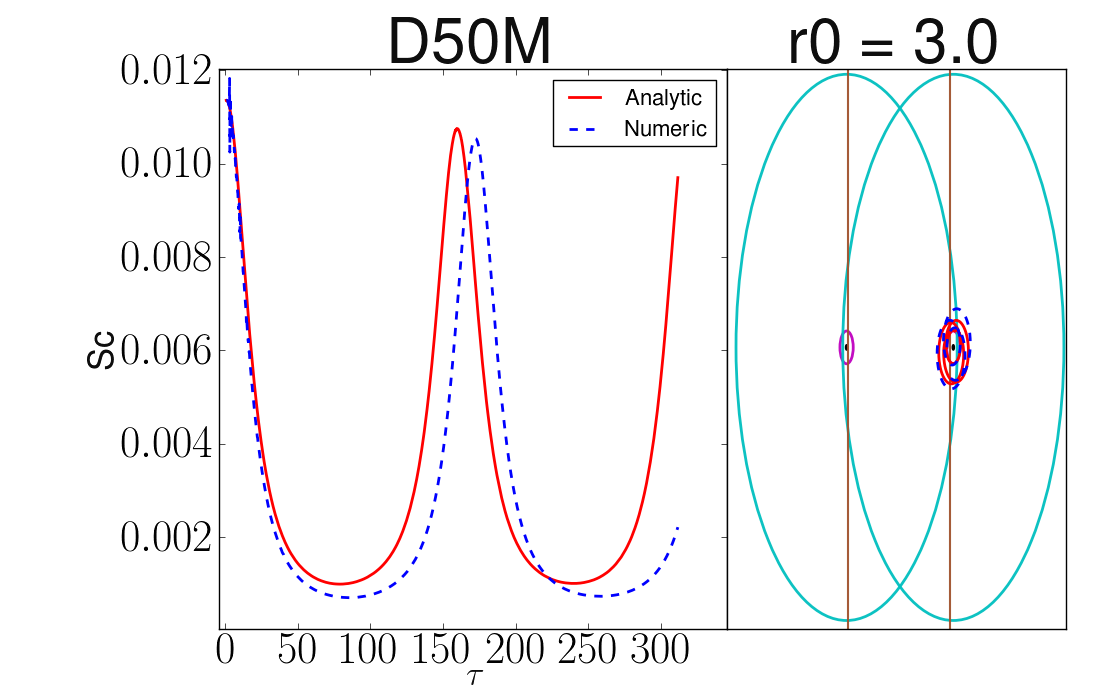}
\includegraphics[width=.30\textwidth]{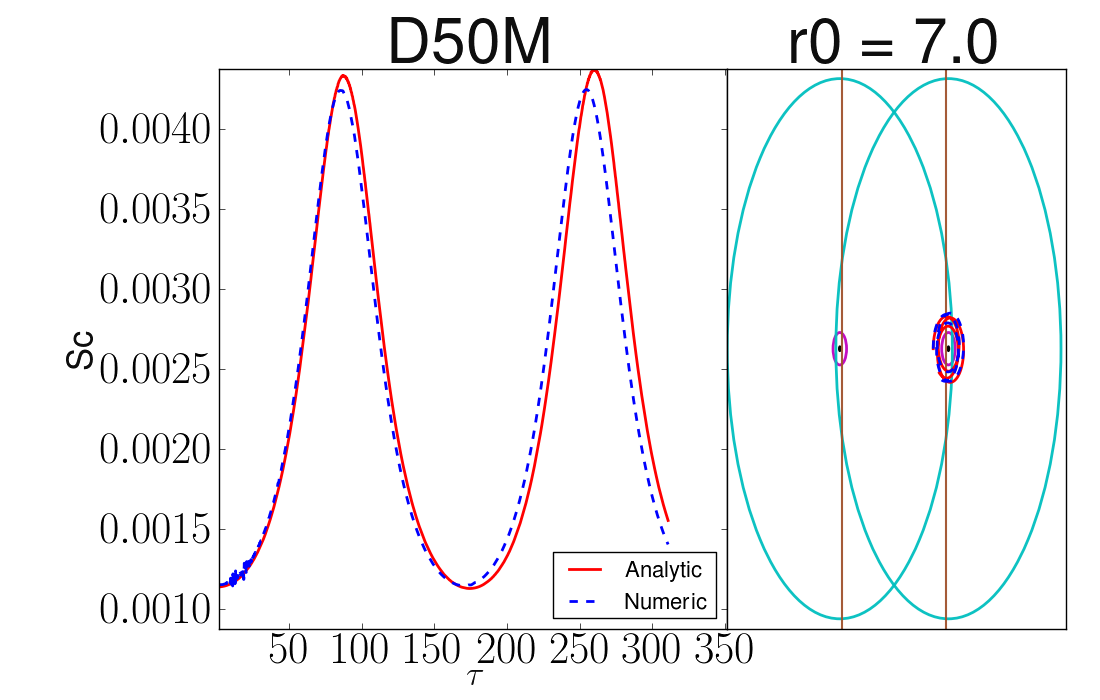}
\includegraphics[width=.30\textwidth]{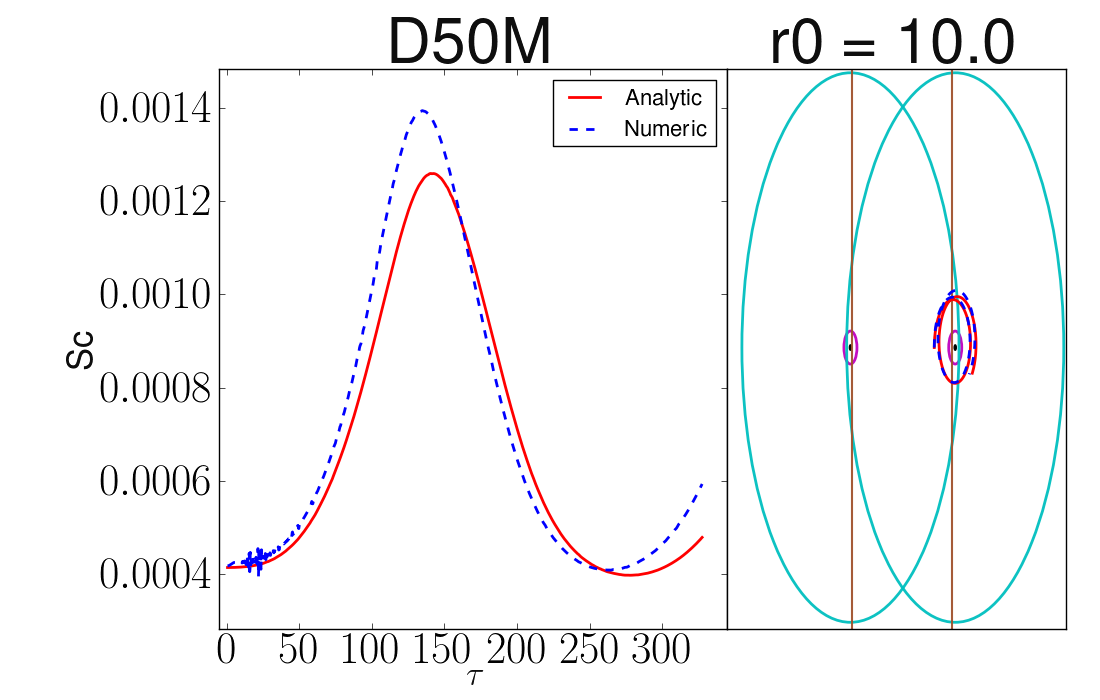}
\includegraphics[width=.30\textwidth]{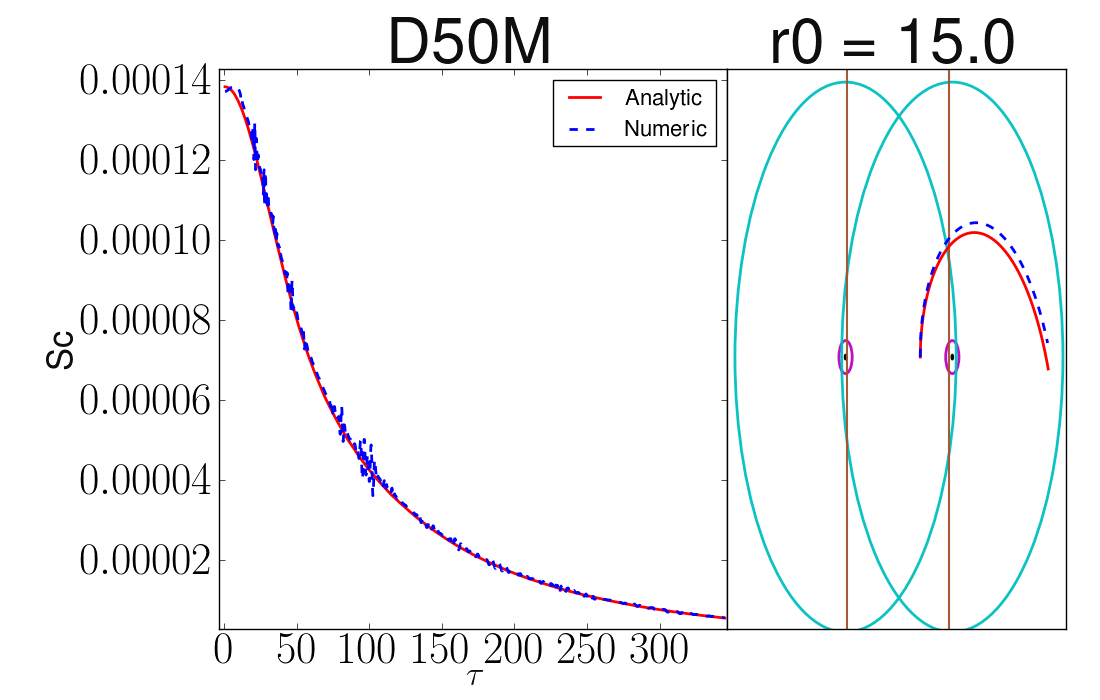}
\includegraphics[width=.30\textwidth]{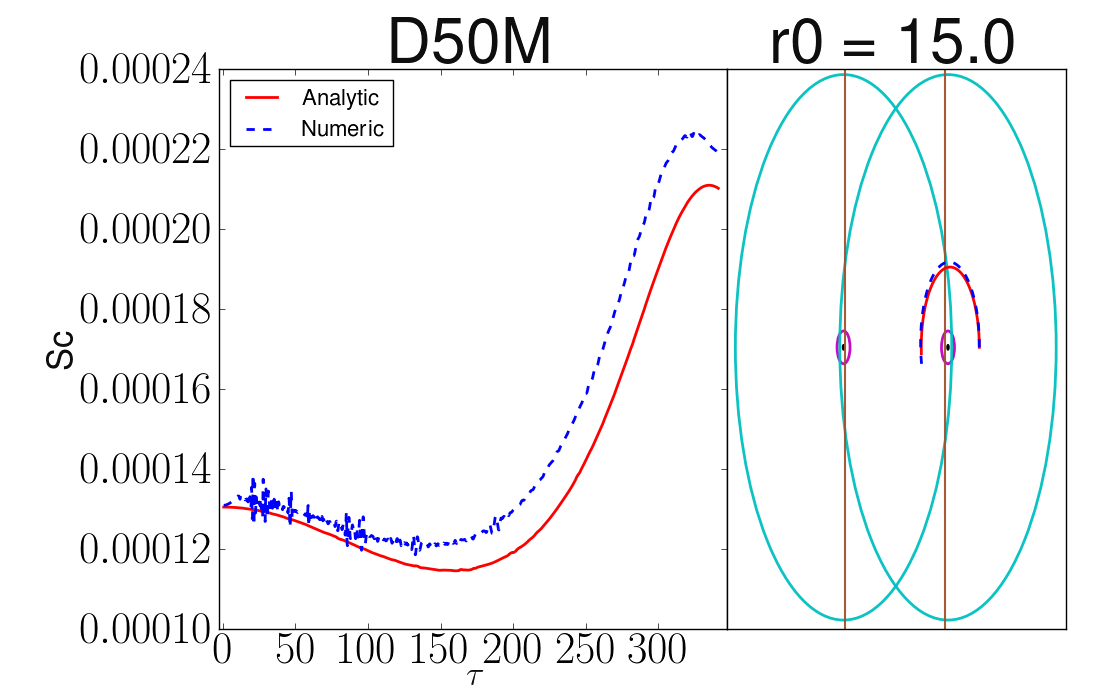}
\includegraphics[width=.30\textwidth]{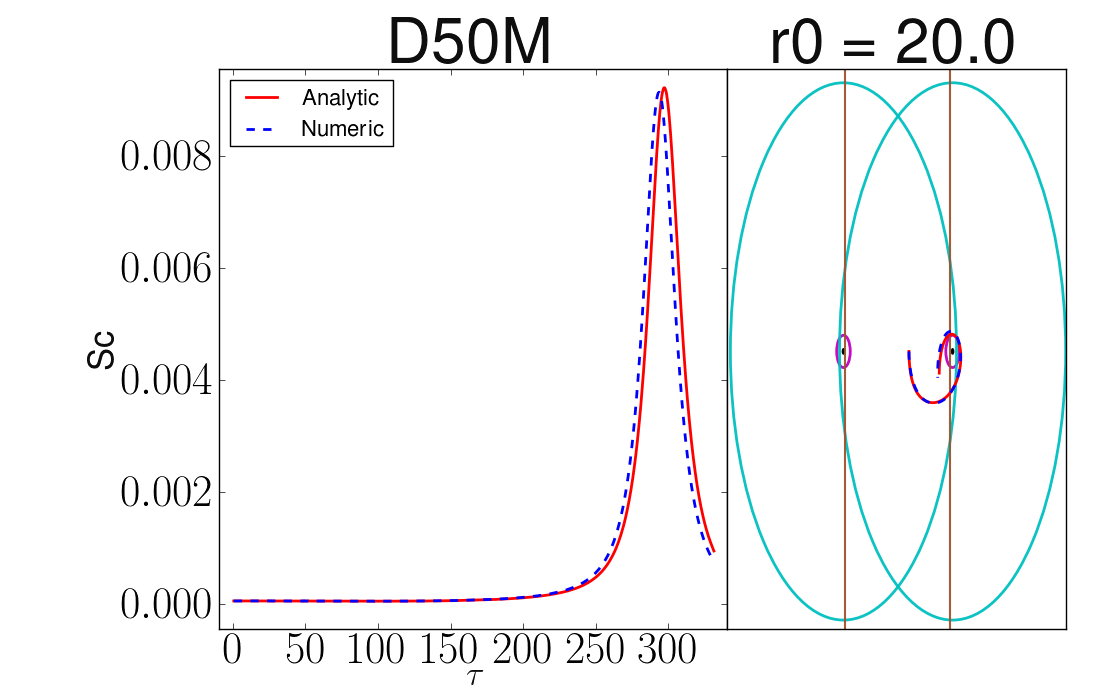}
\includegraphics[width=.30\textwidth]{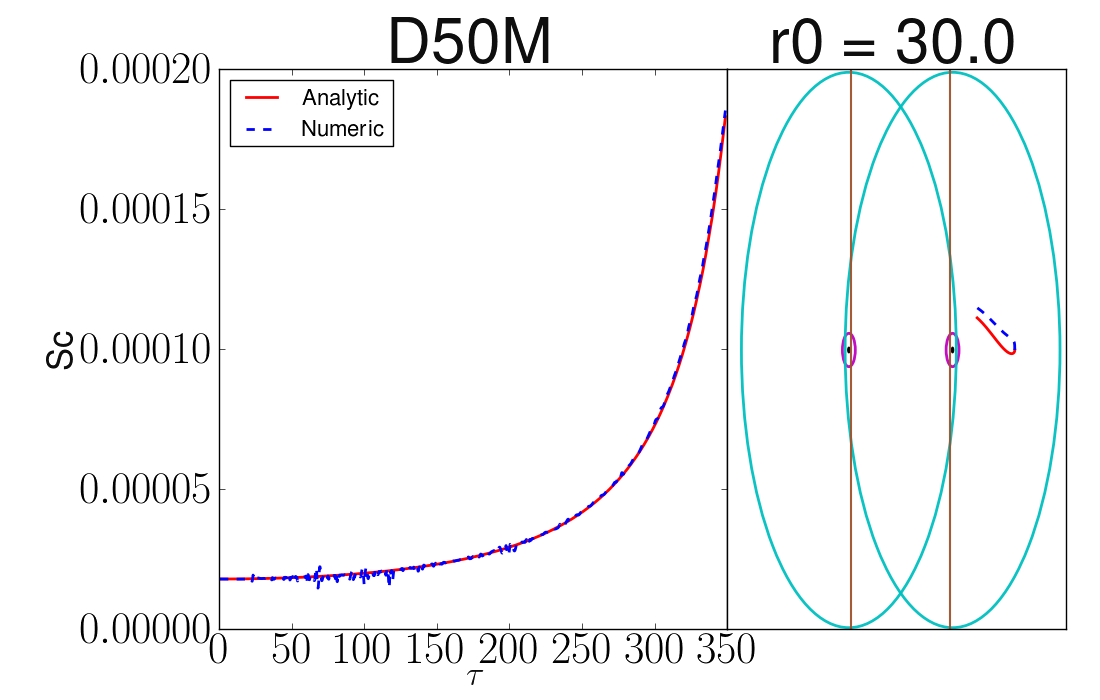}
\includegraphics[width=.30\textwidth]{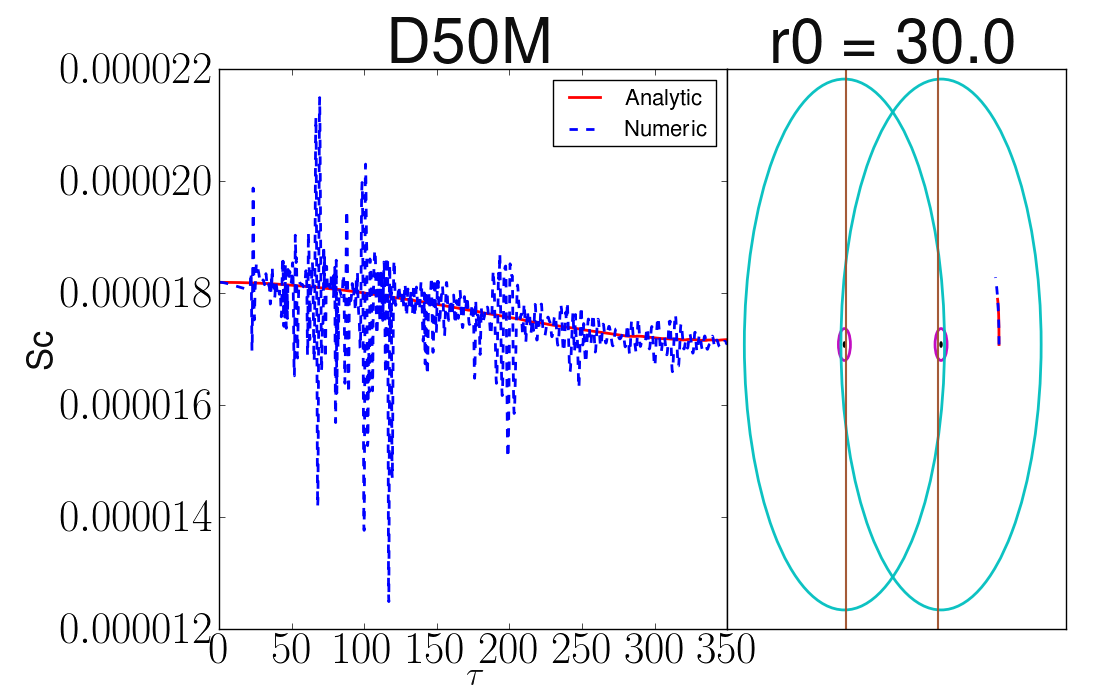}
\includegraphics[width=.30\textwidth]{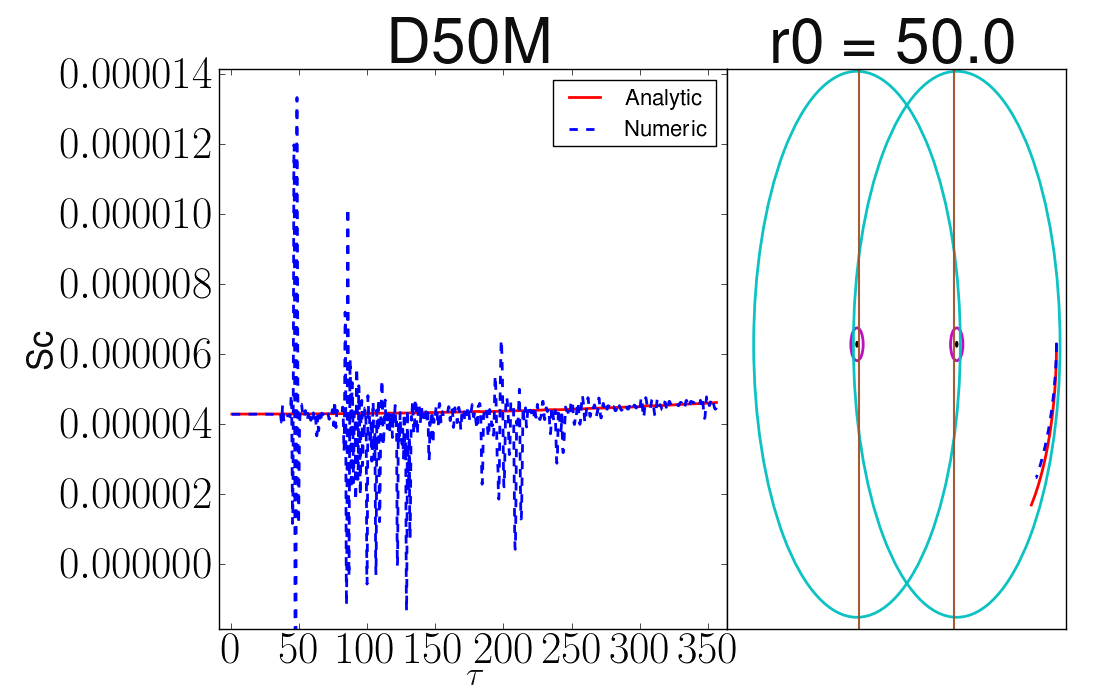}
\includegraphics[width=.30\textwidth]{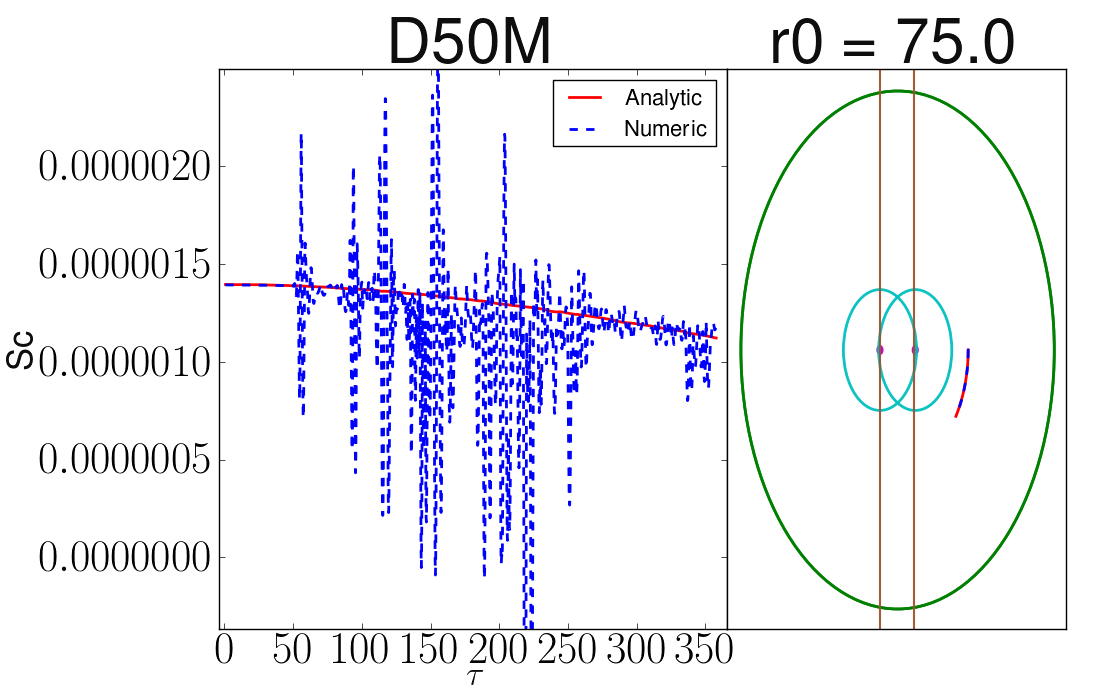}
\includegraphics[width=.30\textwidth]{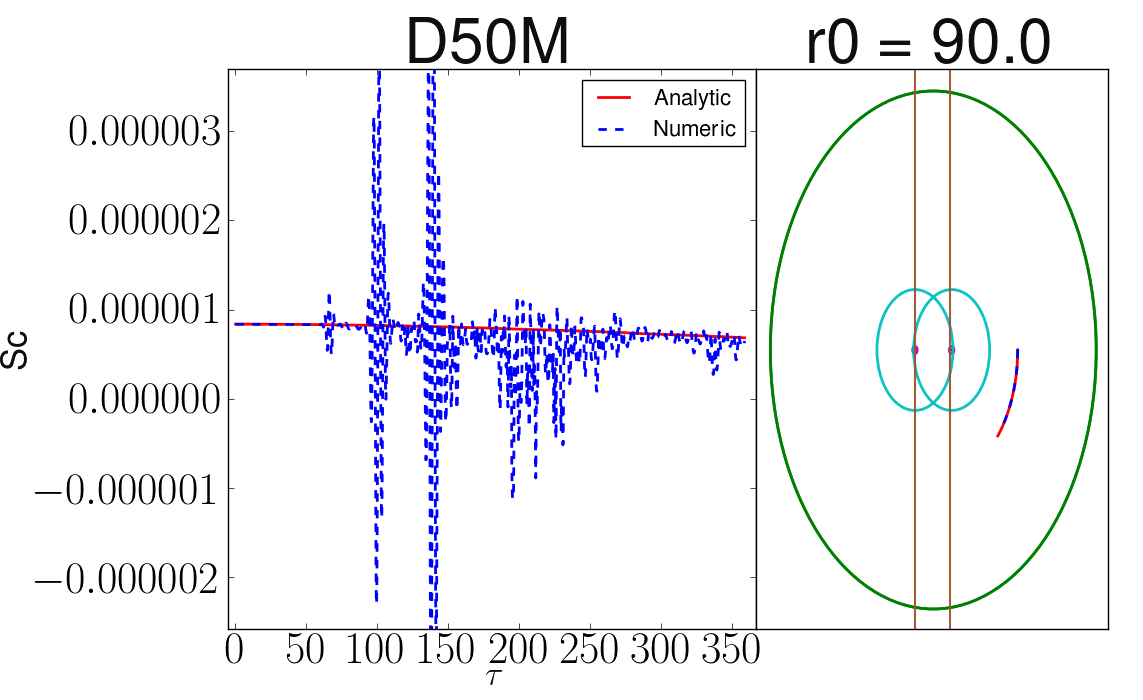}
\includegraphics[width=.30\textwidth]{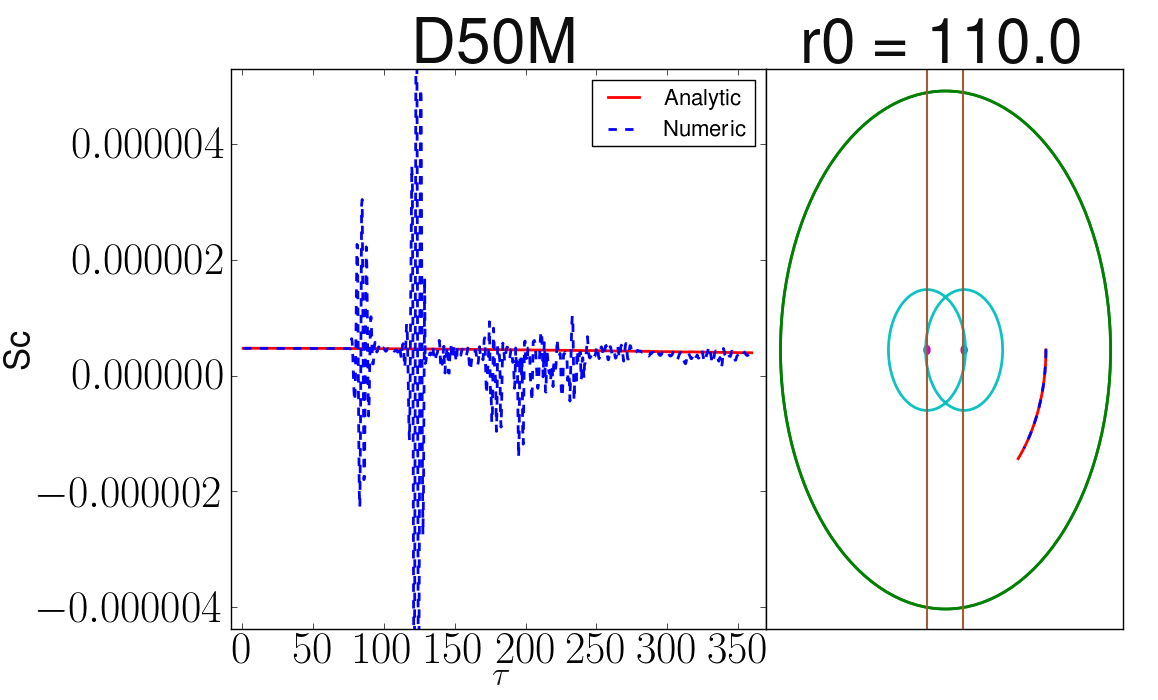}
\includegraphics[width=.30\textwidth]{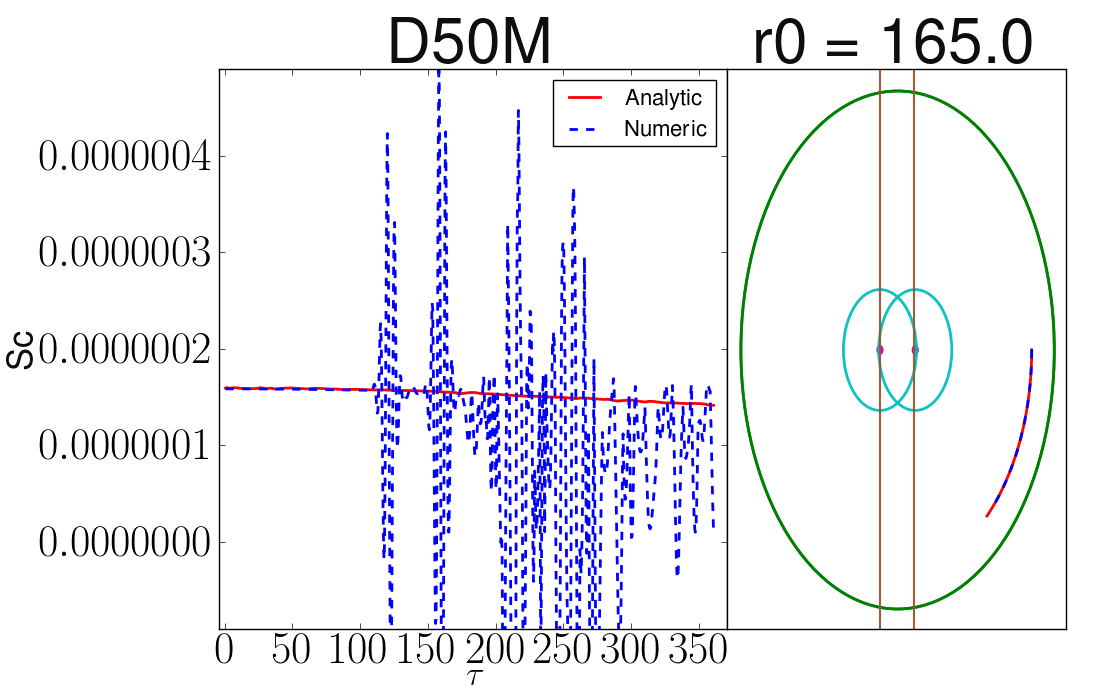}
\includegraphics[width=.30\textwidth]{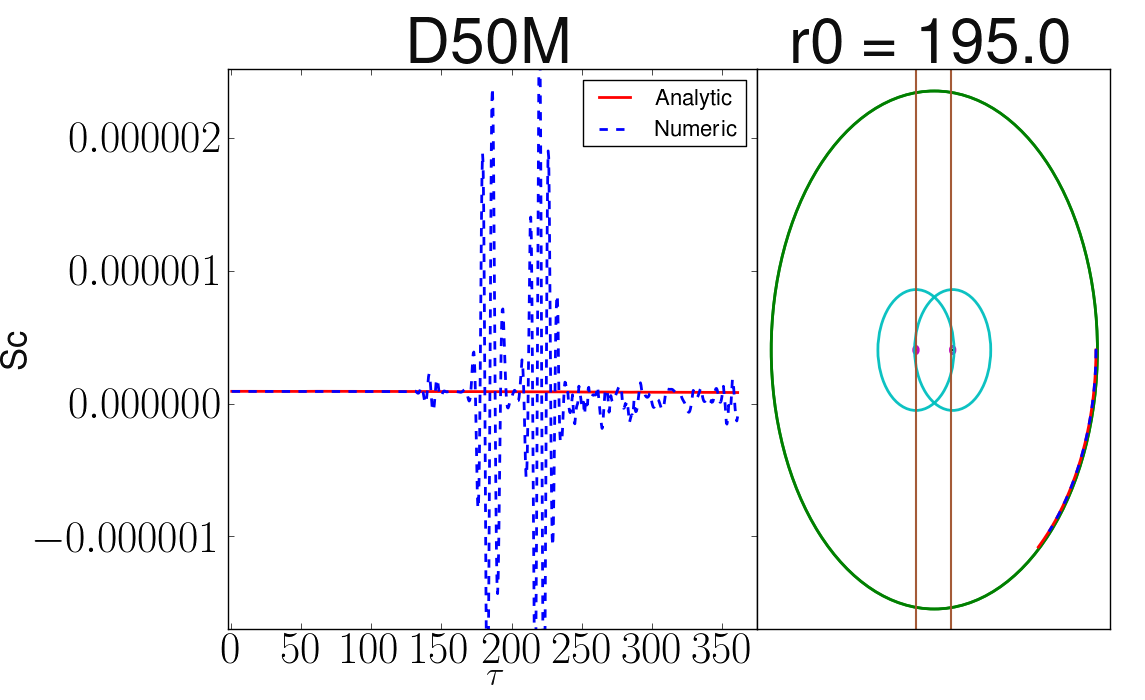}
\includegraphics[width=.30\textwidth]{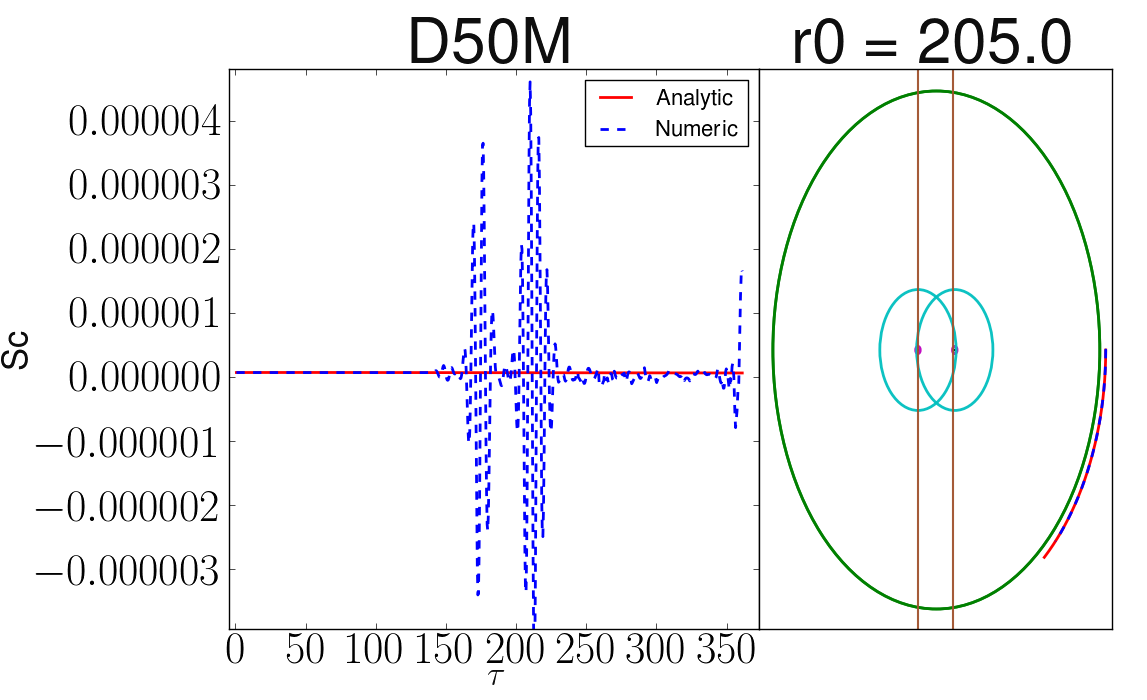}
  \caption{Separation $D=50M$ results. Here we plot the value of the
    largest (in magnitude) curvature eigenvalue (\textbf{Sc}) versus time
    (as evolved using the
    numerical and \analytic metric), as well as plot the coordinate position
    of the geodesics {\it in a corotating frame} (i.e., one where the
    BH positions are nearly fixed) on the right side of each
    panel. The vertical lines and circles in these trajectory plots show
    the location of the various zones described in
    Sec.~\ref{sec:analytic_metric}. The number $r_0$ (normalized by $M$)
    is the initial coordinate distance of the geodesic from the nearest BH.
    For the geodesics close to the BHs, the noise in the
    numerically evolved spacetime is low compared to the magnitude of the
    curvature eigenvalues, the
    opposite is true for the farthest ones.}
  \label{fig:D50}
\end{figure*}

\begin{figure*}
\includegraphics[width=.30\textwidth]{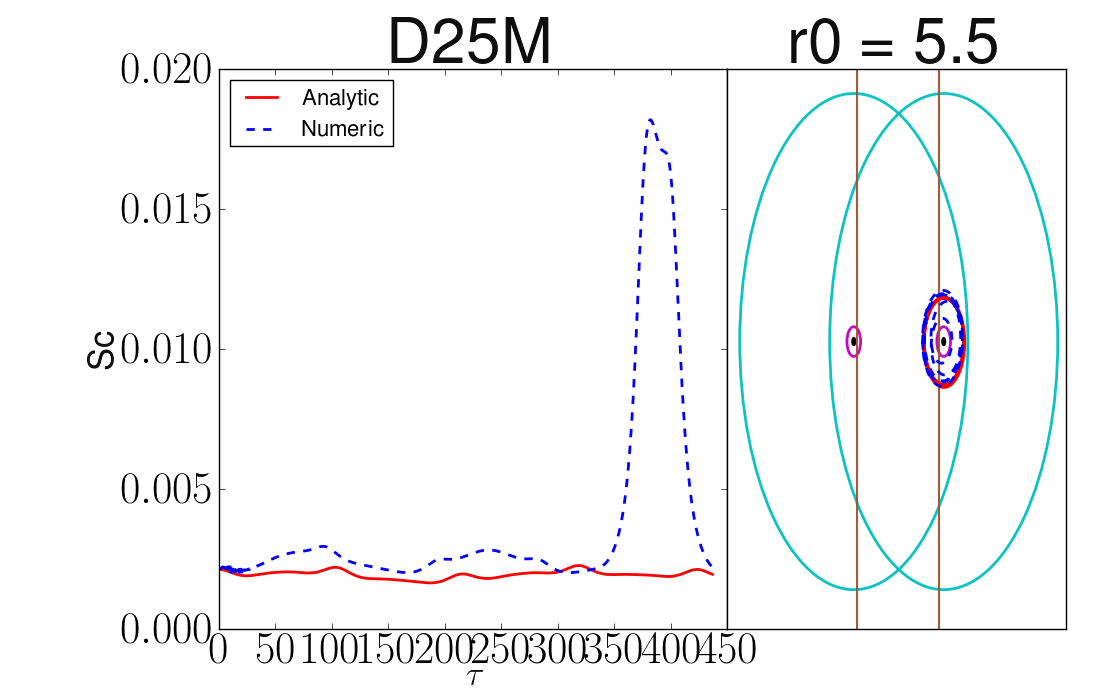}
\includegraphics[width=.30\textwidth]{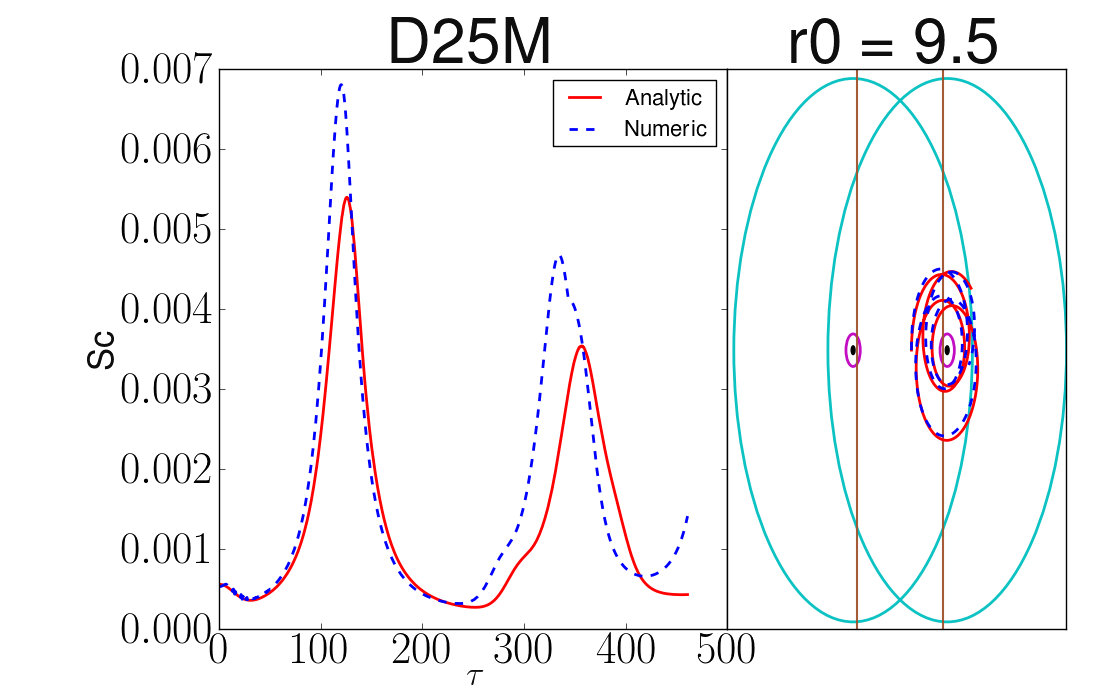}
\includegraphics[width=.30\textwidth]{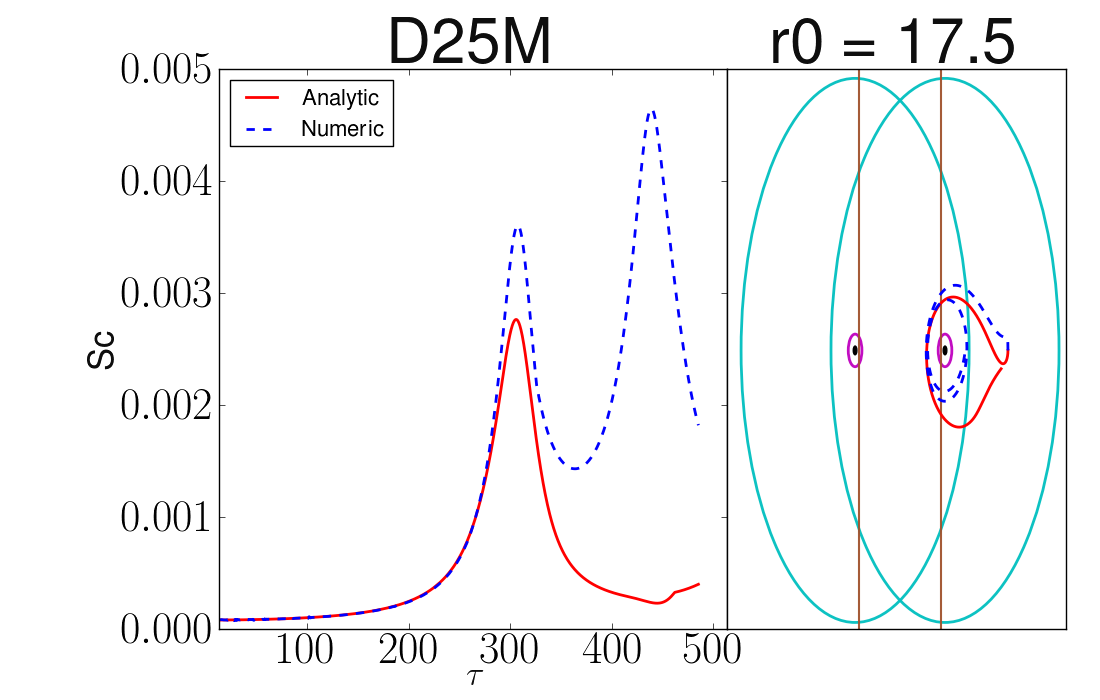}
\includegraphics[width=.30\textwidth]{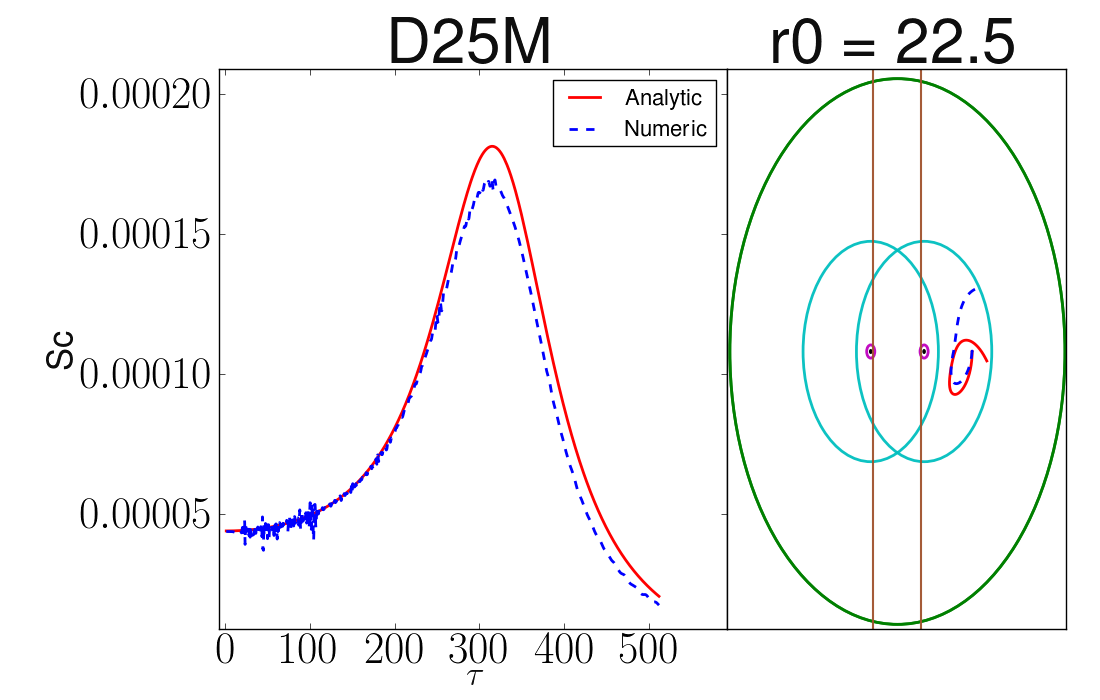}
\includegraphics[width=.30\textwidth]{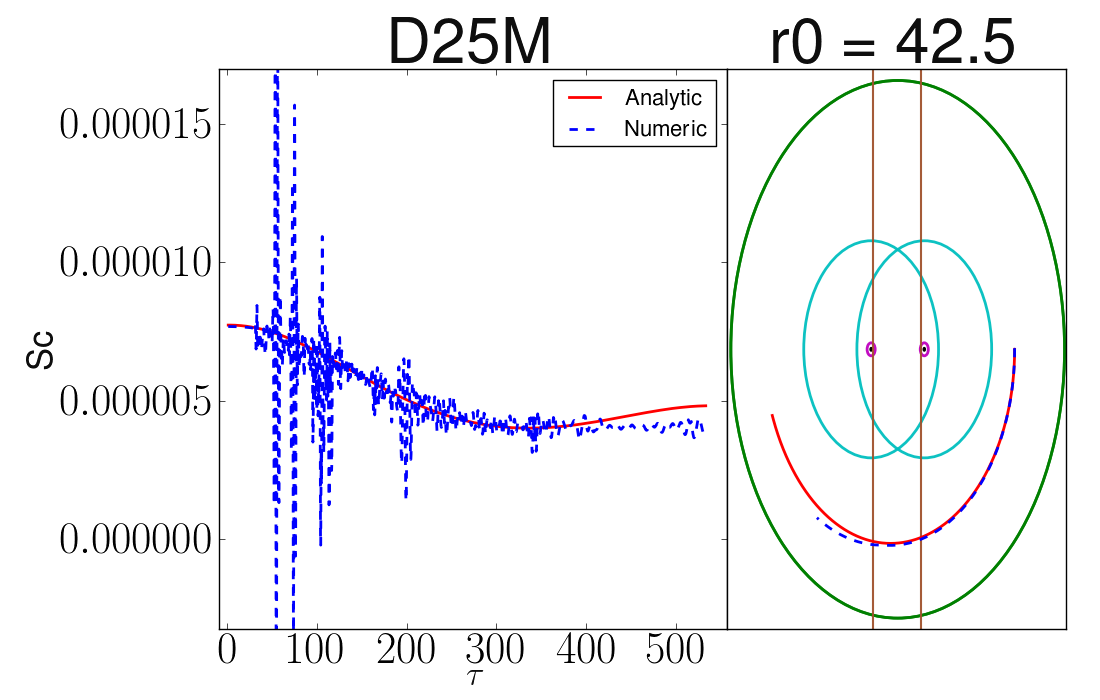}
\includegraphics[width=.30\textwidth]{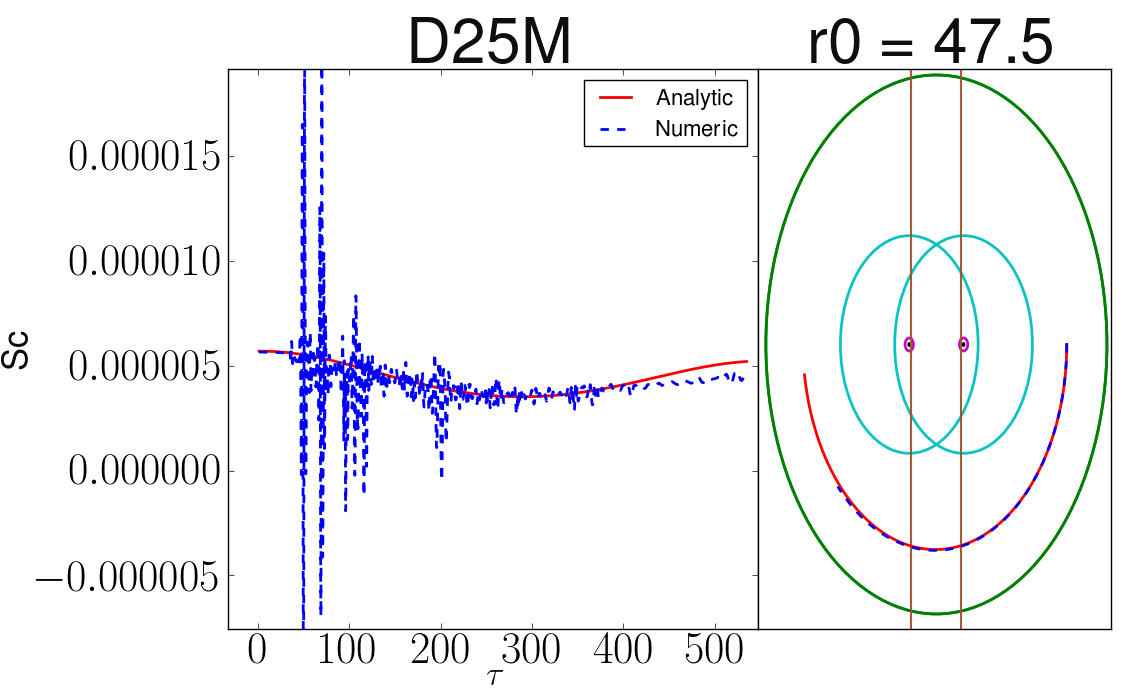}
\includegraphics[width=.30\textwidth]{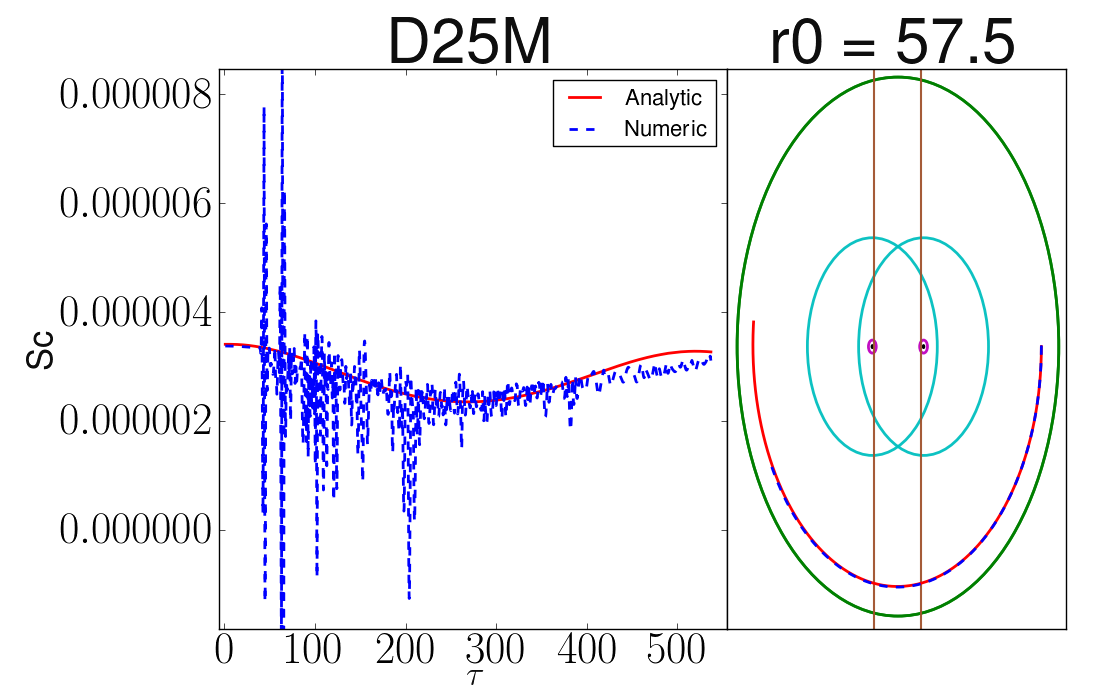}
\includegraphics[width=.30\textwidth]{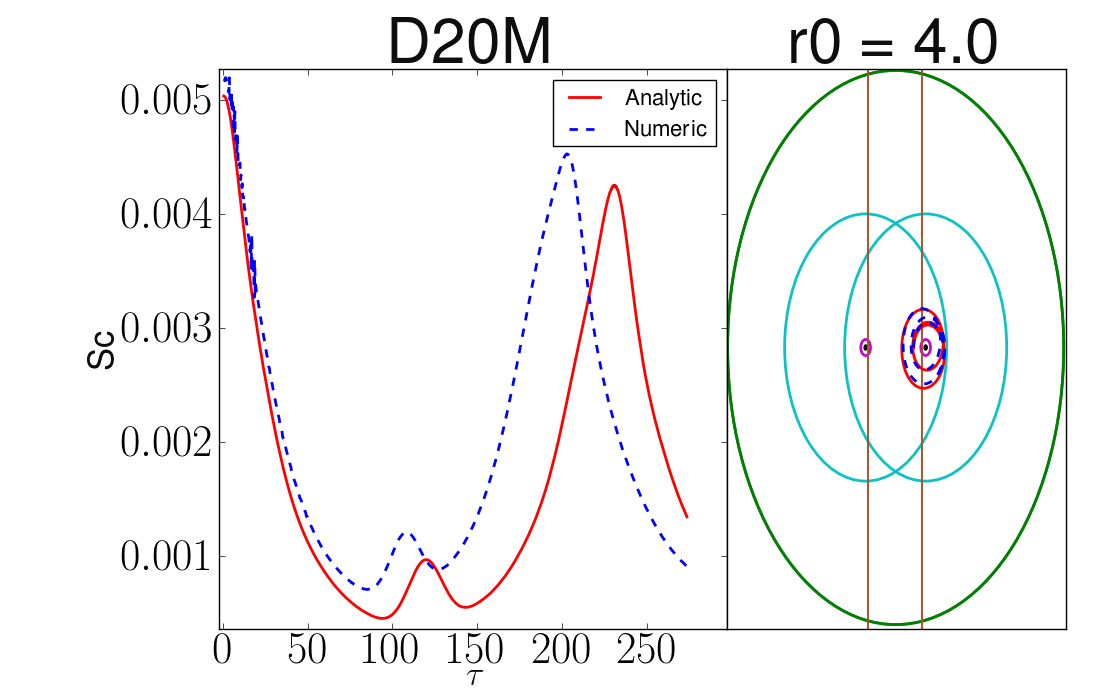}
\includegraphics[width=.30\textwidth]{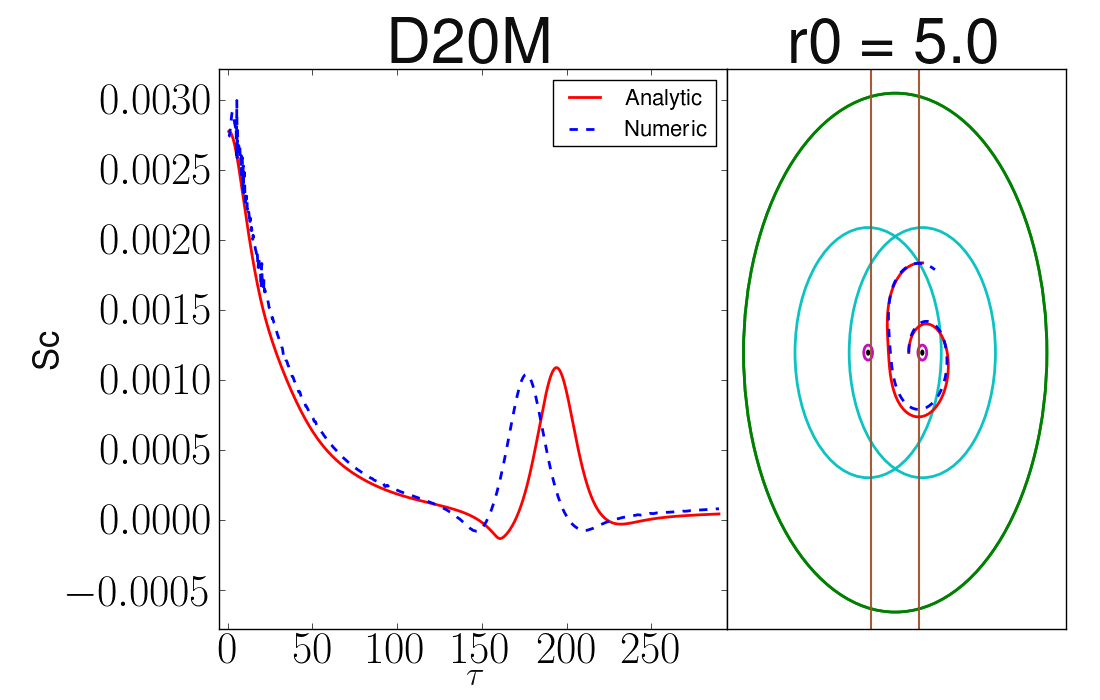}
\includegraphics[width=.30\textwidth]{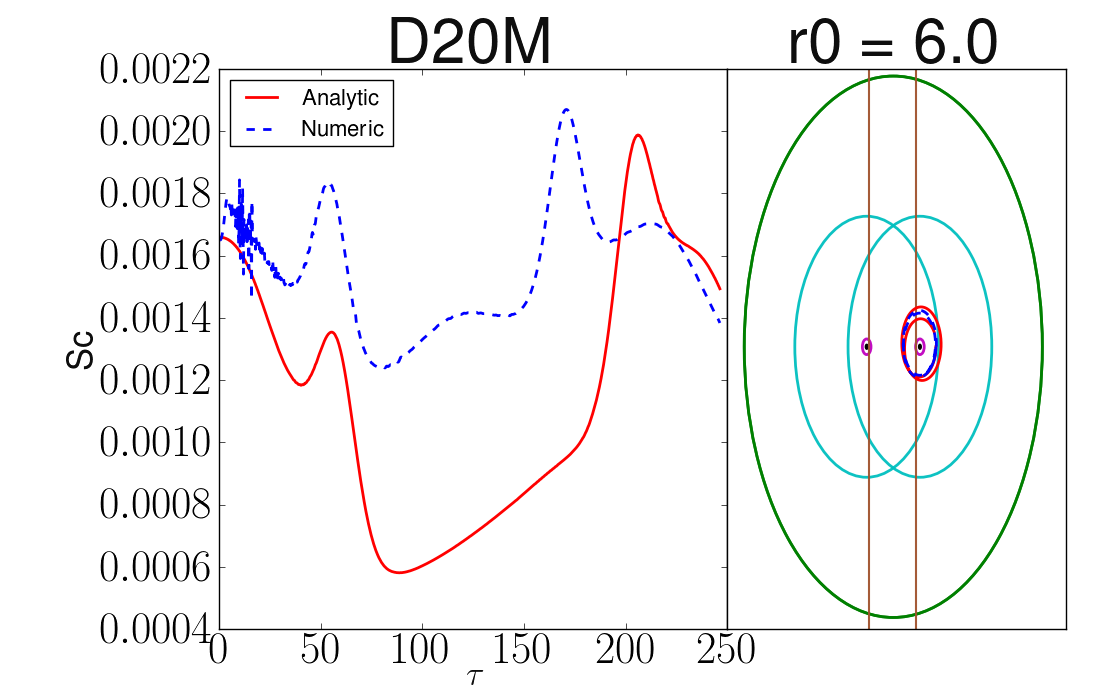}
\includegraphics[width=.30\textwidth]{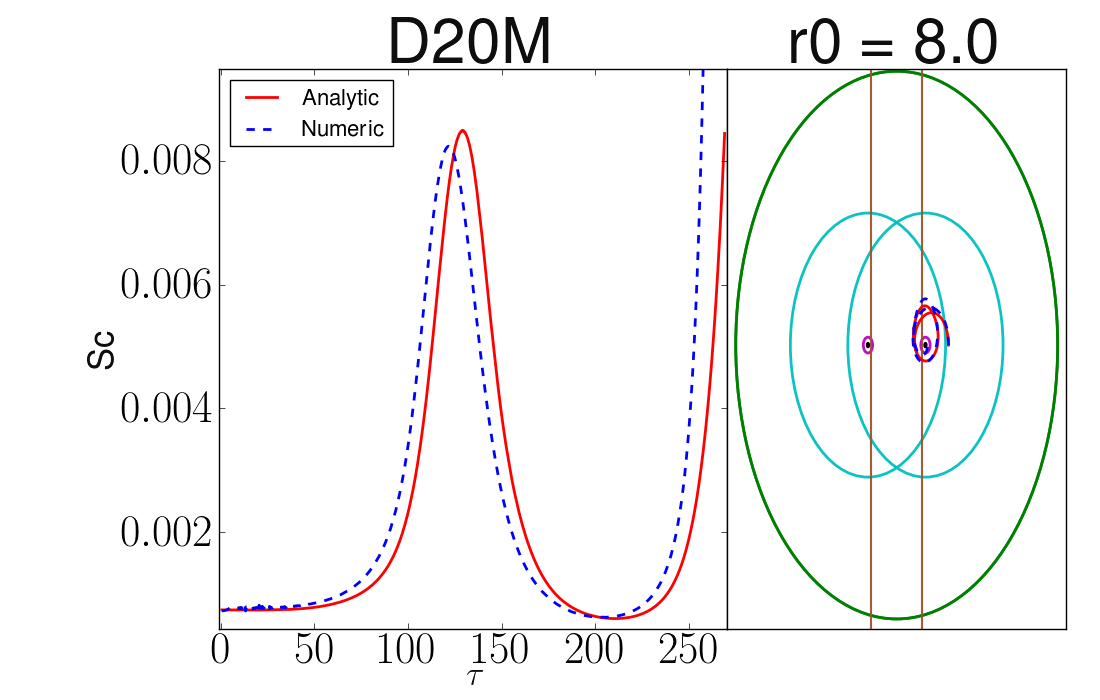}
\includegraphics[width=.30\textwidth]{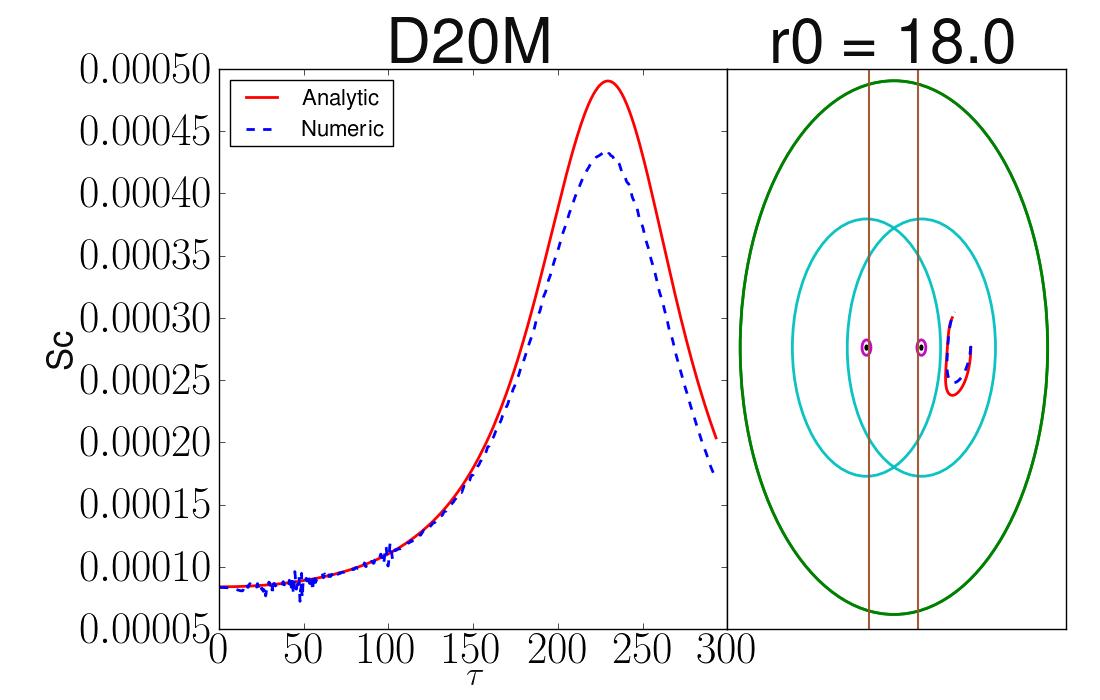}
\includegraphics[width=.30\textwidth]{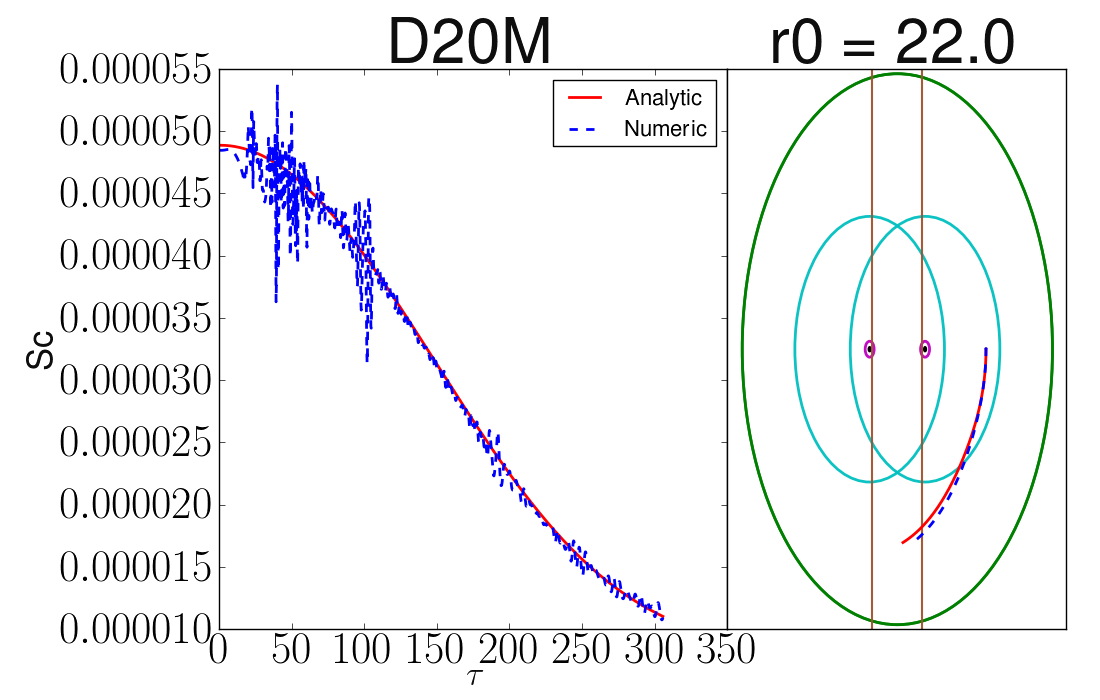}
\includegraphics[width=.30\textwidth]{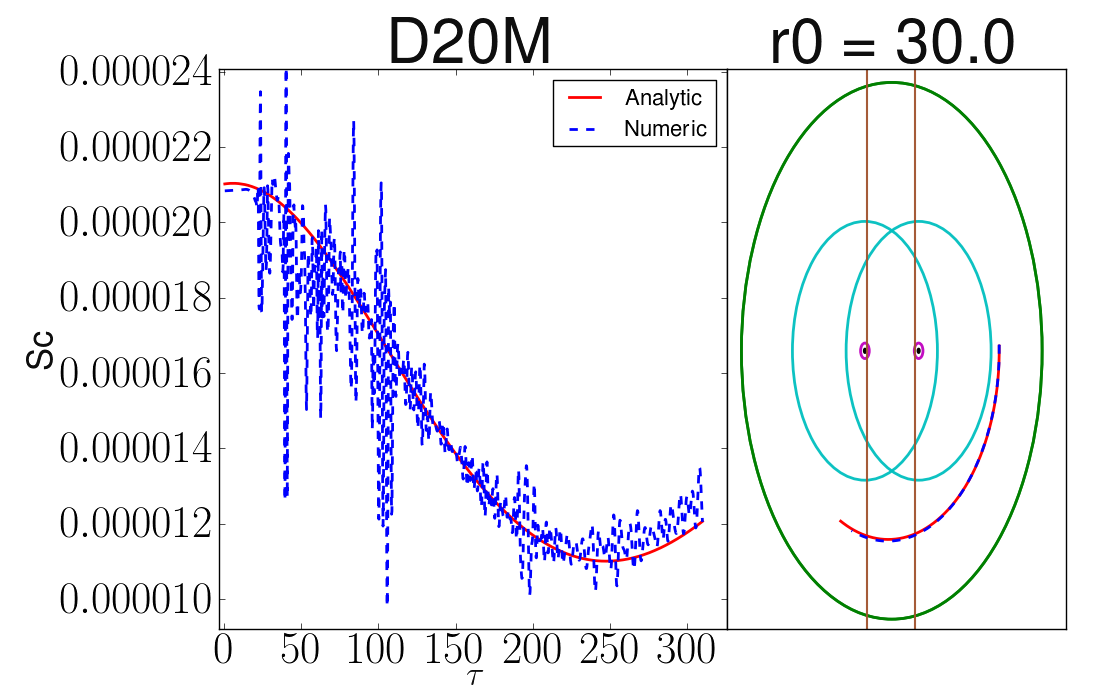}
\includegraphics[width=.30\textwidth]{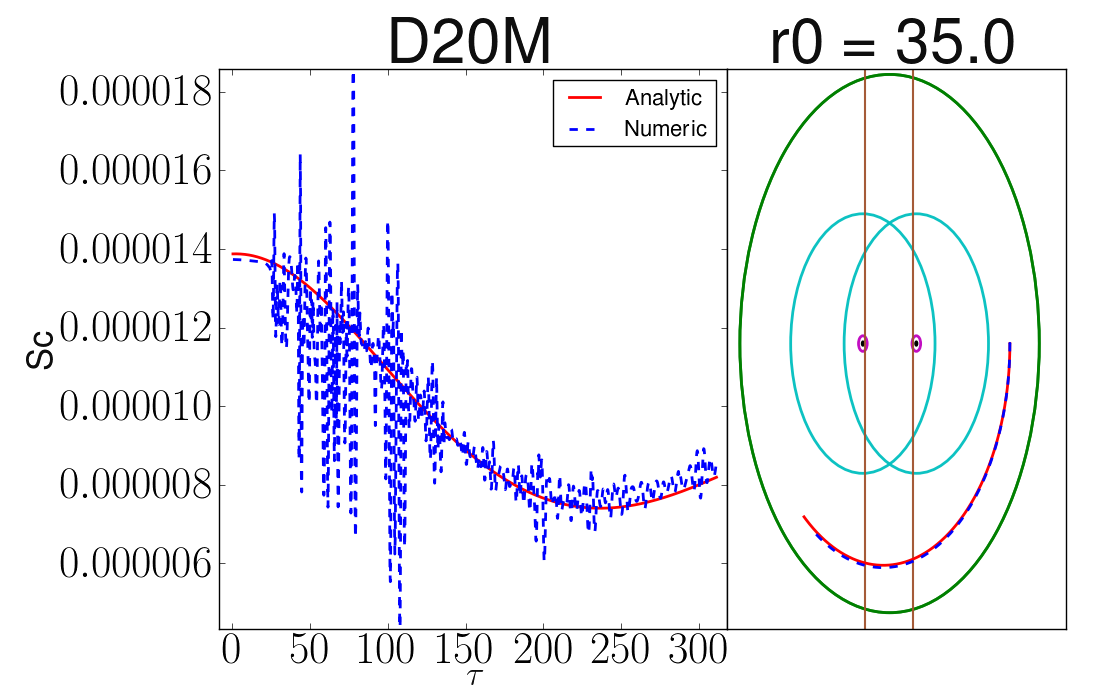}
\includegraphics[width=.30\textwidth]{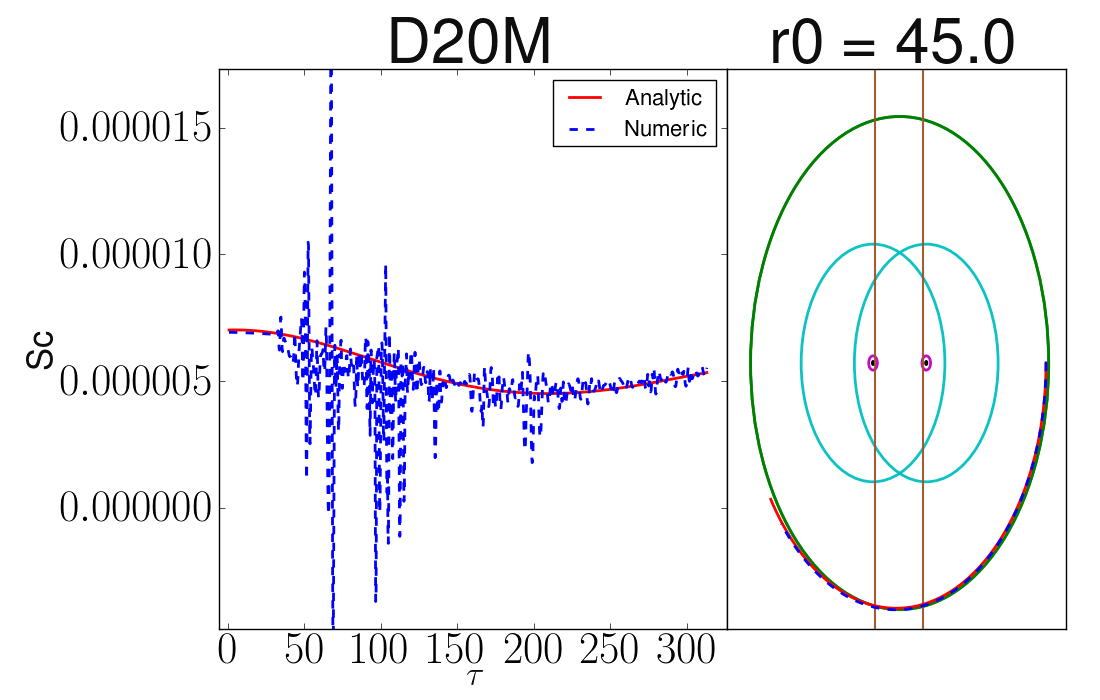}
\includegraphics[width=.30\textwidth]{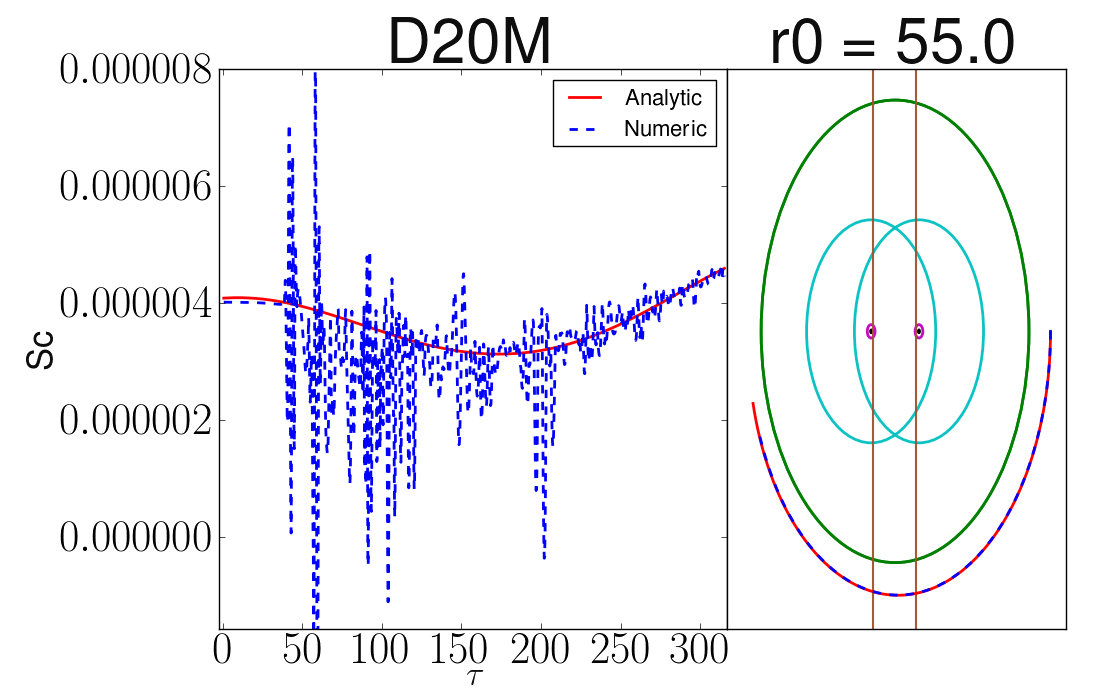}
\includegraphics[width=.30\textwidth]{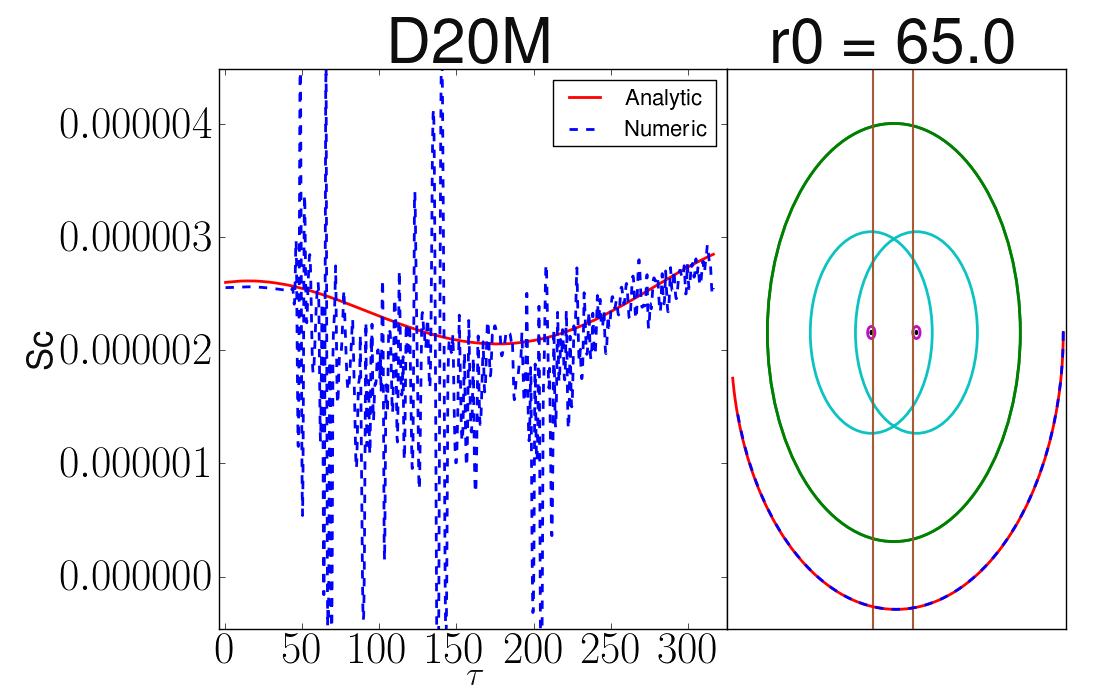}
  \caption{Separation $D=25M$  and $D=20M$ results. Here we plot the value of the
    largest (in magnitude) curvature eigenvalue (\textbf{Sc}) versus time
    (as evolved using the
    numerical and \analytic metric), as well as plot the coordinate position
    of the geodesics {\it in a corotating frame} (i.e., one where the
    BH positions are nearly fixed) on the right side of each
    panel. The vertical lines and circles in these trajectory plots show
    the location of the various zones described in
    Sec.~\ref{sec:analytic_metric}. The number $r_0$ (normalized by $M$)
    is the initial coordinate distance of the geodesic from the nearest BH.
    For the geodesics close to the BHs, the noise in the
    numerically evolved spacetime is low compared to the magnitude of the
    curvature eigenvalues, the opposite is true for the farthest ones.}
  \label{fig:D25andD20}
\end{figure*}

\begin{figure*}
\includegraphics[width=.30\textwidth]{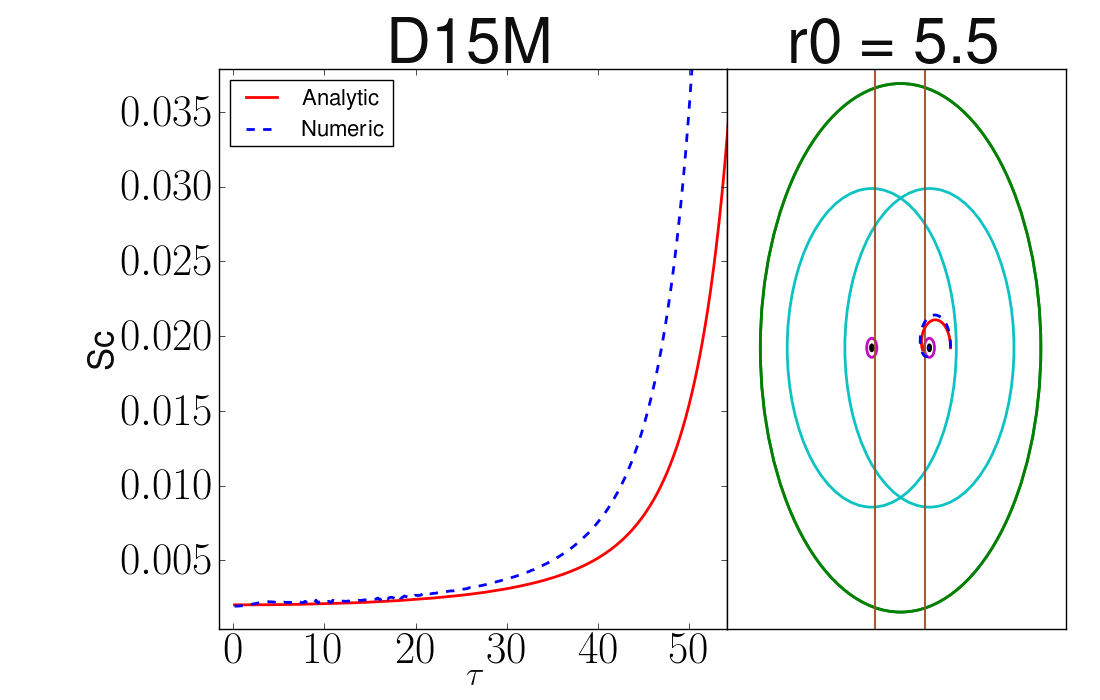}
\includegraphics[width=.30\textwidth]{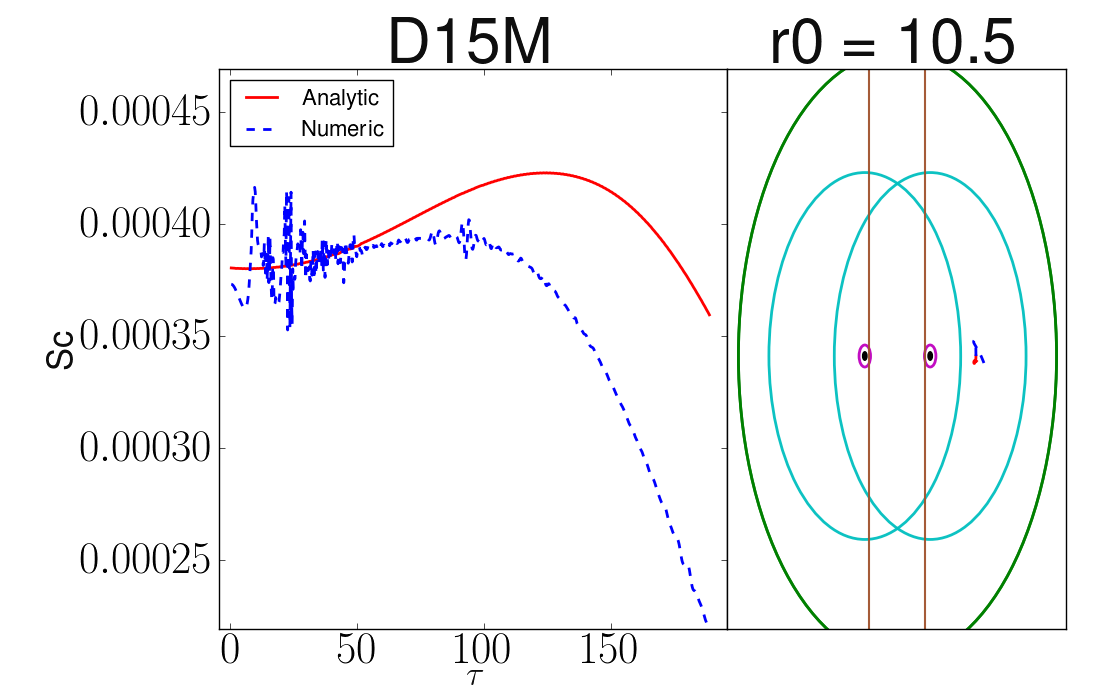}
\includegraphics[width=.30\textwidth]{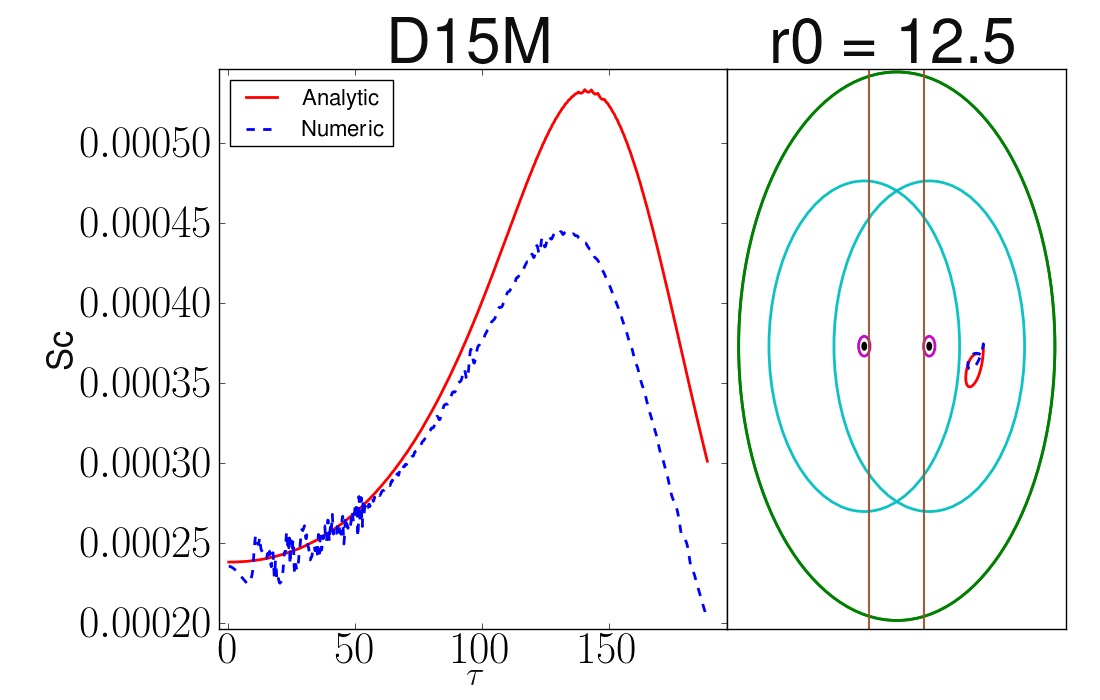}
\includegraphics[width=.30\textwidth]{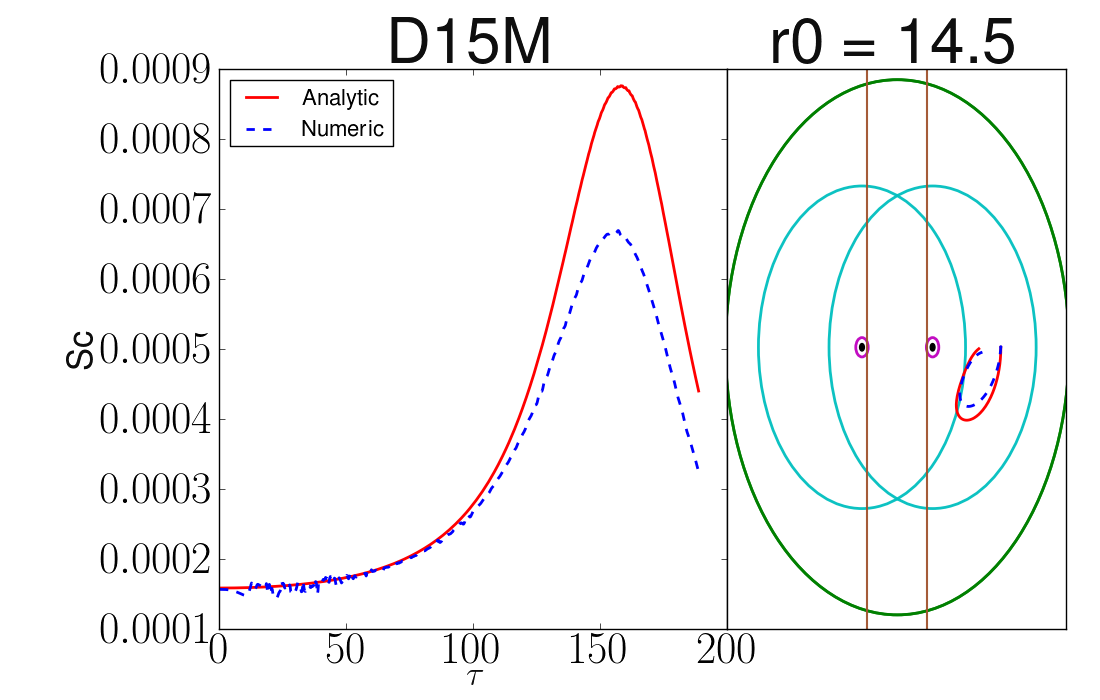}
\includegraphics[width=.30\textwidth]{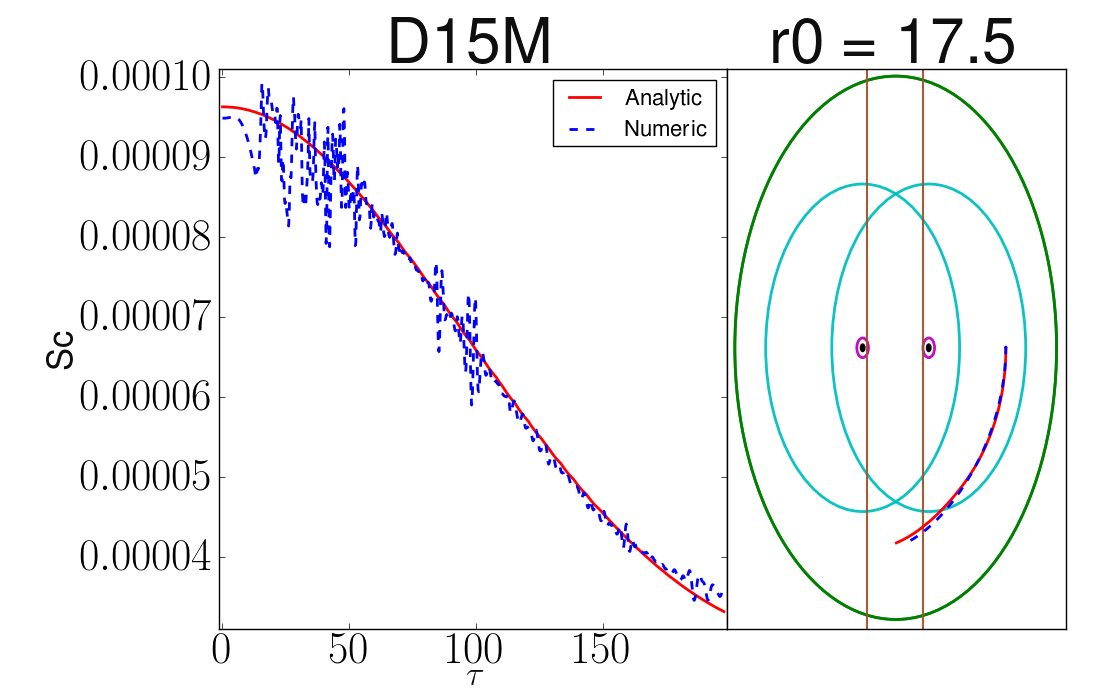}
\includegraphics[width=.30\textwidth]{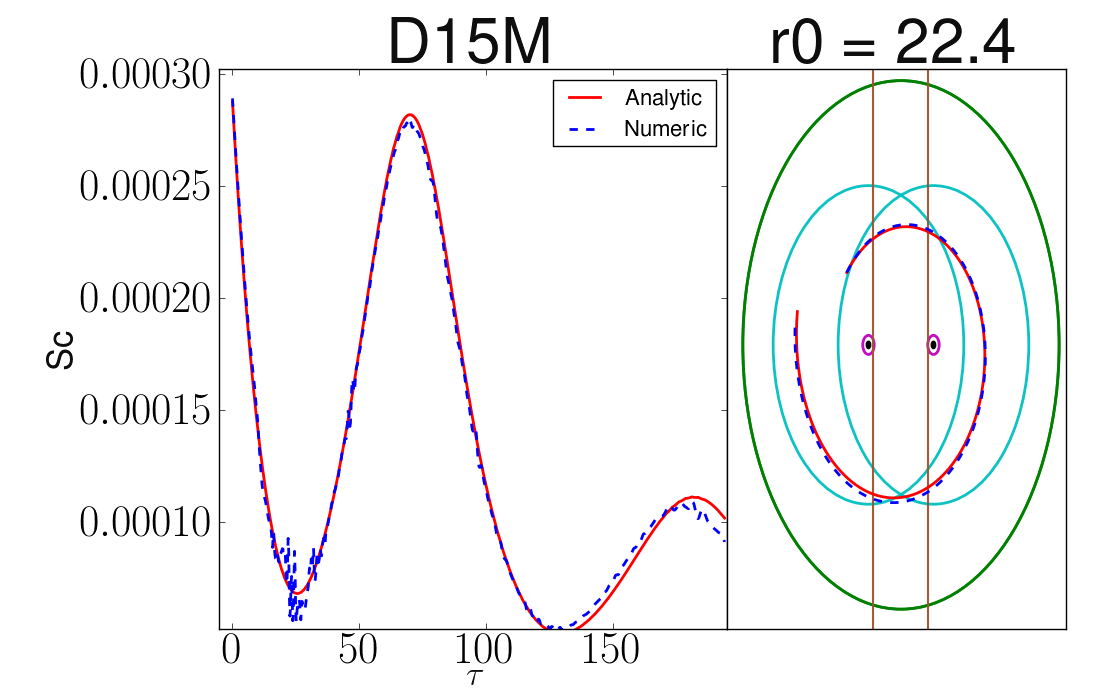}
\includegraphics[width=.30\textwidth]{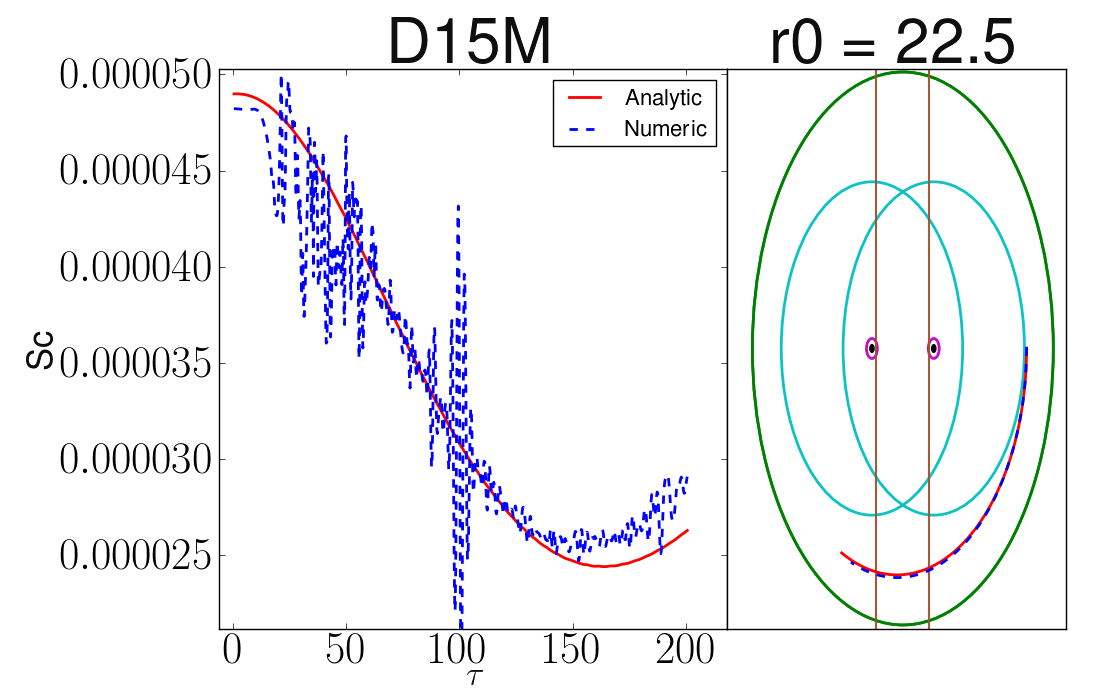}
\includegraphics[width=.30\textwidth]{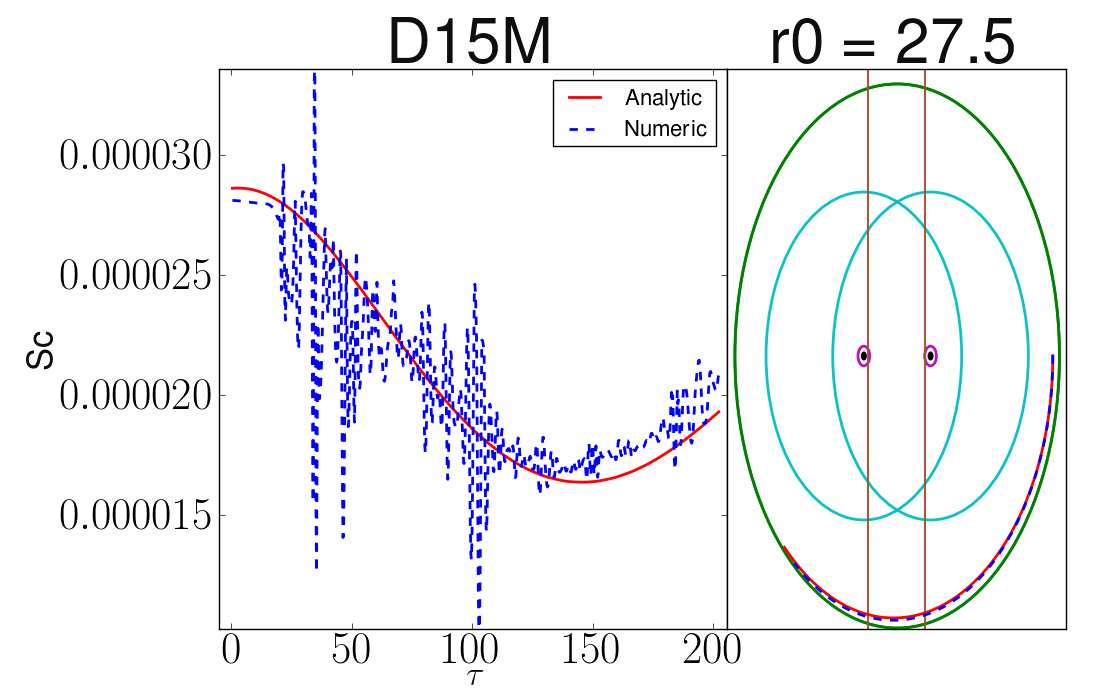}
\includegraphics[width=.30\textwidth]{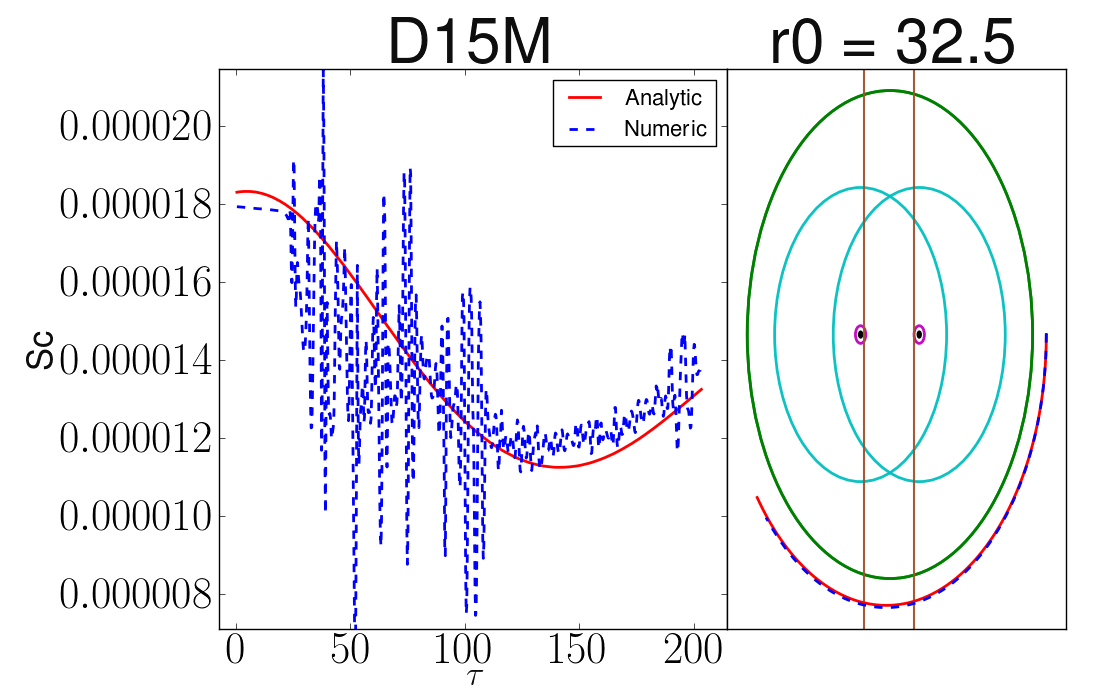}
\includegraphics[width=.30\textwidth]{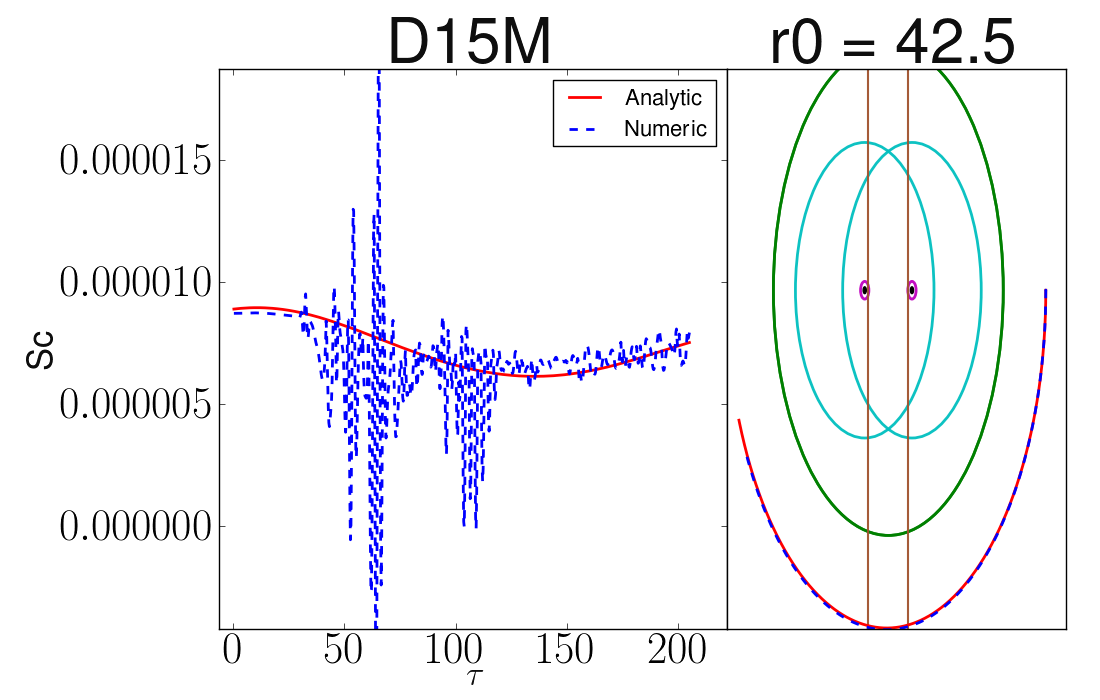}
\includegraphics[width=.30\textwidth]{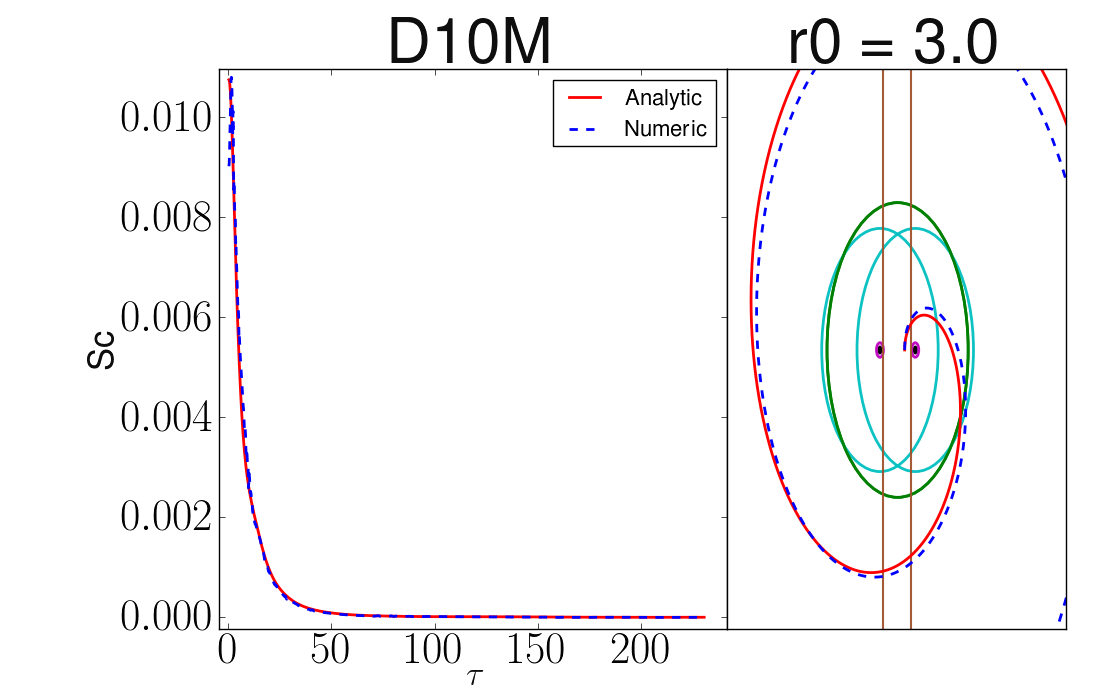}
\includegraphics[width=.30\textwidth]{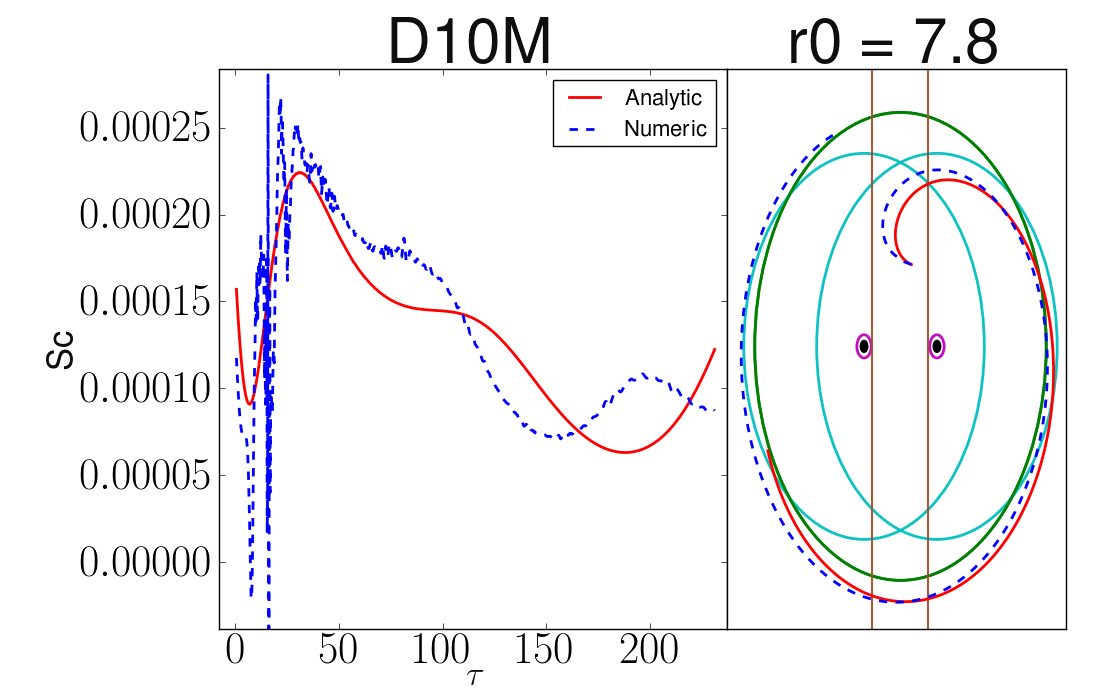}
\includegraphics[width=.30\textwidth]{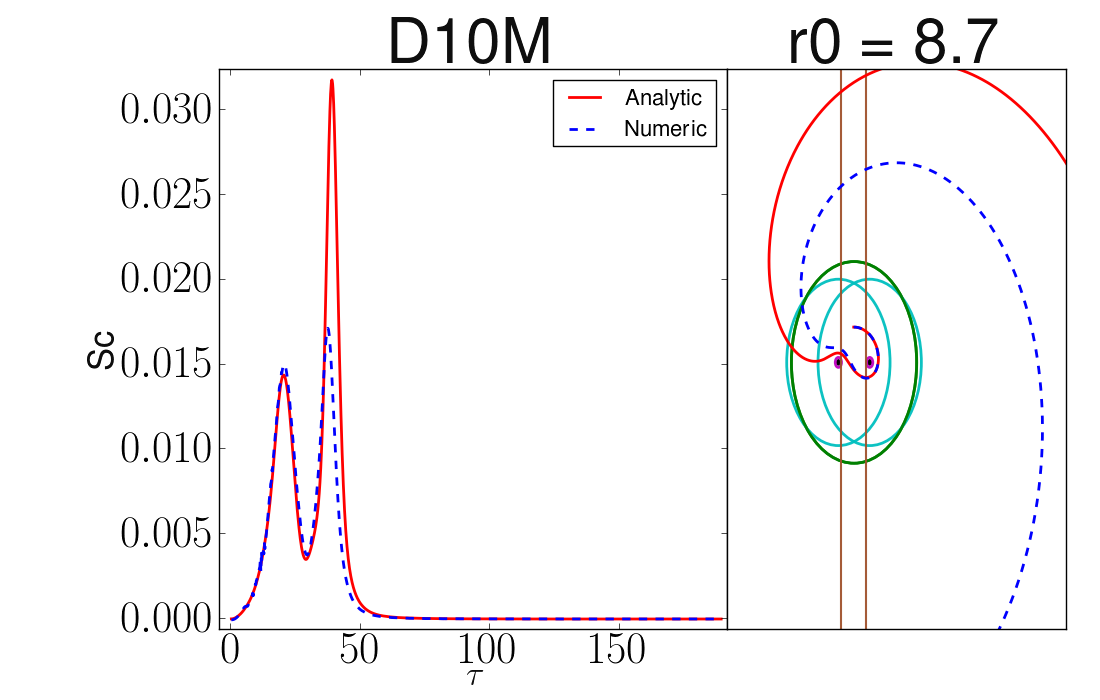}
\includegraphics[width=.30\textwidth]{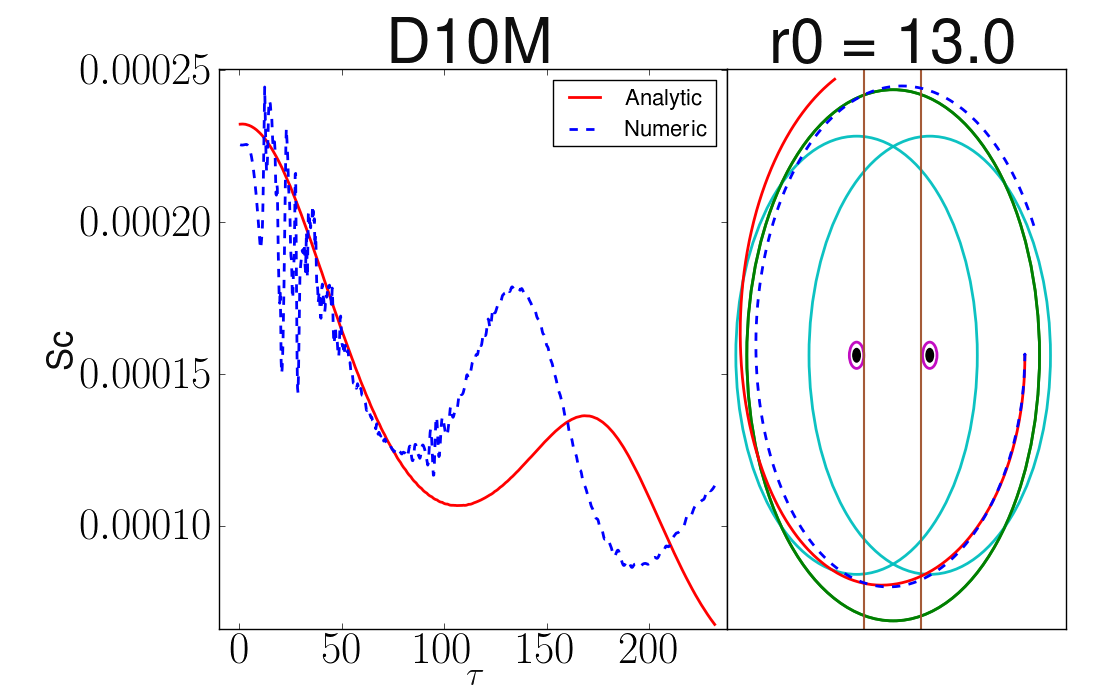}
\includegraphics[width=.30\textwidth]{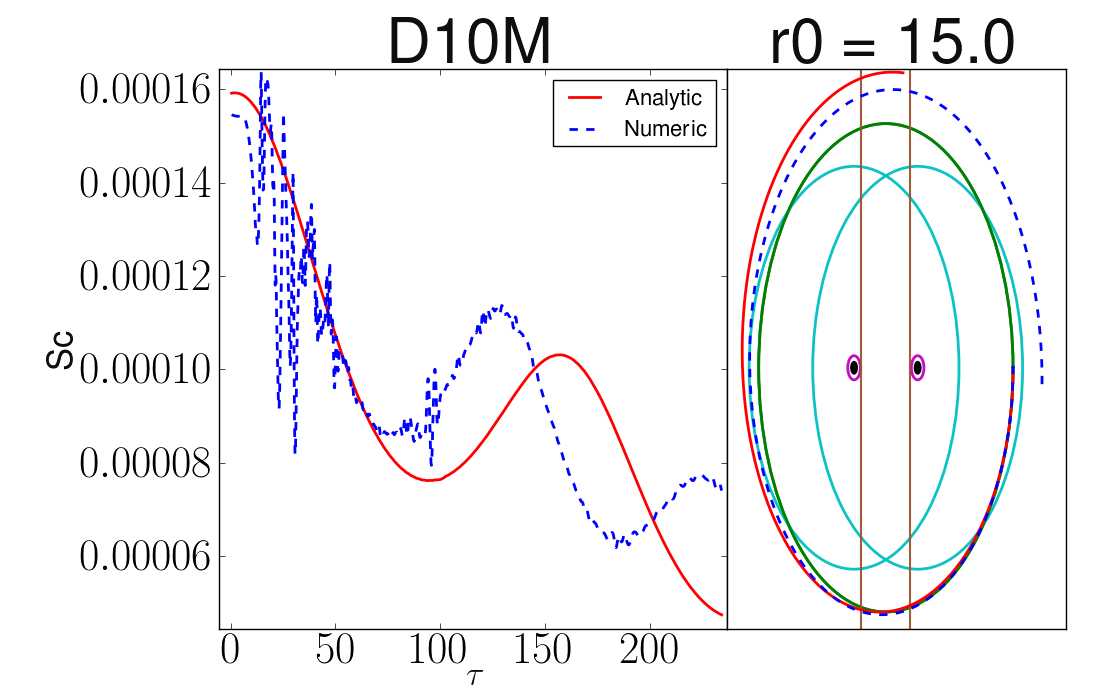}
\includegraphics[width=.30\textwidth]{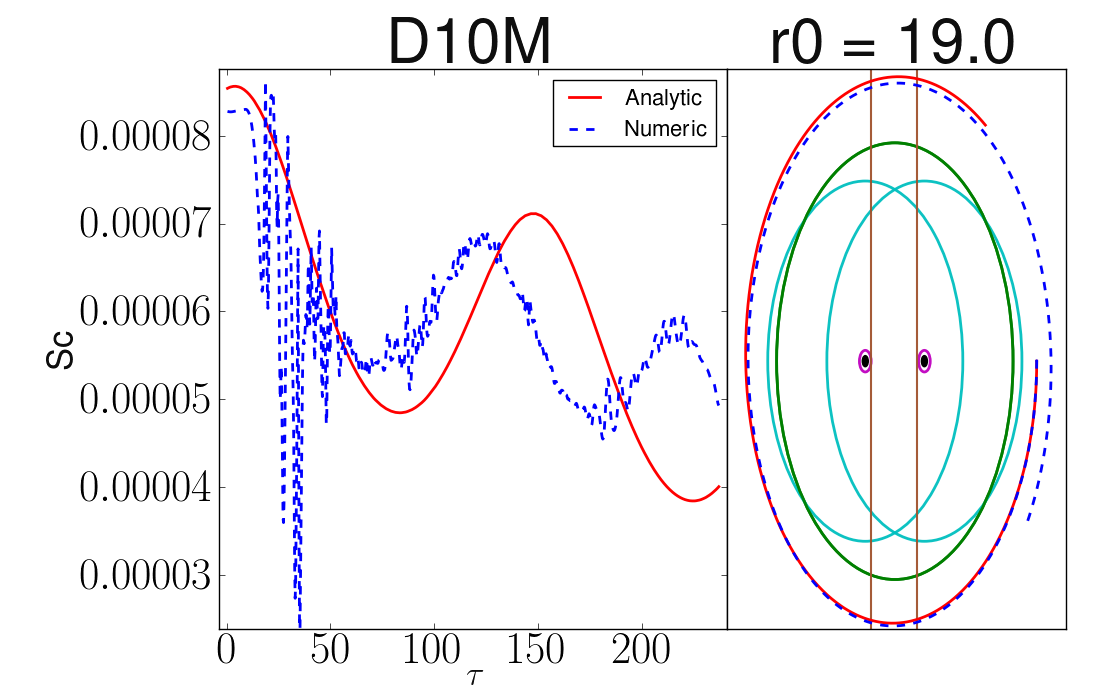}
\includegraphics[width=.30\textwidth]{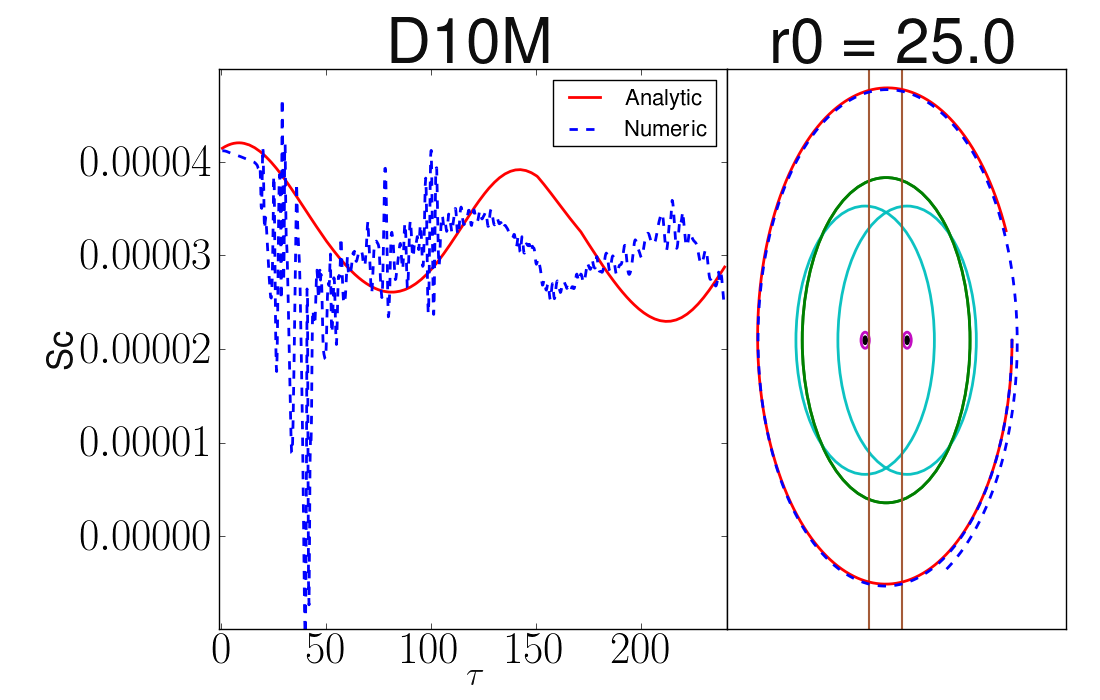}
\includegraphics[width=.30\textwidth]{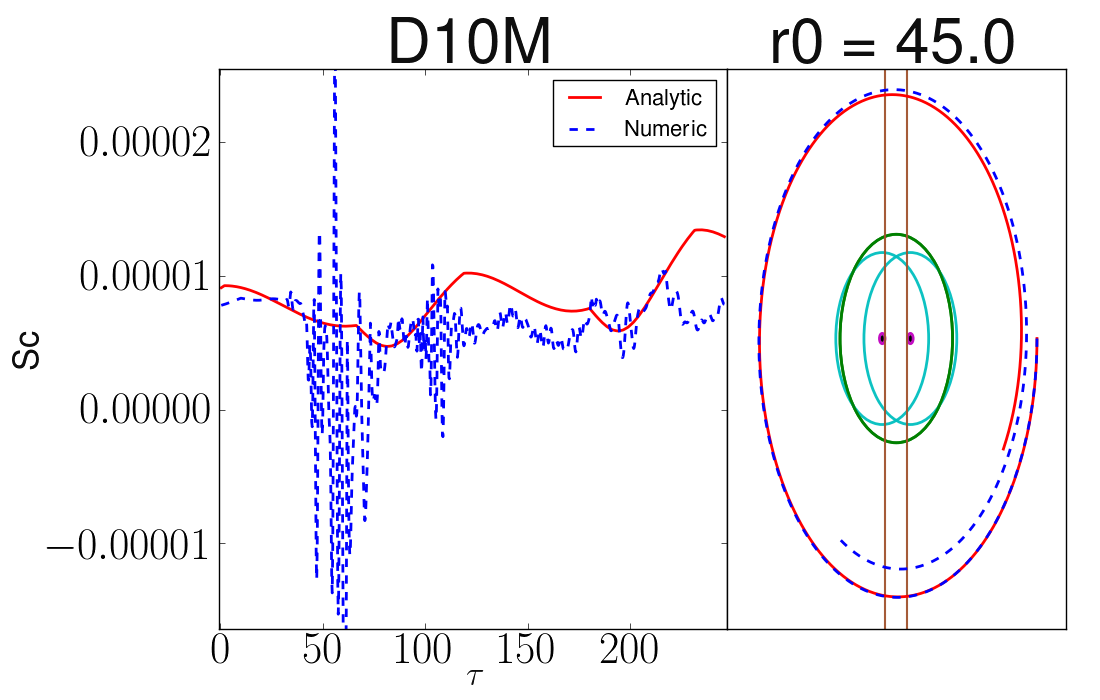}
  \caption{Separation $D=15M$  and $D=10M$ results. Here we plot the value of the
    largest (in magnitude) curvature eigenvalue (\textbf{Sc}) versus time
    (as evolved using the
    numerical and \analytic metric), as well as plot the coordinate position
    of the geodesics {\it in a corotating frame} (i.e., one where the
    BH positions are nearly fixed) on the right side of each
    panel. The vertical lines and circles in these trajectory plots show
    the location of the various zones described in
    Sec.~\ref{sec:analytic_metric}. The number $r_0$ (normalized by $M$)
    is the initial coordinate distance of the geodesic from the nearest BH.
    For the geodesics close to the BHs, the noise in the
    numerically evolved spacetime is low compared to the magnitude of the
    curvature eigenvalues, the opposite is true for the farthest ones.}
  \label{fig:D15andD10}
\end{figure*}

The results from a wide variety of geodesics are shown in
Figs.~\ref{fig:D50}, \ref{fig:D25andD20}, and \ref{fig:D15andD10}.
The figures show one of the curvature eigenvalues (\textbf{Sc}) versus proper time,
$\tau$, for
various starting configurations. The coordinate trajectories of the
geodesics {\it in a corotating frame} are also shown. The boundaries
of the inner, near, far, and buffer zones are denoted by vertical
lines and ellipses.

The noise apparent in the curvature eigenvalues for the geodesics far from the
BHs is due reflections of spurious waves off of the AMR
boundaries (this is the same error associated with high-frequency
oscillations in the waveform seen in numerical evolutions of
BHBs using AMR-based codes). At far distances, this
noise is larger in magnitude than the curvature eigenvalues. See, for example, the
$r_0\gtrsim 100M$ curves for the $D=50M$ configuration in Fig.~\ref{fig:D50}.

At a separation of $D=50M$ in Fig.~\ref{fig:D50},
one would expect very good agreement
between the \analytic metric and the numerical one. Quantitatively,
there is remarkably good agreement when the geodesics are about
$20M \lesssim r_0 \lesssim 100M $ from the BHs.
Closer than this, there are
small, but noticeable differences, and farther than this, there is
some evidence significant differences, but in those cases the noise is
significantly larger than the curvature eigenvalues themselves.

We find that initial conditions that
lead to nearly circular geodesics for one metric do lead to nearly
circular geodesics for the other. The best agreement here are for
geodesics in the outer regions of the inner-to-near-zone buffer
regions and the near zone.

At a binary separation of $D=25M$ (see Fig.~\ref{fig:D25andD20}),
the disagreement between the analytical and numerical results  when the geodesics are close are exacerbated.
Good agreement between the curvature eigenvalues is still apparent in the outer part
of the inner-to-near zone buffer region and the near zone,
although at late times these 
geodesics show deviations as shown for the $r_0>40M$ cases.
At a binary separation of $D=10M$ (see Fig.~\ref{fig:D15andD10}),
there are noticeable differences in the curvature
eigenvalues for almost all geodesics.
However, examining the $D=15M$ geodesics (see
Fig.~\ref{fig:D15andD10}) shows something perhaps
surprising. The geodesics in the outer part
of the inner-to-near zone buffer region and the near zone are
remarkably good. Here, the deviations for far geodesics also start to
deviate at later times. It seems these deviations are dependent
on the binary separation. One may have expected these geodesics to be
substantially worse than the $D=25M$ and $D=20M$ analogs,
but we do not see this.

\begin{figure*}
  \includegraphics[width=.32\textwidth]{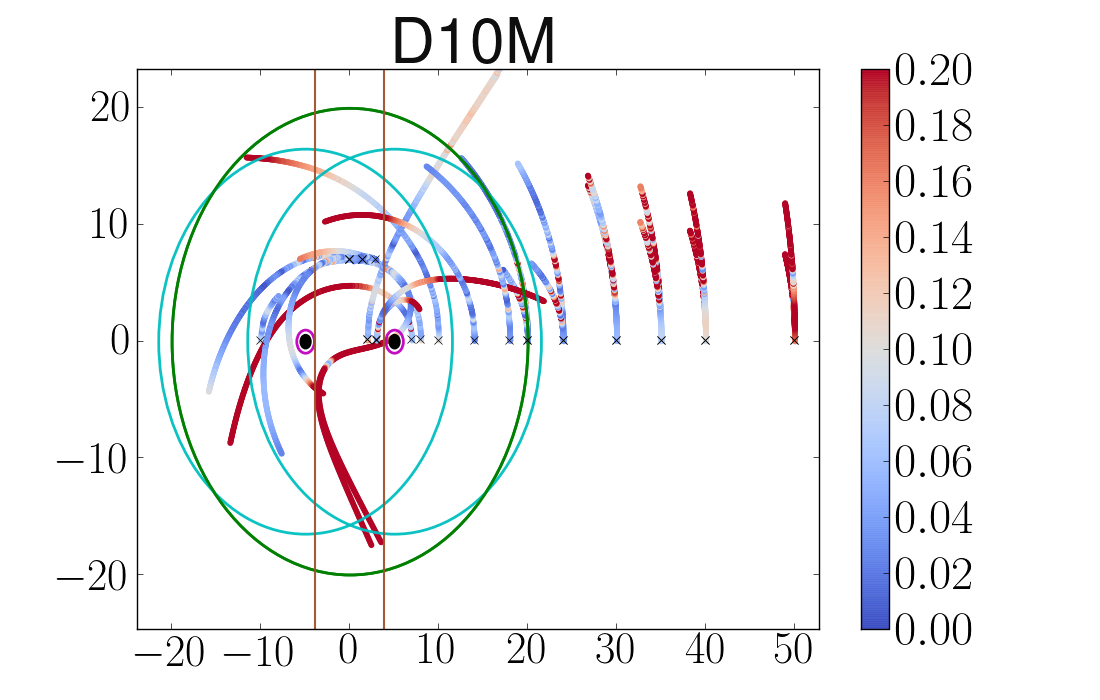}
  \includegraphics[width=.32\textwidth]{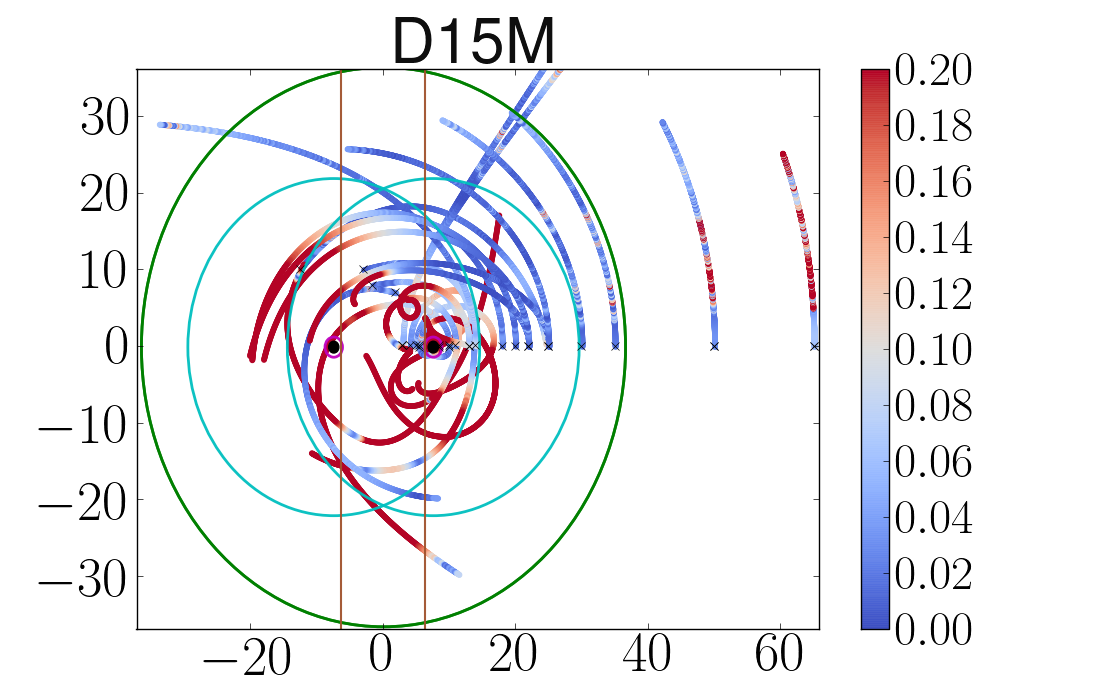}
  \includegraphics[width=.32\textwidth]{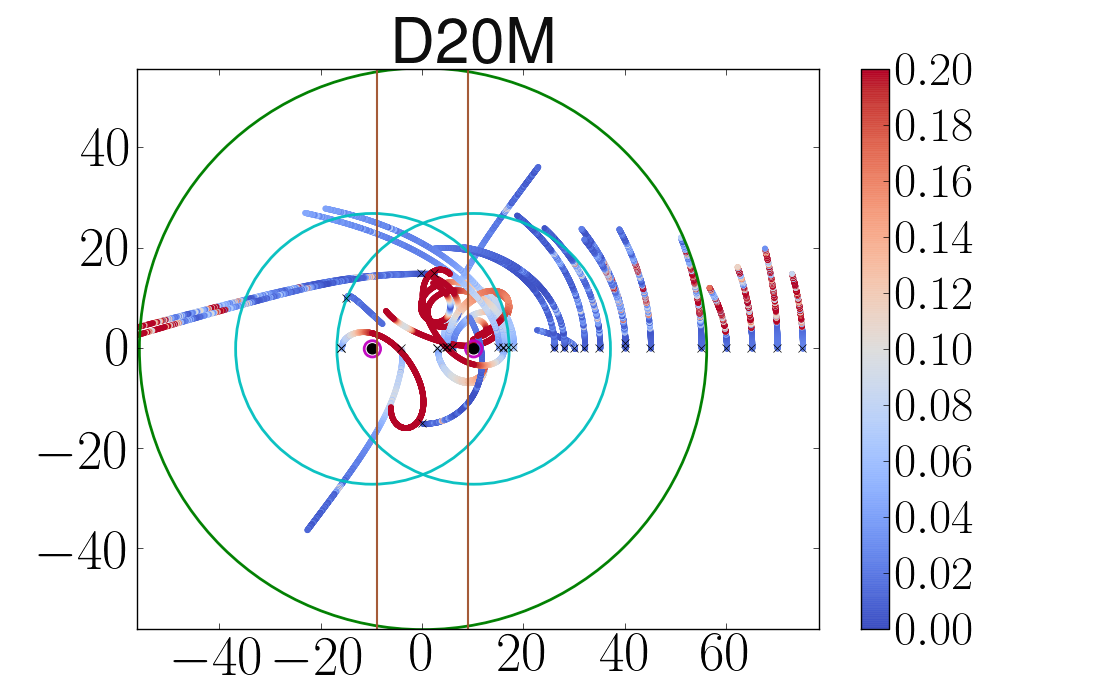}
  \includegraphics[width=.32\textwidth]{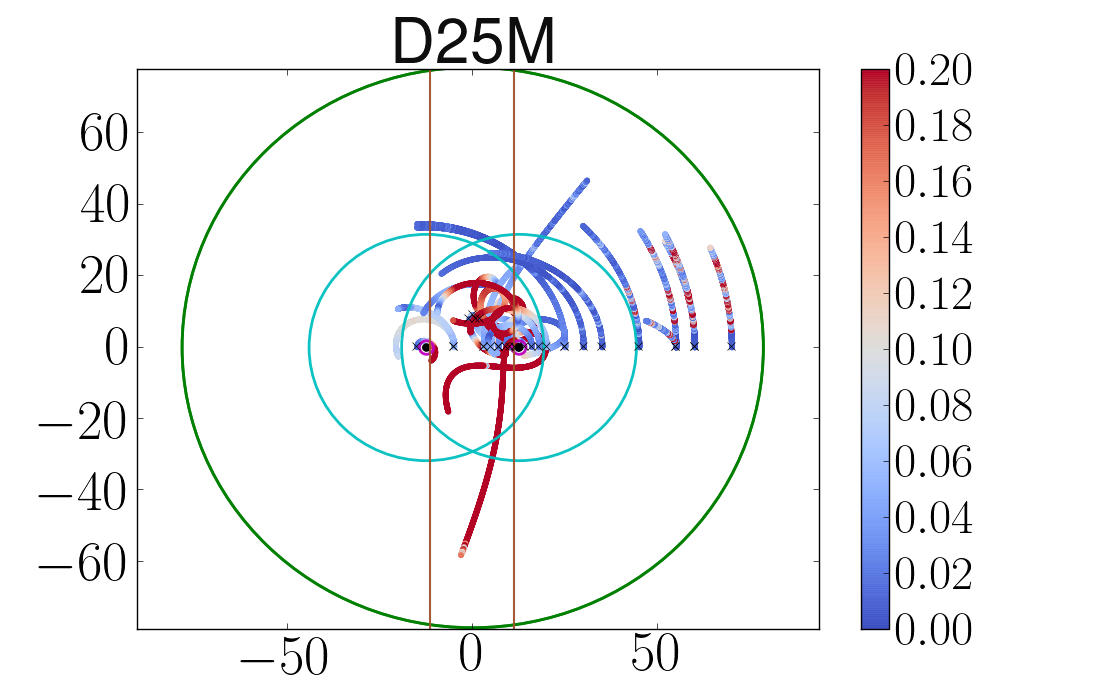}
  \includegraphics[width=.32\textwidth]{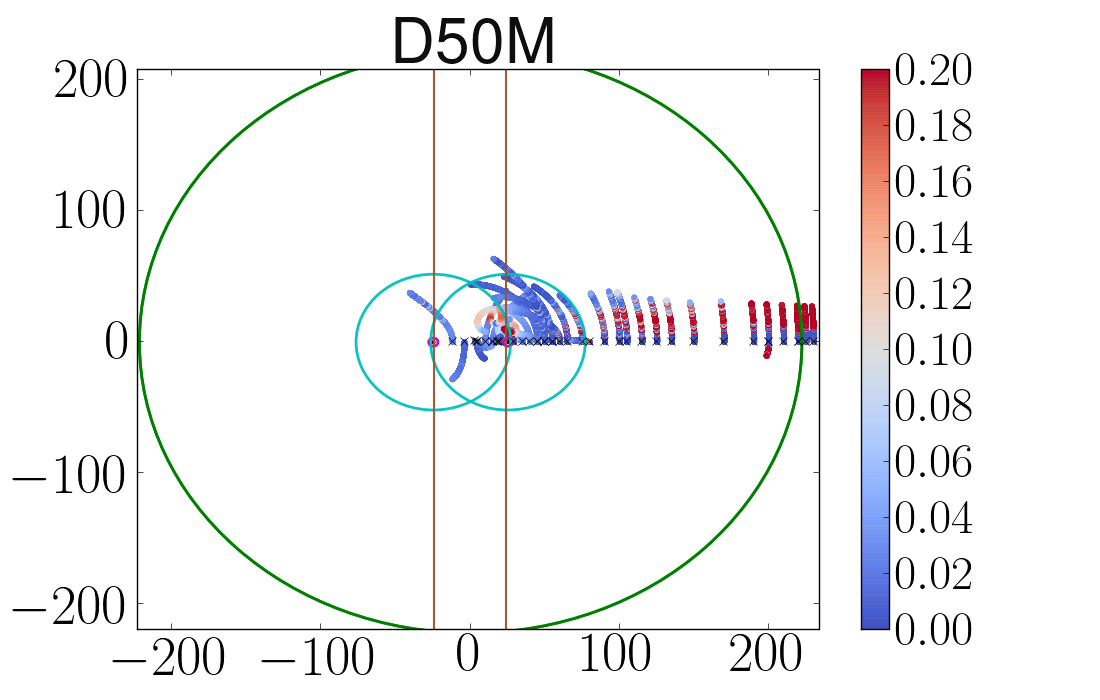}

 \includegraphics[width=.32\textwidth]{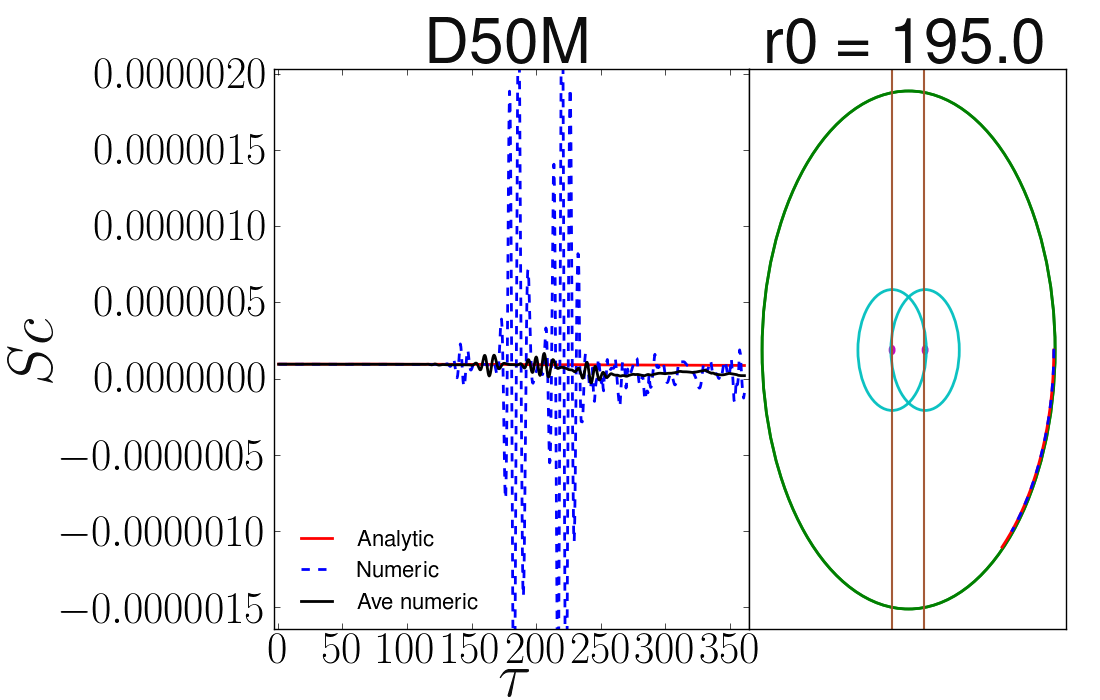}
  \includegraphics[width=.32\textwidth]{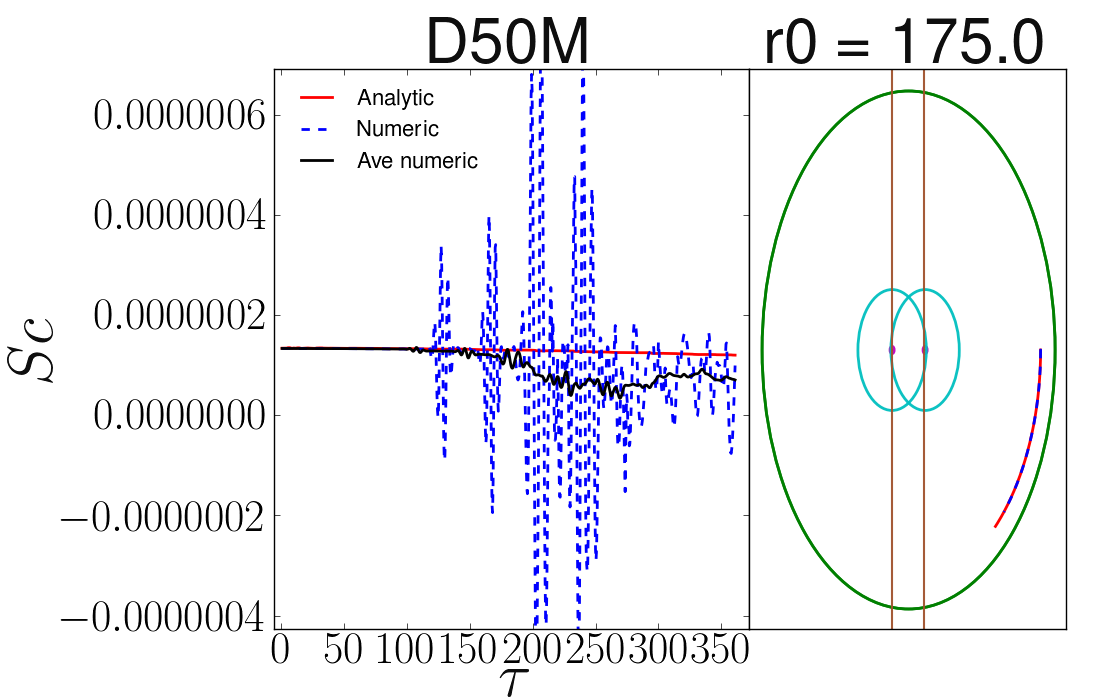}
  \includegraphics[width=.32\textwidth]{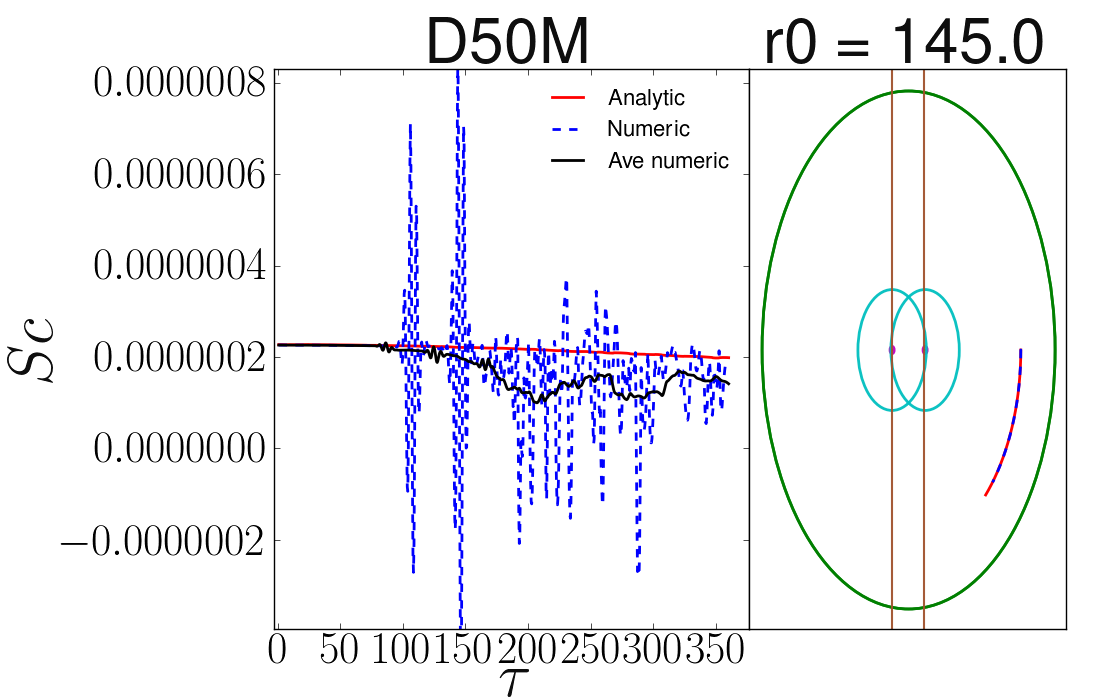}
  \includegraphics[width=.32\textwidth]{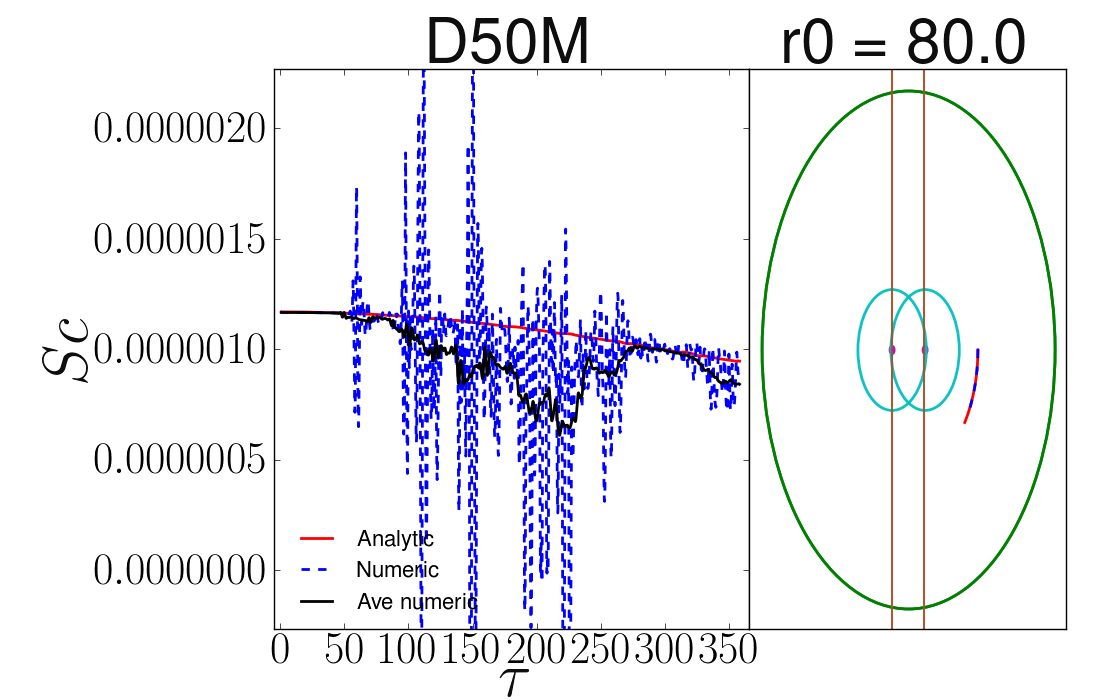}

  \caption{(Top two rows) A summary of the results. The plots show the trajectories of
    the geodesics {\it in a non-corotating frame}. The color scale gives
    the relative differences between the curvature eigenvalues (\textbf{Sc}) as
    calculated using the numerical (and smoothed by a running average)
    and analytic metrics. Note that the
    color changes from blues to reds at 10\% relative difference.
    (Bottom two rows)
    Plots showing
    curvature eigenvalues as calculated using the
    analytical and numerical metrics, as well as a running average of the
    latter. There are hints here of systematic differences between the
    analytical and numerical results. However, as can be seen, the noise
    is much larger than these differences. 
  }
  \label{fig:summary_and_smooth}
\end{figure*}

In Fig.~\ref{fig:summary_and_smooth}, we show the relative difference between the
curvature eigenvalues calculated using the numerical and analytic metrics.
Here, we use a running average to smooth out the noise.
Note that the color scale changes from blue to red at a 10\% relative
difference. From these plots, we can see that the buffer zone between
the inner and near zones, as well as the near zone itself shows the
smallest relative errors. The near-to-far zone buffer region (no plot
shows the far zone) is generally worse, as least in terms of relative
errors, than the near zone. The large relative differences seen for
the $D=50M$ case may be due to noise, but as seen in
Fig.~\ref{fig:summary_and_smooth}, there are hints of systematic differences between
the analytic and numerical spacetimes. Note that in
Fig.~\ref{fig:summary_and_smooth}, we plot the geodesics {\it in a non-corotating
frame}. The reason for this is, that while plotting in a corotating
frame allows us to see which zones the geodesics pass through, it also
gives a false sense of how far in (quasi) inertial coordinates
the geodesics actually traveled. 
By comparing the plots in Fig.~\ref{fig:summary_and_smooth} with
Figs.~\ref{fig:D50}--\ref{fig:D15andD10}, one can get a more accurate
of the actual motion of each geodesic.

\begin{figure*}
  \includegraphics[width=.32\textwidth]{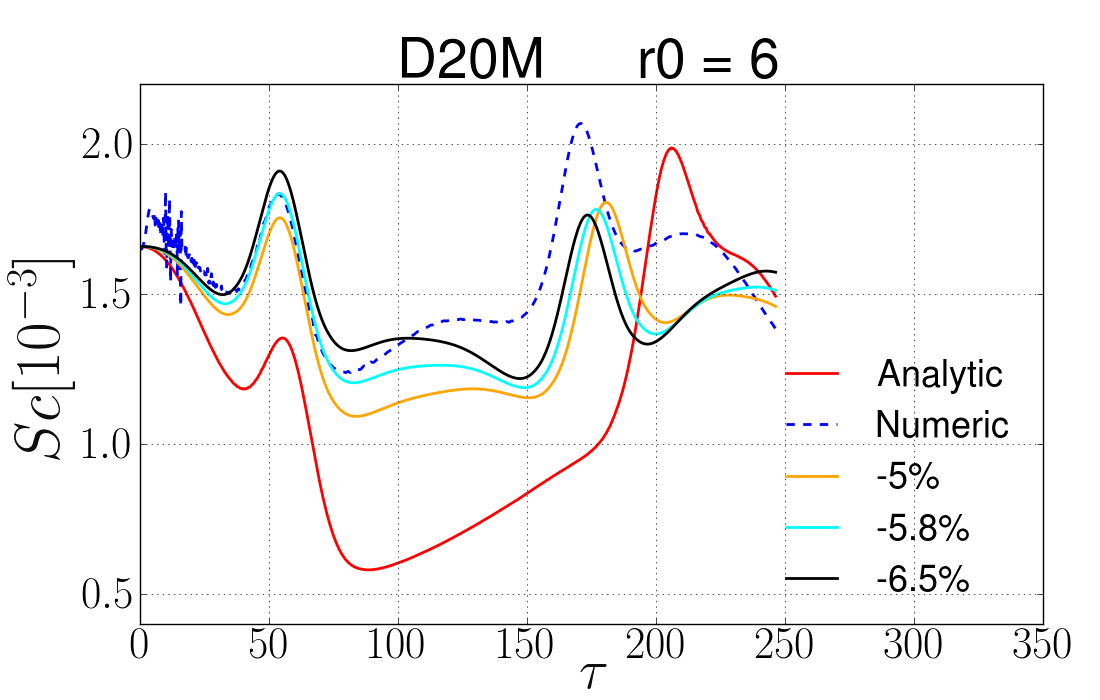}
  \includegraphics[width=.32\textwidth]{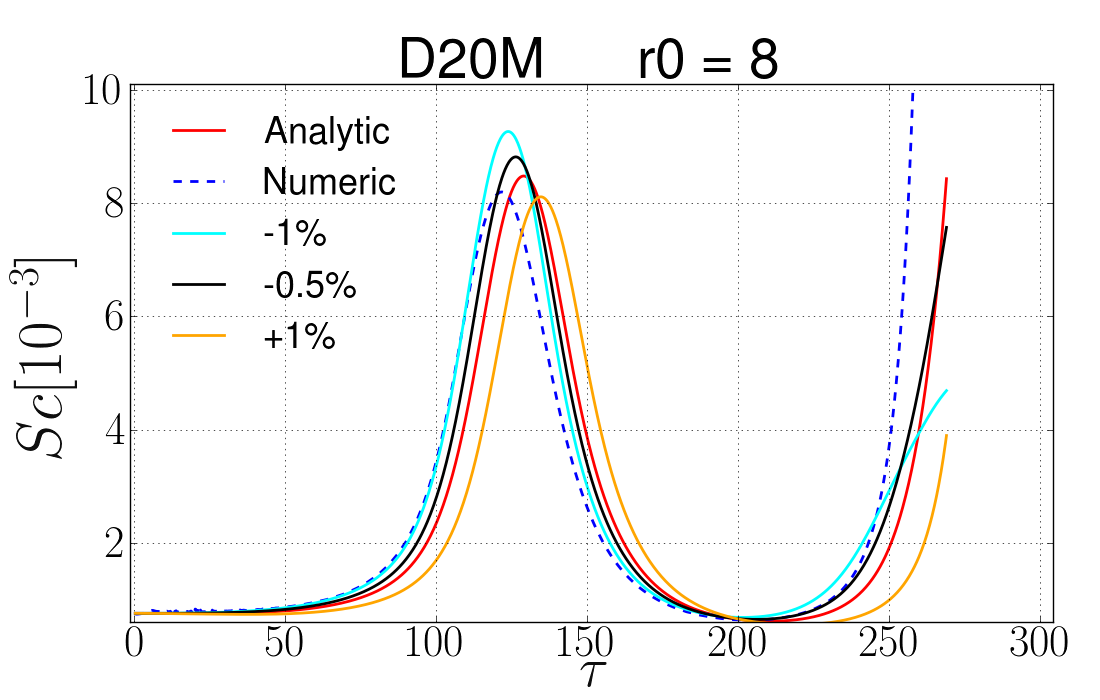}
  \includegraphics[width=.32\textwidth]{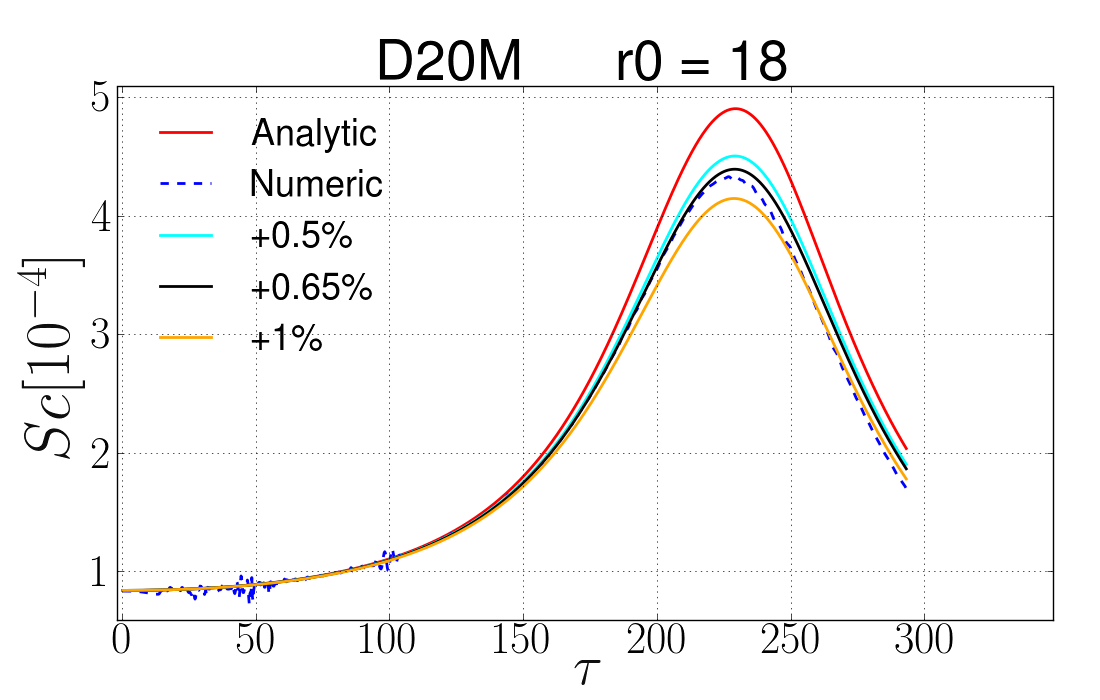}
  \includegraphics[width=.32\textwidth]{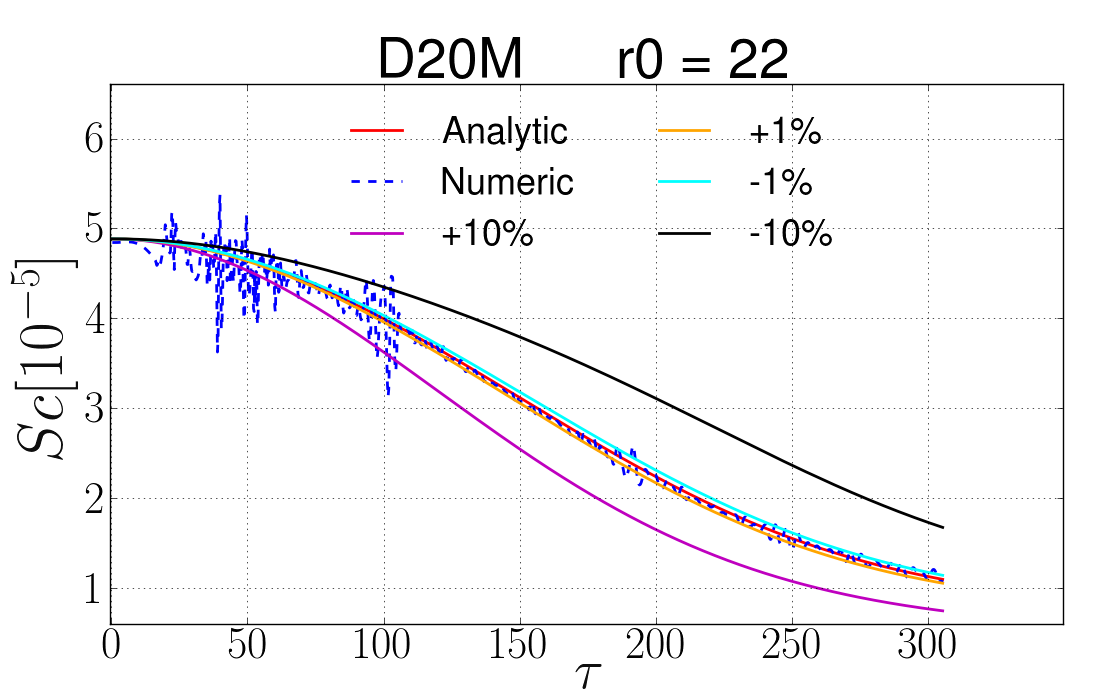}
  \includegraphics[width=.32\textwidth]{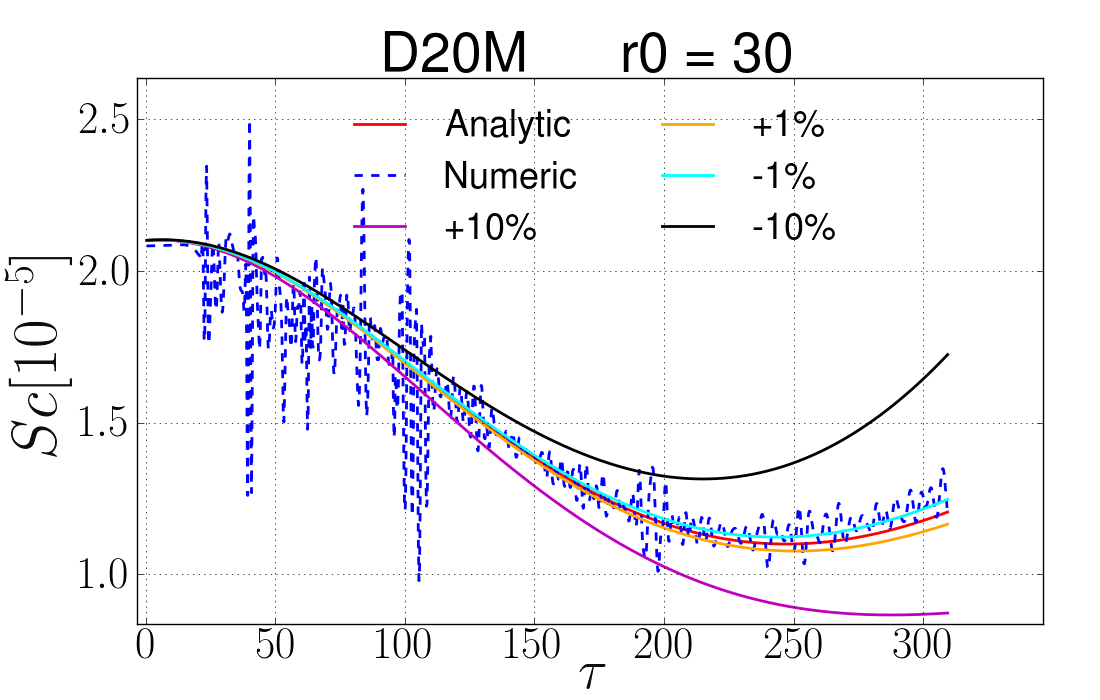}
  \includegraphics[width=.32\textwidth]{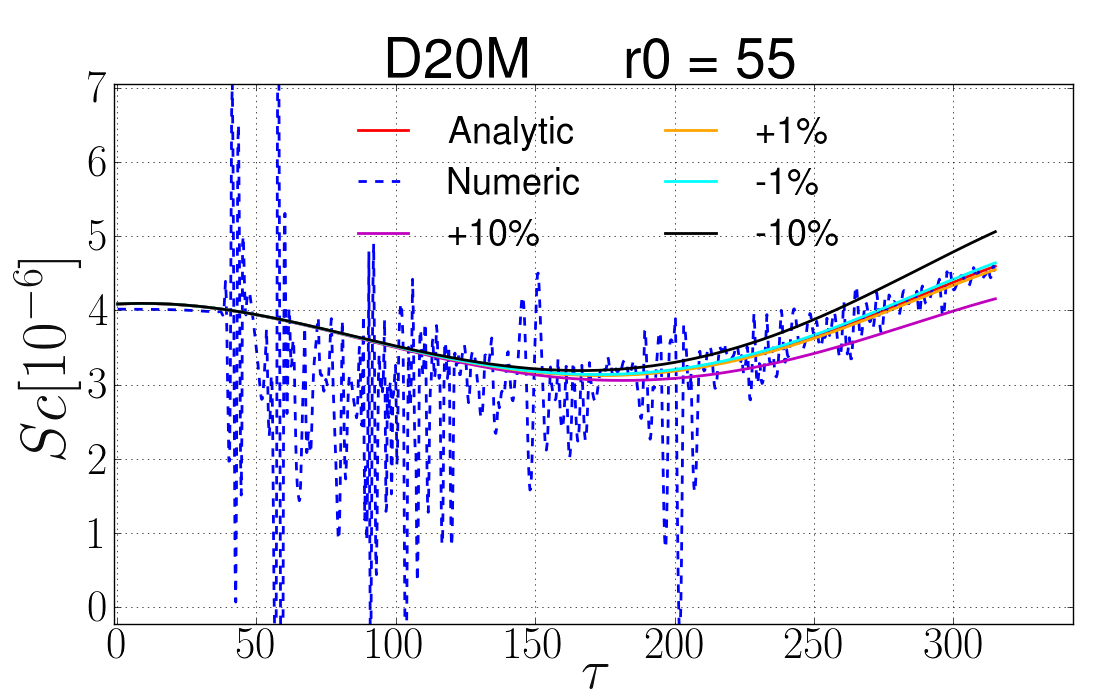}
  \includegraphics[width=.32\textwidth]{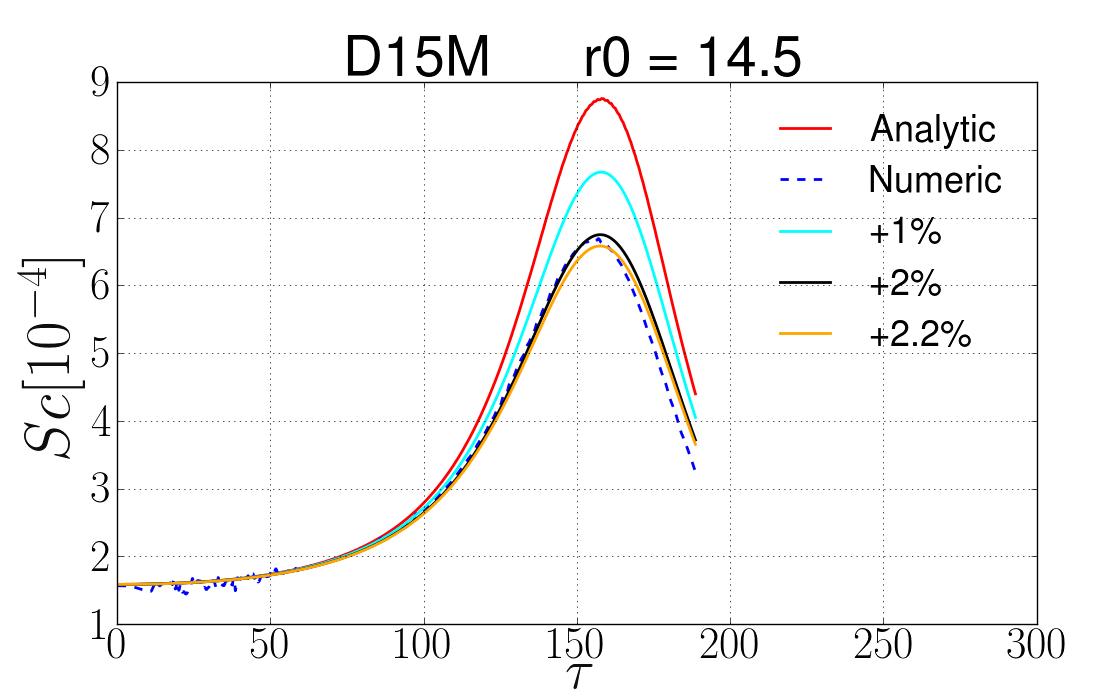}
  \includegraphics[width=.32\textwidth]{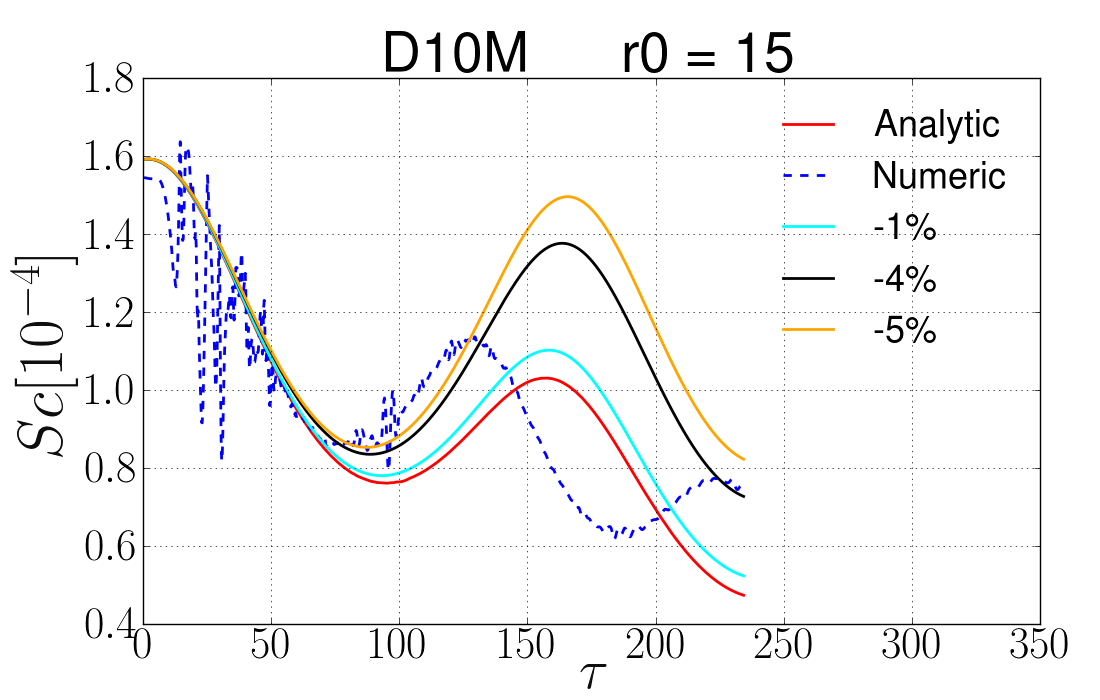}
  \caption{Curvature eigenvalues (\textbf{Sc}) for numerical and  analytical
    spacetimes. For the analytical spacetime, we perturb the initial
    velocity of geodesics by the factors shown in the graphs.
    The dotted blue curve is the numerical result
    with velocity associated with the unperturbed analytic geodesic.
    As can be seen, the larger the value of the
    eigenvalues (i.e., geodesic deviation) the larger the effect of a
    $\pm10\%$ perturbation. On the other hand, with small perturbations,
    we were able to find geodesics in the analytical spacetime that
    closely matched the dynamics (time dependence the eigenvalue) of the
    numerical one for geodesics farther than $r_0 \sim 10M$ from the BHs.}
\label{fig:stability_test}
\end{figure*}

\subsection{Comparing first and second-order matched spacetime}

There are two versions of the analytic metric presented above. The
standard one, known as the second-order metric, uses higher-order PN
terms in the near zone, and matches the $\ell=2$ and $\ell=3$
multipoles in the inner zone. The first-order metric, which we will
explore below, uses lower-order PN terms and only matches the $\ell=2$
multipoles in the inner zone.

\begin{figure*}
\includegraphics[width=.24\textwidth]{Revisedb50geo2.png}
\includegraphics[width=.24\textwidth]{Revisedb50geo3.png}
\includegraphics[width=.24\textwidth]{Revisedb50geo7.png}
\includegraphics[width=.24\textwidth]{Revisedb50geo9.png}

\includegraphics[width=.24\textwidth]{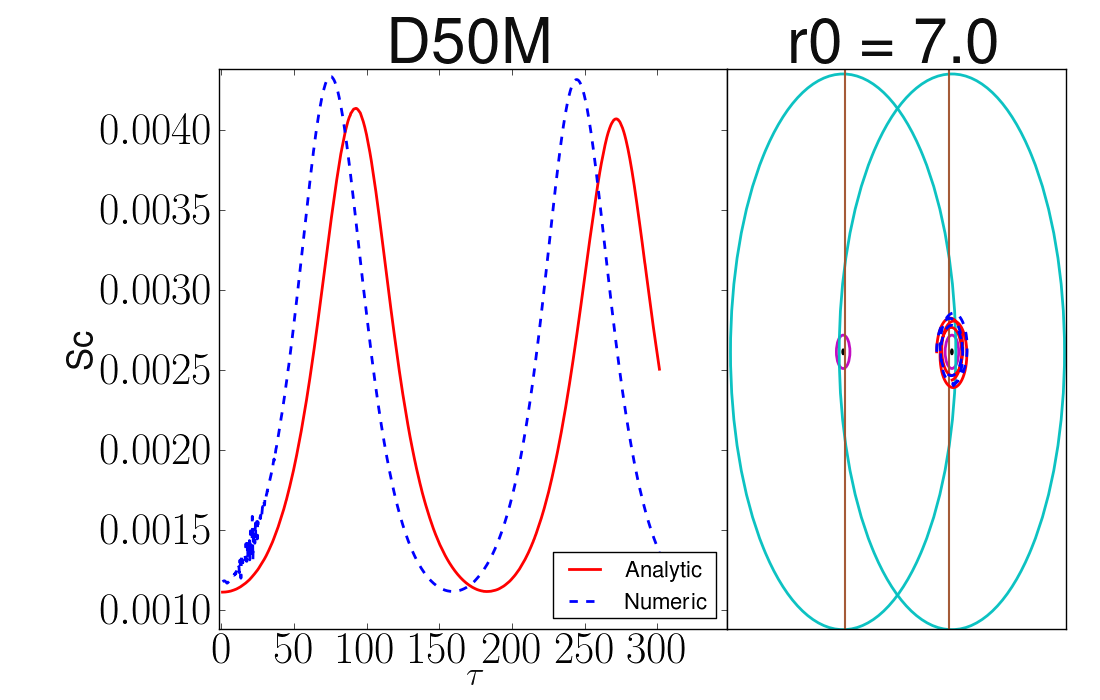}
\includegraphics[width=.24\textwidth]{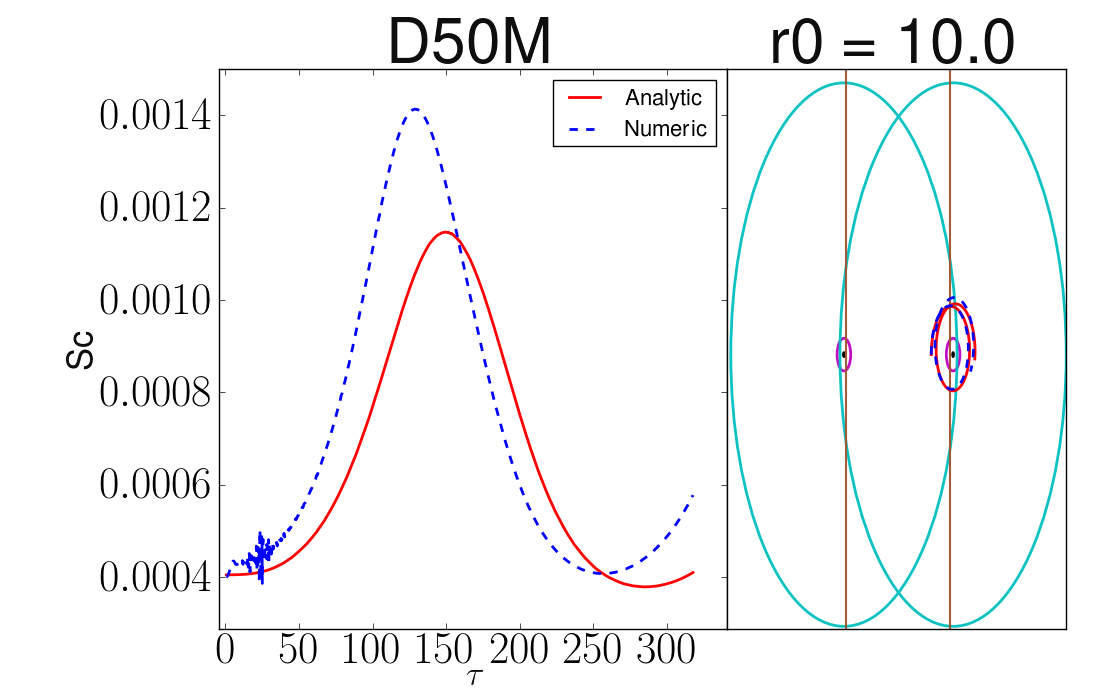}
\includegraphics[width=.24\textwidth]{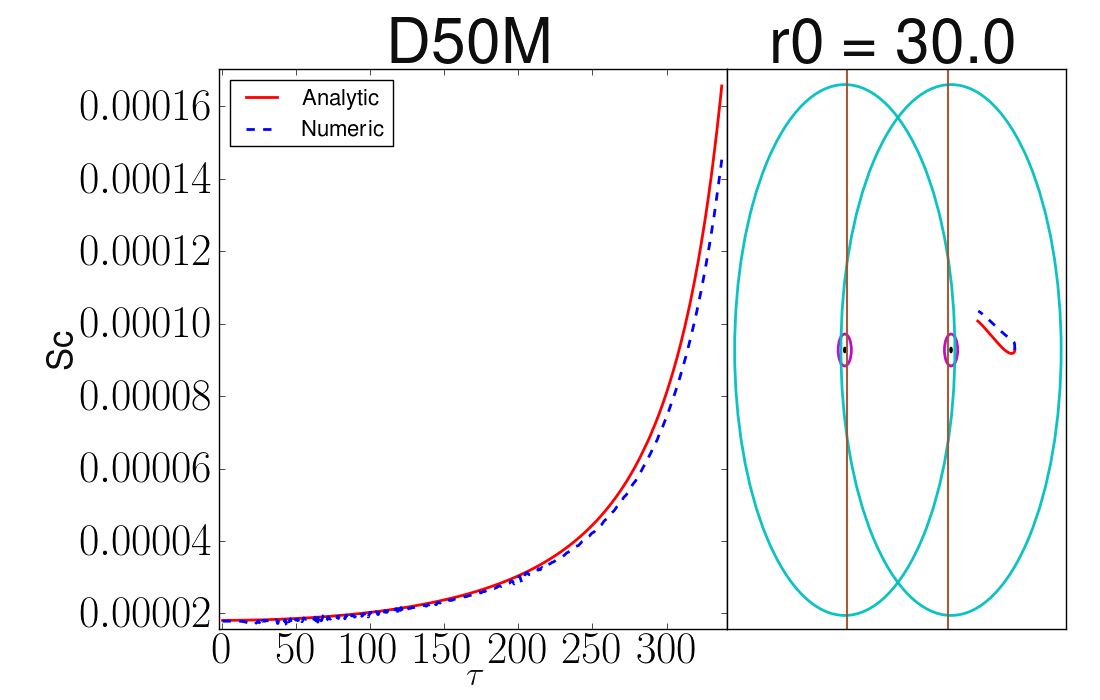}
\includegraphics[width=.24\textwidth]{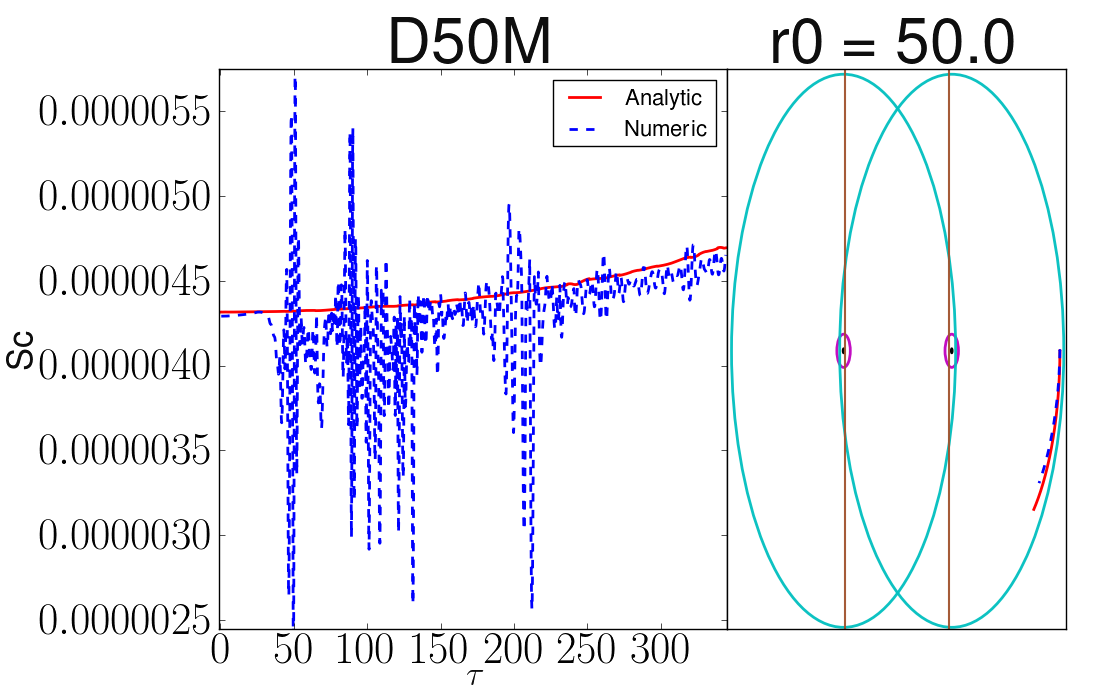}
\caption{A comparison of how well the curvature eigenvalues of the
  second-order metric and first-order metric agree with the
  eigenvalues of the associated numerical metrics for the $D=50M$ case.
  The top row shows
  the second-order results (which were previously shown in
  Fig.~\ref{fig:D50}). The bottom row shows the first-order results. Note
  that at larger distances from the black holes the two results are
  comparable, while at closer distances the second-order results are
qualitatively better.}\label{fig:D501st}
\end{figure*}

\begin{figure*}
\includegraphics[width=.24\textwidth]{Revisedb20geo4.png}
\includegraphics[width=.24\textwidth]{Revisedb20geo6.png}
\includegraphics[width=.24\textwidth]{Revisedb20geo7.png}
\includegraphics[width=.24\textwidth]{Revisedb20geo9.png}

\includegraphics[width=.24\textwidth]{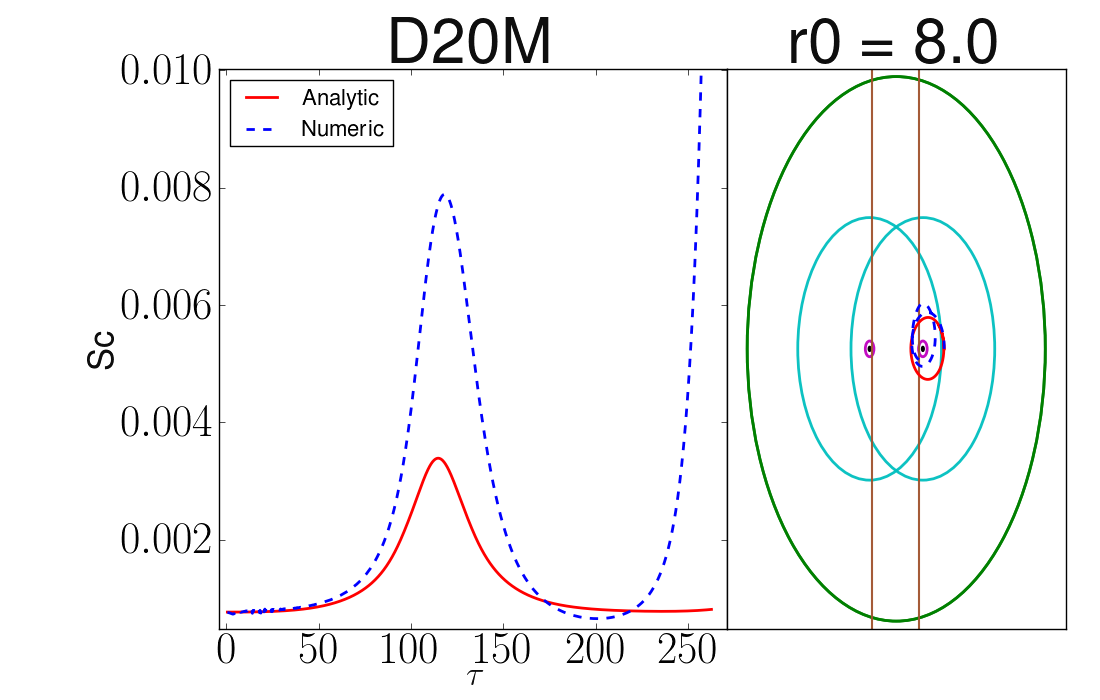}
\includegraphics[width=.24\textwidth]{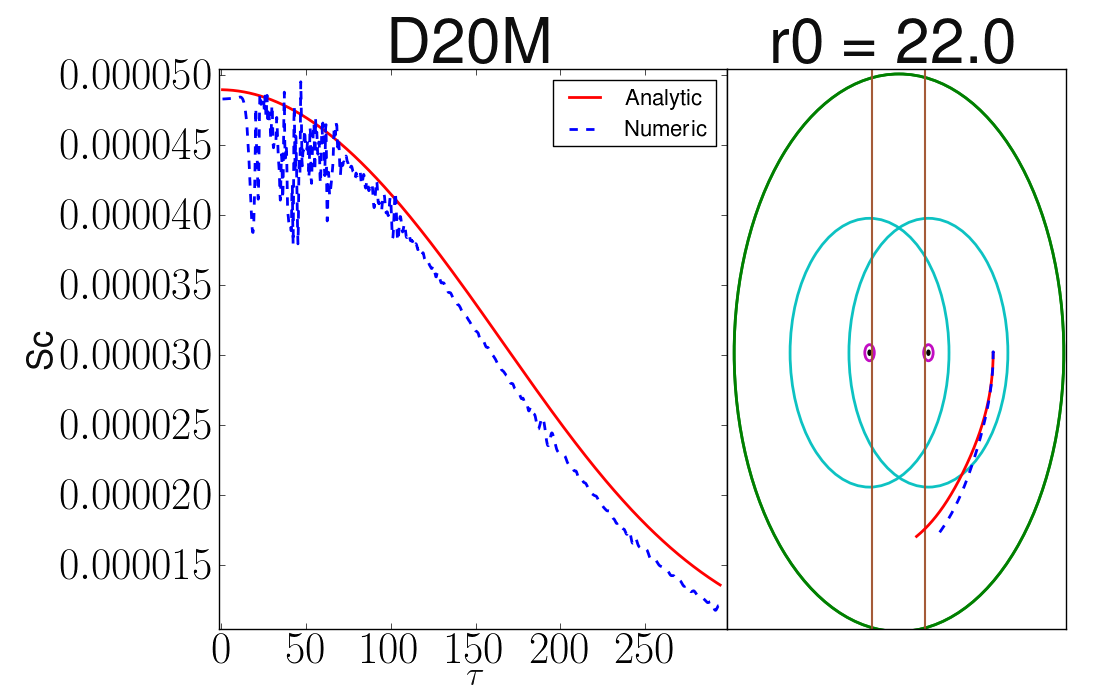}
\includegraphics[width=.24\textwidth]{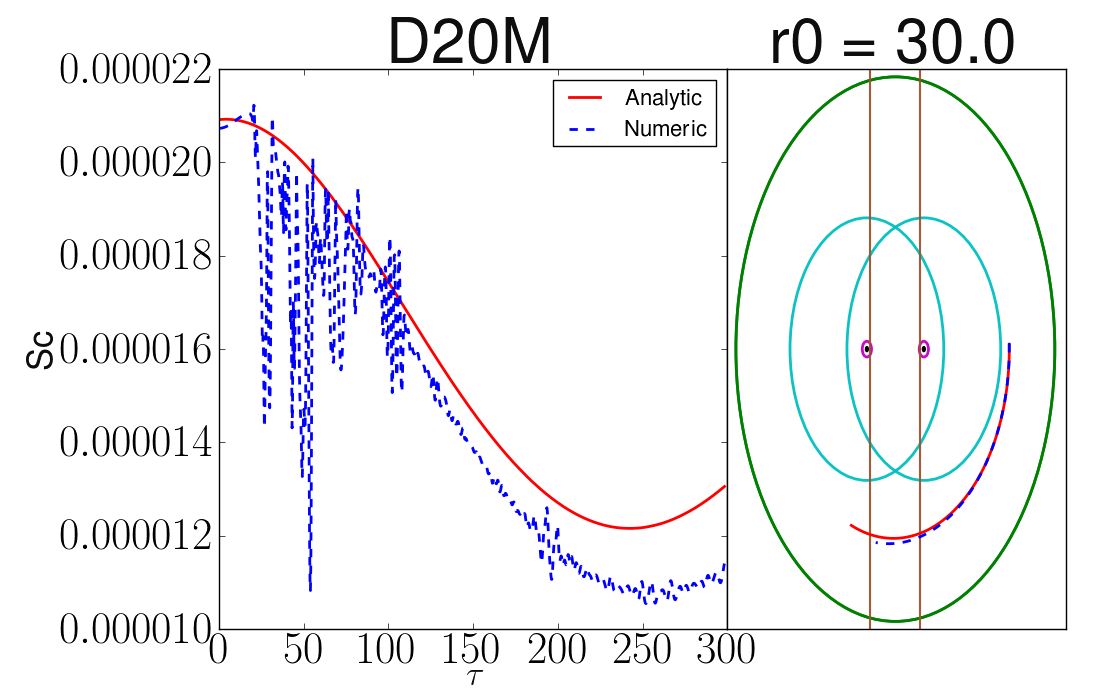}
\includegraphics[width=.24\textwidth]{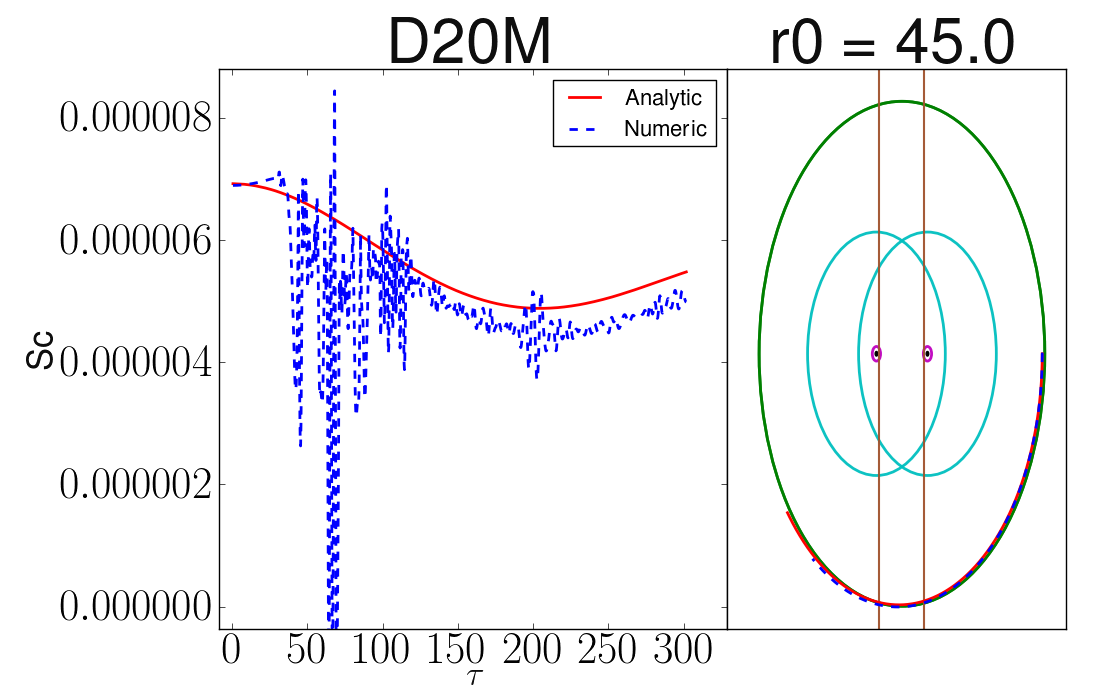}
\caption{A comparison of how well the curvature eigenvalues of the
  second-order metric and first-order metric agree with the
  eigenvalues of the associated numerical metrics for the $D=20M$ case.
  The top row shows
  the second-order results (which were previously shown in
  Fig.~\ref{fig:D25andD20}). The bottom row shows the first-order results. Unlike
  for the $D=50M$ case, there are significant differences between the
  analytical and numerical scalars for the first-order metric even at
  larger distances from the black holes.
}\label{fig:D201st}
\end{figure*}

\begin{figure*}
\includegraphics[width=.24\textwidth]{Revisedb15geo2.png}
\includegraphics[width=.24\textwidth]{Revisedb15geo6.png}
\includegraphics[width=.24\textwidth]{Revisedb15geo9.png}
\includegraphics[width=.24\textwidth]{Revisedb15geo10.png}

\includegraphics[width=.24\textwidth]{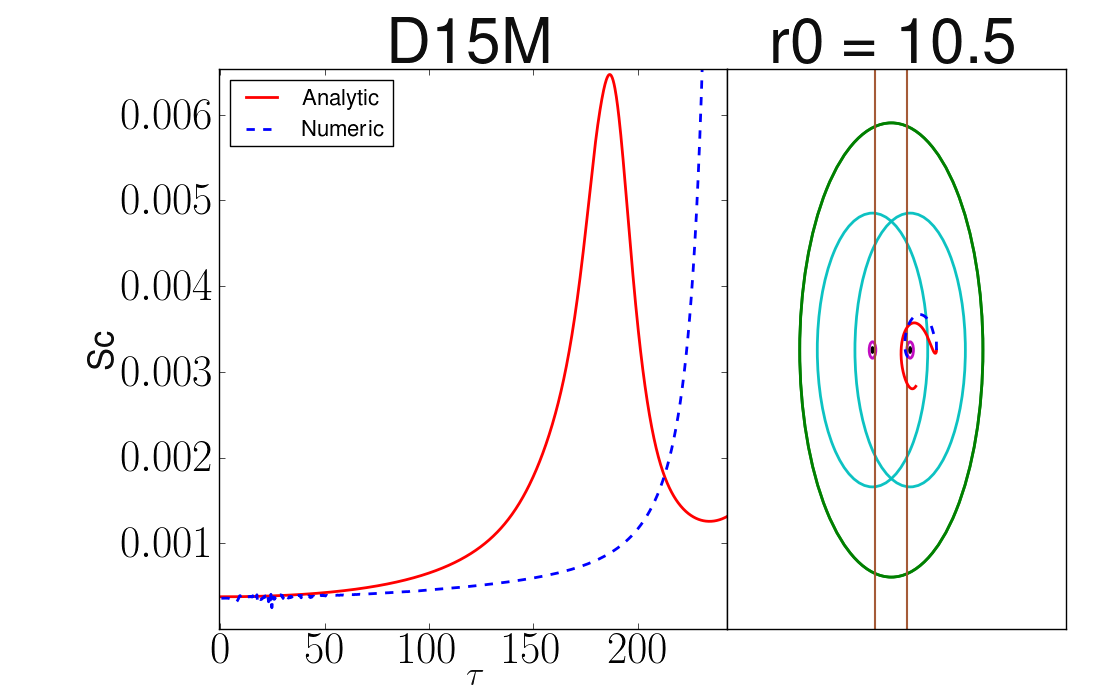}
\includegraphics[width=.24\textwidth]{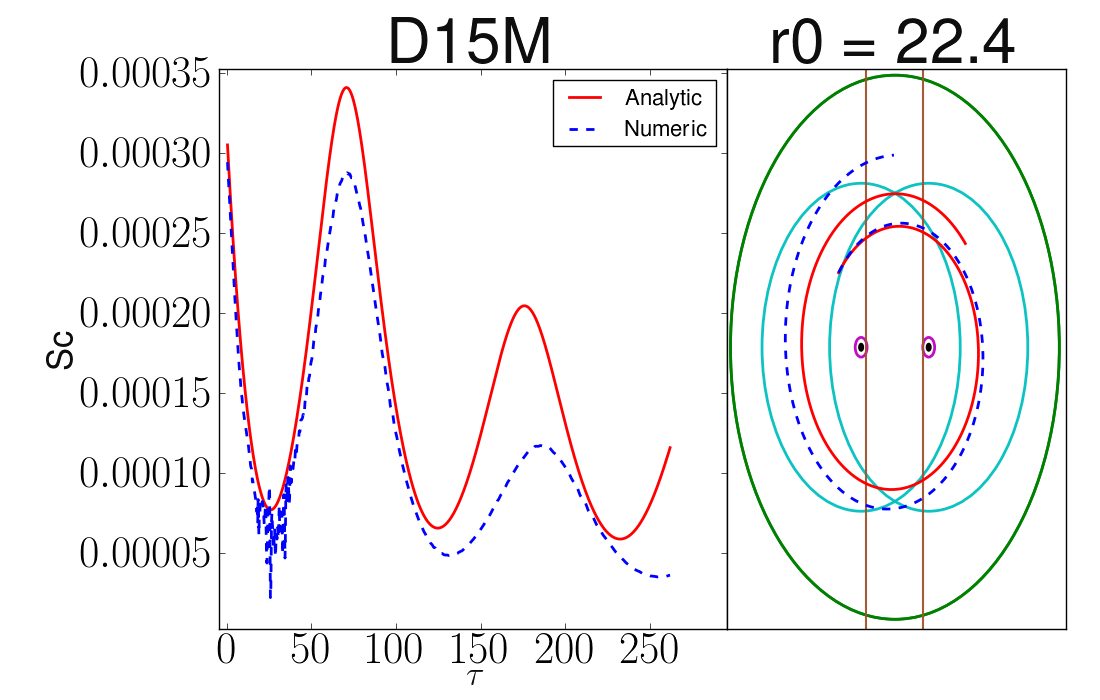}
\includegraphics[width=.24\textwidth]{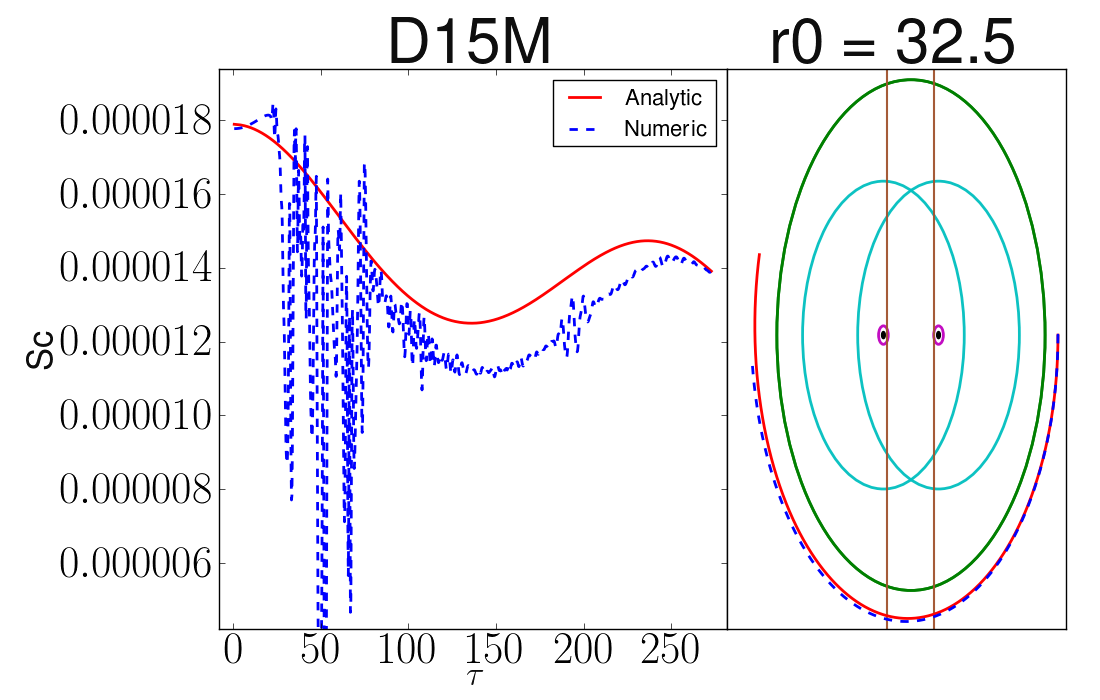}
\includegraphics[width=.24\textwidth]{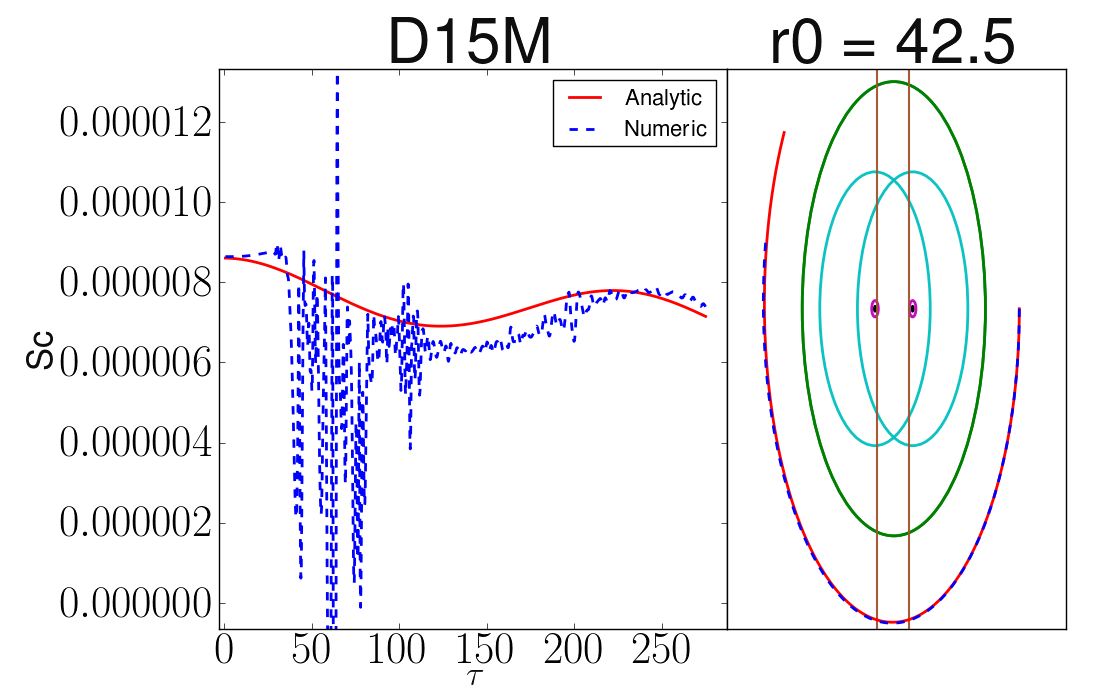}
\caption{A comparison of how well the curvature eigenvalues of the
  second-order metric and first-order metric agree with the
  eigenvalues of the associated numerical metrics for the $D=15M$ case.
  The top row shows
  the second-order results (which were previously shown in
  Fig.~\ref{fig:D15andD10}). The bottom row shows the first-order results. As with
  the $D=20M$ case, there are significant differences between the
  analytical and numerical scalars for the first-order metric even at
  larger distances from the black holes. 
}\label{fig:D151st}
\end{figure*}

\begin{figure*}
\includegraphics[width=.24\textwidth]{Revisedb10geo2.png}
\includegraphics[width=.24\textwidth]{Revisedb10geo6.png}
\includegraphics[width=.24\textwidth]{Revisedb10geo8.png}
\includegraphics[width=.24\textwidth]{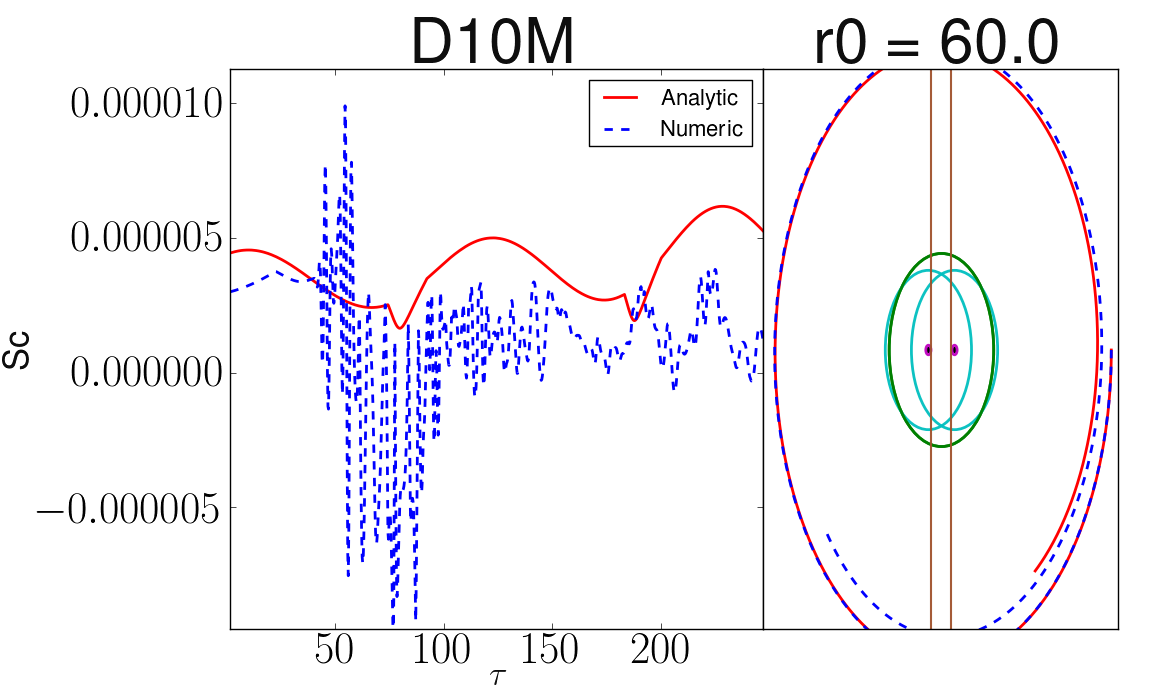}

\includegraphics[width=.24\textwidth]{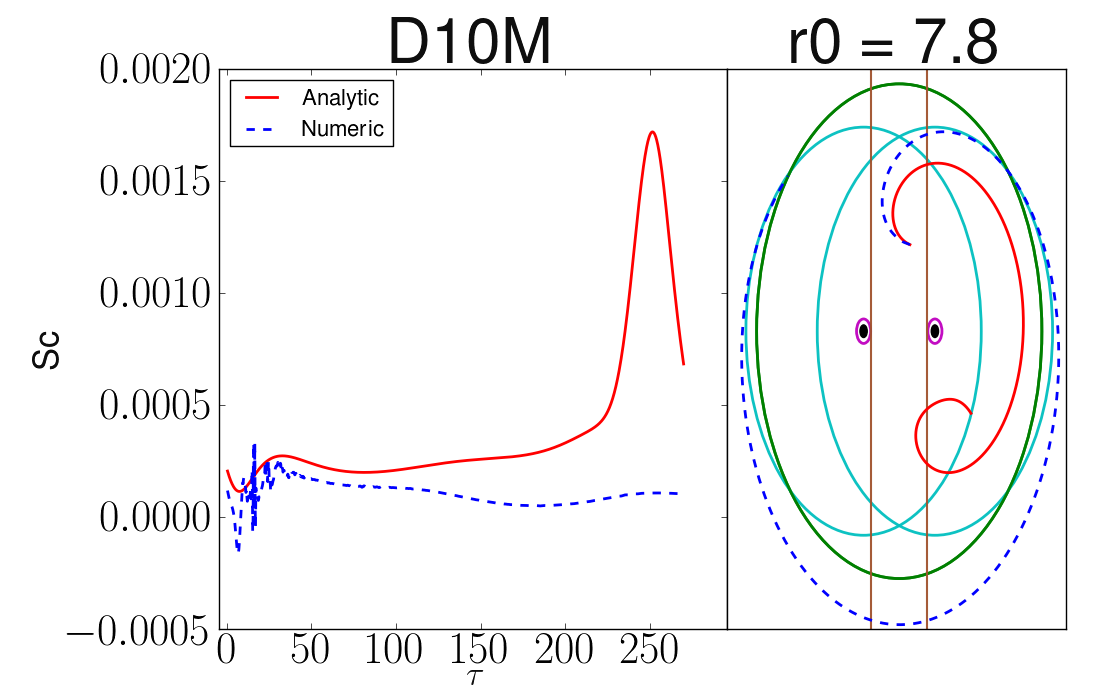}
\includegraphics[width=.24\textwidth]{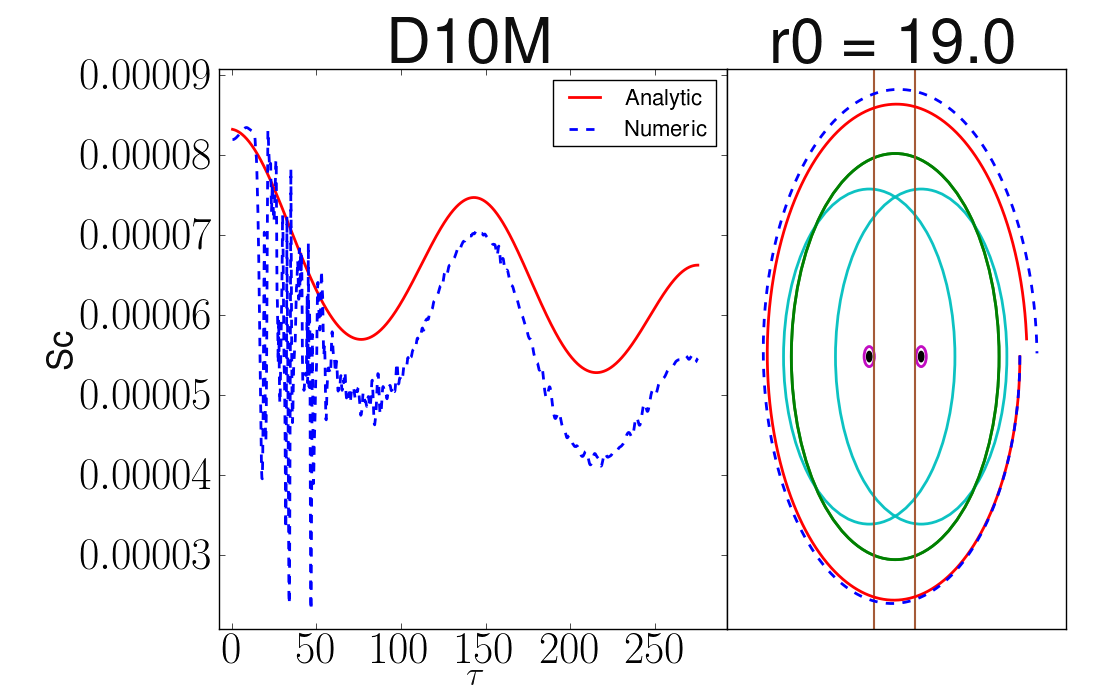}
\includegraphics[width=.24\textwidth]{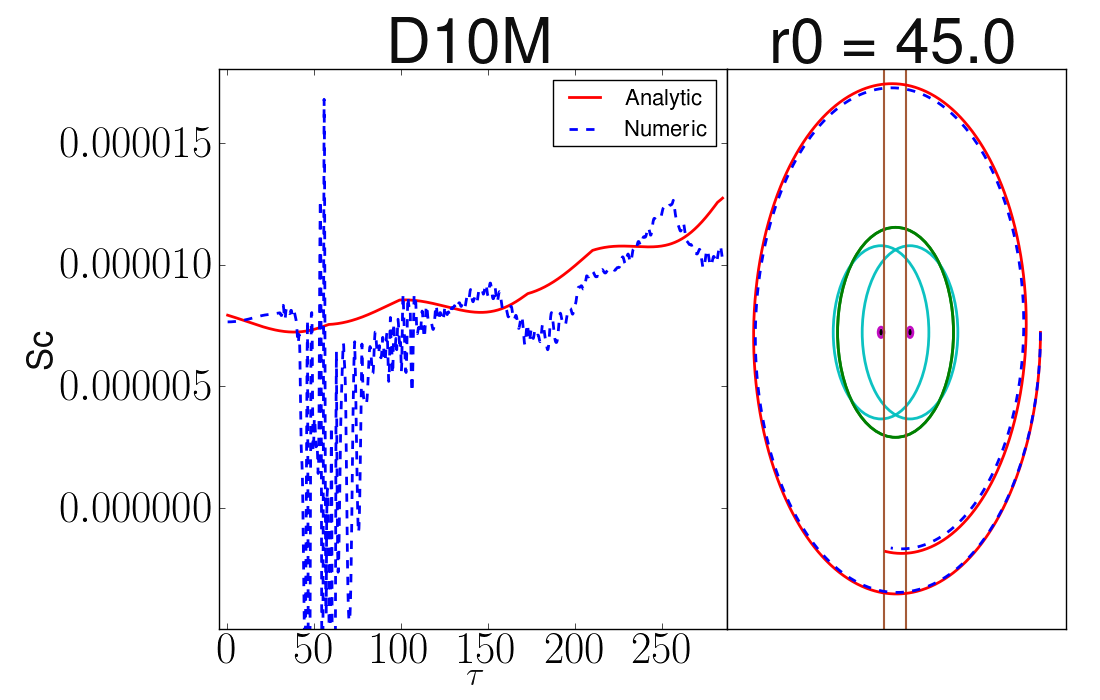}
\includegraphics[width=.24\textwidth]{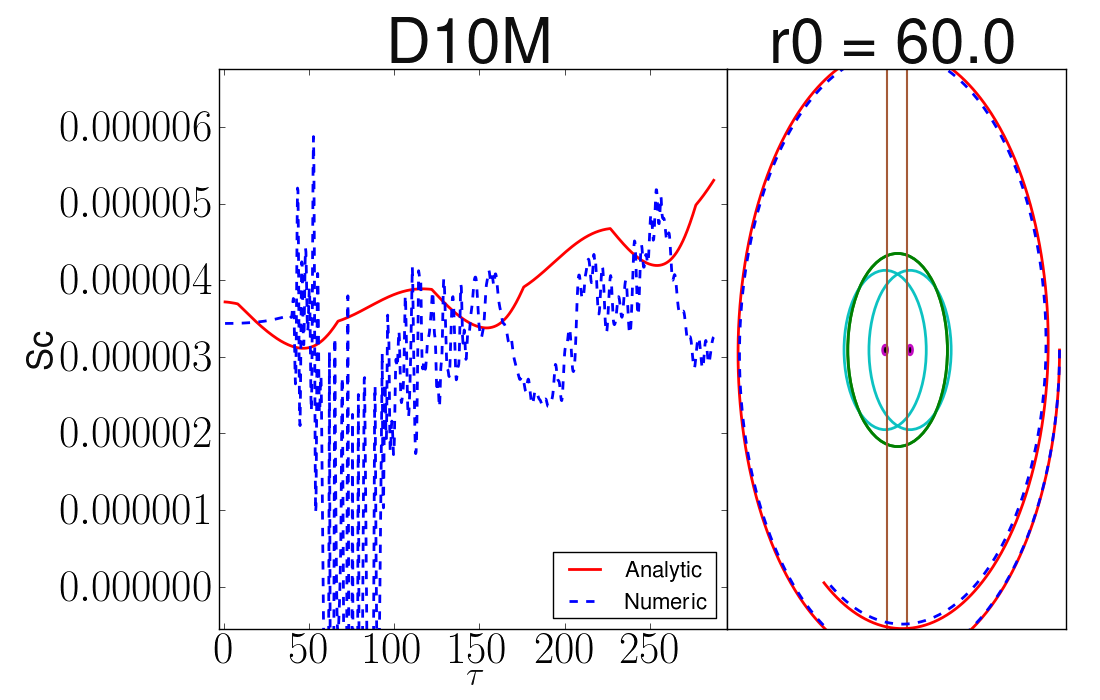}
\caption{A comparison of how well the curvature eigenvalues of the
  second-order metric and first-order metric agree with the
  eigenvalues of the associated numerical metrics for the $D=10M$ case.
  The top row shows
  the second-order results (which were previously shown in
  Fig.~\ref{fig:D15andD10}). The bottom row shows the first-order results. 
  Here, we do not see a significant improvement of the second-order
  metric over the first-order one.
}\label{fig:D101st}
\end{figure*}

Thus, we expect the second-order metric to be superior to the
first-order one. In this section, we repeat our calculations comparing
analytic metric to numerically evolved ones, but this time using the
first-order analytic metric. Again, we plot results for $D=50M$,
$D=20M$, $D=15M$, and $D=10M$ in Figs.~\ref{fig:D501st},
\ref{fig:D201st}, \ref{fig:D151st}, and \ref{fig:D101st}. As in the
sections above, we
compare the curvature eigenvalues for from the analytic metric with
the eigenvalues obtained by numerically evolving the analytic metric
using the CCZ4 system. Thus we compare the second-order analytic
eigenvalues with those obtained by numerically evolving the
second-order metric and compare the first-order analytic eigenvalues
with those obtained by numerically evolving the first-order metric.  As expected,
for $D\geq15M$, the second-order curvature eigenvalues more closely
match the associated numerical ones than the first-order eigenvalues
do.
For both metrics, the general
trend for $D\geq 15M$ is that the first and second order results both
become better at larger distances from the black holes and larger
black-hole separations. The $D=10M$ results appear to be equally
inaccurate for the first and second-order metrics. Our method is thus
able to distinguish between a lower-accuracy and a higher-accuracy
metric. Thus, we expect it will be a useful testing ground for
developing still higher-accuracy analytic metrics.

\section{Discussion}\label{sec:discussion}

In order to conclude if there are important systematic differences
between the numerical and analytic metrics, we need to consider the
possibility that the analytical metric is better approximated by an
exact solution that does not agree exactly with the analytic metric on
$\Sigma_0$. In such a case, one would expect that the appropriate
initial conditions for the geodesics in the exact spacetime are not
identical to those for the analytic one. But since small perturbations
in the initial conditions of a geodesic can lead to significant
differences on secular timescales (e.g., fall into one BH or the
other, bounded versus unbounded, etc.), we considered here only
geodesics that did not fall into the BHs or escape to large radii.

To see how small differences in the initial affect the geodesics we
presented above (i.e., the stability of the above geodesics), we
perturbed the initial velocities of a set of included geodesics by up
to 10\%. The results several geodesics are shown in
Fig.~\ref{fig:stability_test} for the $D=20M$ case.
We find that the effect of a $\sim 10\%$ perturbation is
smaller for the farther out geodesics. We also find that a
perturbation of $\lesssim 1\%$ seems to be sufficient to get
reasonable agreement between the geodesics in the numerical and 
analytic spacetimes for geodesics farther than $r_0 \sim 10M$ from the
BHs. However, for the closer geodesic, the agreement
is much poorer than for further out ones, which matches the general
trend seen in Fig.~\ref{fig:D25andD20}. Indeed, for the $r_0=6M$ case,
no perturbation is able to reproduce the behavior of the numerical
geodesic past $\tau \sim 75M$. To further support the argument that
small differences in the scalars can be removed by small changes
on the initial conditions of the geodesics but large differences
cannot be removed, we examined geodesics a distance of $\sim15M$ from
the black holes for the $D=15M$  and $D=10M$ case. Here we see that
no perturbation of the $D=10M$ geodesic's initial conditions will lead
to qualitative agreement between the analytical and numerical
eigenvalues. On the other hand, for the $D=15M$ very good agreement is
achieved.

The fact that reasonable agreement between the analytical and
numerical eigenvalues can be achieved by perturbing the analytical
geodesics indicates a limitation of our basic method in that it may
overemphasize the differences between to similar spacetimes. Large
differences in the eigenvalues, like the ones seen in the $D=20M$ case
near the BHs (and $D=10M$ everywhere) seem to be indicative of
significant differences between the two spacetimes.
On the other hand, where the differences are small, a
given geodesic in one spacetime may behave nearly identically to one
in the other, just with slightly different initial conditions.
Consequently, one may expect that small differences in the eigenvalues
will have little effect on, among others, gas dynamics.

One final note concerns the potential usefulness of our analysis at
late times. The issue is that small differences in the trajectories
generally grow on secular timescales. Thus the numerical and
analytical eigenvalues represent curvature terms at increasingly
different points of the spacetimes. For example,
in the $D=25M$ case (see Fig.~\ref{fig:D25andD20})
for the farthest geodesics, we see differences between the numerical
and analytic eigenvalues after about $\tau=400M$. From this point on,
the geodesics will start taking different paths, and the scalars will
disagree more and more, even though the two spacetimes are quite
close, as is evident by the early time agreement of the scalars and
the fact that the geodesics do not get significantly closer to either
black hole.  At
close separations, these effects are larger and happen earlier. For
example the $r_0=17.5M$ case shows significant deviations after
$\tau=250M$, and the $r_0=9.5M$ shows significant differences after
$\tau=75M$. The net effect is,  the
closer the two spacetimes  are to each other (in the vicinity of the geodesic), the longer in time
the analysis is valid.

\section{Conclusion}

In this paper, we introduced a new method for comparing the geodesic
dynamics of two spacetimes. We used this method to compare the
dynamics of recently developed analytical metrics that approximate the
metric from an inspiraling black hole binary with fully nonlinear
numerical evolutions of the Einstein equations. We find that the agreement
in the dynamics between the two spacetimes is generally better for
more separated binaries. Close to the black holes, as one might
expect, we see the largest differences. Interestingly, we see that
these differences scale in a highly nonlinear way with separation,
with the $D=10M$ spacetime showing much larger differences than the
$D=15M$ one. On the other hand, even for the $D=50M$ case, there are
measurable differences in the geodesic deviation scalars between the
analytical and numerical spacetimes for geodesics farther than
$r_0 \sim100M$ from the black holes.

\begin{acknowledgments}
We thank Manuela Campanelli for many valuable discussions.
JS acknowledges support from the Fulbright PhD Program.
YZ
gratefully acknowledges the National Science Foundation (NSF) for
financial support from Grants No.\
PHY-1607520, No.\ PHY-1707946, No.\ ACI-1550436, No.\ AST-1516150,
No.\ ACI-1516125, No.\ PHY-1726215. 
HN acknowledges support from JSPS KAKENHI Grant No.\ JP16K05347 and No.\ JP17H06358.
This work used the Extreme
Science and Engineering
Discovery Environment (XSEDE) [allocation TG-PHY060027N], which is
supported by NSF grant No. ACI-1548562.
Computational resources were also provided by the NewHorizons and
BlueSky Clusters at the Rochester Institute of Technology, which were
supported by NSF grants No.\ PHY-0722703, No.\ DMS-0820923, No.\
AST-1028087, and No.\ PHY-1229173.
\end{acknowledgments}

\bibliographystyle{apsrev4-1}
\bibliography{references}
\end{document}